\documentstyle[preprint,eqsecnum,aps]{revtex}

\def\addcontentsline#1#2#3{\relax}

\begin{document}
\draft
\title{
Exact ground-state correlation functions\\
of the one-dimensional strongly correlated electron models\\
with the resonating-valence-bond ground state
}
\author{Masanori Yamanaka$^1$,
Shinsuke Honjo$^1$, Yasuhiro Hatugai$^2$, and Mahito Kohmoto$^1$}
\address{
$^1$Institute for Solid State Physics, University of Tokyo,
7-22-1, Roppongi, Minato-ku, Tokyo 106, Japan
}
\address{
$^2$Department of Applied Physics, University of Tokyo,
7-3-1, Hongo, Bunkyo-ku, Tokyo 113, Japan
}
\maketitle
\begin{abstract}
We investigate the one-dimensional strongly correlated electron models
which have the resonating-valence-bond state as the exact ground state.
The correlation functions are evaluated exactly
using the transfer matrix method
for the geometric representations
of the valence-bond states.
In this method,
we only treat matrices with small dimensions.
This enables us to give analytical results.
It is shown that the correlation functions decay
exponentially with distance.
The result suggests that there is a finite excitation gap,
and that the ground state is insulating.
Since the corresponding non-interacting systems may be
insulating or metallic,
we can say that the gap originates from strong correlation.
The persistent currents of the present models are also investigated
and found to be exactly vanishing.
\end{abstract}

\small

\newpage
\section{Introduction}
\label{sec:intro}

Strongly interacting electron systems have been always
one of the most important subjects in condensed matter physics.
Although rigorous results and exact solutions
are useful, they are rare.
Recently, Brandt and Giesekus \cite{BG92} introduced
model of strongly interacting electrons
on $d$-dimensional ($d \geq 2$) perovskite-like lattices
in which the exact ground-state wave functions
were obtained for a certain range of the parameters.
Mielke \cite{Mi92} showed that the exact ground state can be obtained
in similar models on a general class of line graphs.
Following the line of Brandt and Giesekus,
models in which the exact ground state can be obtained were constructed
by several authors \cite{St93,Ta93a,Ta93b}.
These models are conveniently described
by the cell construction of Tasaki \cite{Ta93a}
which will be reviewed in Sec.\ \ref{sec:ham}
(also see Appendix\ \ref{app:models} and Ref.\ \cite{Ta93b}).
Tasaki \cite{Ta93b} proved the uniqueness of the ground states
in this class of models.
Not only the ground state but also the singlet-pair correlation function
in a model on a tree was obtained \cite{Ta93a}.
Bares and Lee \cite{BL93} performed a detailed analysis
for one of the models of Strack \cite{St93}.
They proved the uniqueness of the ground state
and exactly evaluated the equal-time correlation functions
by a transfer matrix method.

It was pointed out \cite{Ta93a,BL93} that the exact ground states
have the resonating-valence-bond (RVB) structure \cite{LP49,PWA74}.
It is the so-called hopping-dominated RVB states \cite{TK92}
which is different from the tunneling-dominated RVB states.
The latter have been studied intensively
in connection to the high $T_c$ superconductivity
\cite{PWA87,ABZH87,IC87,BS88,LDA88,RK88,MK88}.
Tasaki and Kohmoto \cite{TK92} studied the difference of the mechanism
that causes the resonance in the hopping-dominated RVB states
and the tunneling-dominated RVB states.

In this paper
we shall exactly evaluate the equal-time correlation functions
of one-dimensional models (Model A, B, and C)
which will be defined in Sec.\ \ref{sec:corr}.
One of the models of Strack which was studied by Bares and Lee
is called Model B in this paper.
We shall use the transfer matrix method
for the geometric representations
of the valence-bond states \cite{BS88,MK88,KSHMS,FM88},
which is different from that of Bares and Lee.
In one dimension, the extension of the formalism to other models
which will not be included in this paper is cumbersome
but essentially straightforward.
It is crucial that we do not have to treat large matrices
as needed in the method of Bares and Lee.
(In one of the models by Strack, for example, we only treat 3$\times$3
transfer matrices, while Bares and Lee needed those of 16$\times$16.)
This enables us to give the completely analytical solution
with finite amount of efforts.

It is shown that all the correlation functions decay
exponentially with distance.
The result suggests that the existence of a finite excitation gap.
It is expected that
the excitation gap is originated from the structure of the ground state
which is described by a collection of local spin singlets.
The filling factor of the ground state corresponds to
that of a band insulating state or metallic one
in the non-interacting system.
The properties of the ground state and the gap are
completely different from that in the non-interacting system.
The existence of the excitation gap is expected to be
a general feature of this class of models.
In a certain range of parameters,
Model B includes one of the models by Strack \cite{St93,BL93},
where we reproduce the results of Bares and Lee.
In a limit of parameters,
it corresponds to a kind of the Kondo lattice regime
in the sense that there are one localized electron
and one conduction electron per a unit cell,
where the ground state is described by a collection of a local
singlet between them.
The persistent currents \cite{BY61,BIR,LDDB,CWBKGK,MCB}
are also calculated
and turned out to be vanishing.

The plan of this paper is as follows:
In Sec.\ \ref{sec:ham}, we review the cell construction of the models.
In Sec.\ \ref{sec:georep} we describe the geometric representation
of the correlation functions in arbitrary dimensions.
In Sec.\ \ref{sec:corr},
we perform a detailed analysis of the one-dimensional models
using the method in Sec.\ \ref{sec:georep}.
In Sec.\ \ref{sec:persistent}, the absence of the persistent currents
is shown.
Section \ref{sec:summary} is a summary.
The reader who is interested only in the physical results
may take a look at Sec.\ \ref{sec:ham} and \ref{sec:summary}
for general properties of systems in one dimension,
and then read Sec.
\ref{sec:modelahami}, \ref{subsec:modeladis},
\ref{sec:modelbhami}, \ref{subsec:modelbdis},
\ref{sec:modelchami}, and \ref{subsec:modelcdis}
as examples.
The results for the correlation functions are shown
in each end of subsubsection in Sec.\ \ref{sec:corr}.

\newpage
\section{Cell Construction and the Ground State}
\label{sec:ham}

Let us first introduce the solvable models in arbitrary dimension
by following the construction
in Refs.\ \cite{Ta93a} and \cite{Ta93b}.
In the present paper,
we only consider the translation invariant lattices\footnote{
\footnotesize
We can construct more generalized models
whose lattices are not translation invariant.
See Ref.\ \cite{Ta93b} for such models.
}.
The lattice is constructed from identical cells $C_n$
with $n=1$, $2$, $\cdots$, $N$,
where $N$ is the number of the cells.
The cell $C_n$ is a finite set of sites,
where each site $r \in C_n$
($r$ $=1,2,$ $\cdots$, $R$, where\footnote{\footnotesize
Throughout the present paper,
$\left\vert S \right\vert$ denotes the number of elements in a set $S$.
}
$R$ $= \vert C_n \vert$)
is either
a site with infinitely large on-site Coulomb repulsion
(a $U = \infty$ site or a $d$-site)
which can have at most one electron, or a $U = 0$ site
(or a $p$-site) which can have at most two electrons
with opposite spins.
The full lattice $\Lambda_N$ ($ = \cup_{n=1}^N C_n$)
is constructed by starting from
the lattice $\Lambda_1=C_1$,
and adding cells $C_2$, $C_3$, $\cdots$, $C_N$ successively.
When we add a new cell $C_{n+1}$ to the lattice $\Lambda_n$
($ = \cup_{i=1}^n C_i$),
we identify some of sites in $C_{n+1}$ with sites in $\Lambda_n$
in a one-to-one manner.
We note that a cell is not a unit cell
in the sense of crystalliography.
We denote sites in the full lattice $\Lambda_N$
by $x$ ($x$ $=1,2,$ $\cdots$, $\left\vert \Lambda_N \right\vert$).
A site $x$ may belong to several different cells.
The correspondence between $x$ ($\in \Lambda_N$)
and $r$ ($\in C_n$)
is given by
\begin{equation}
x=f(n,r),
\label{eq:sitemap}
\end{equation}
where $f(n,r)$ depends on the model under consideration.
See Fig.\ \ref{fig:sitemap} for the correspondence.

For a cell $C_n$, we associate a cell Hamiltonian
\begin{equation}
H_n= \sum_{\sigma = \uparrow, \downarrow}
\alpha_{n,\sigma}^{\phantom{\dagger}} {\cal P}_n^{\phantom{\dagger}}
\alpha_{n,\sigma}^{\dagger},
\label{eq:cellham}
\end{equation}
with
\begin{equation}
\alpha_{n,\sigma}^{\phantom{\dagger}}
= \sum_{r=1}^{R}
\lambda_r^{(n)} c_{r,\sigma}^{\phantom{\dagger}},
\label{eq:alpha}
\end{equation}
where $\lambda_r^{(n)}$
are nonvanishing complex coefficients\footnote{\footnotesize
When a magnetic field is applied, $\lambda_r^{(n)}$ are complex.
The proof of the uniqueness of the ground state
in Ref.\ \cite{Ta93b} holds also in this case.
}
and are chosen independently in each cell\footnote{\footnotesize
When a site $x$ belongs to more than one cells,
$\lambda_x^{(n)}$ can be chosen independently in each cell.
}.
(In Sec.\ \ref{sec:persistent},
we can impose the twisted boundary condition
by making use of this property.
In Secs.\ \ref{sec:georep} and \ref{sec:corr},
we only consider cases where all the cells have the same $\lambda_r$.
Thus we have translationally invariant systems
and we drop the suffix $n$ in $\lambda_r^{(n)}$ there.)
Here $c_{r,\sigma}^{\phantom{\dagger}}$ and $c_{r,\sigma}^{\dagger}$
are the annihilation and the creation operators, respectively,
of an electron at site $r$
with spin $\sigma =\uparrow, \downarrow$.
They satisfy the standard anticommutation relations
$\{ c^{\dagger}_{r, \sigma}, c^{\phantom{\dagger}}_{s, \tau} \}$
$=\delta_{r,s}\delta_{\sigma, \tau}$ and
$\{ c^{\dagger}_{r, \sigma}, c^{\dagger}_{s, \tau} \}$
$=\{ c_{r, \sigma}, c_{s, \tau} \}$
$=0$.
The projection operator which eliminates a double occupancy
on $d$-sites is
\begin{equation}
{\cal P}_n = \prod_{r \in C_{n; U=\infty}}
\left(1 - n_{r,\uparrow} n_{r,\downarrow}\right),
\label{eq:projec}
\end{equation}
where $C_{n; U=\infty}$ is the set of $U=\infty$ sites in $C_n$
and $n_{r,\sigma} = c_{r,\sigma}^{\dagger}
c_{r,\sigma}^{\phantom{\dagger}}$
is the number operator.
This represents the infinitely large on-site Coulomb repulsion.
The full Hamiltonian is
\begin{equation}
H = \sum_{n=1}^N H_n.
\label{eq:ham1}
\end{equation}

We rewrite this Hamiltonian into the ``standard form''. From
the identities \cite{BG92,Ta93b}
\begin{eqnarray}
\begin{array}{@{\,}ll}
c^{\phantom{\dagger}}_{r, \sigma}
{\cal P}_n^{\phantom{\dagger}}
c^{\dagger}_{s, \sigma}
=
-{\cal P}_n^{\phantom{\dagger}}
c^{\dagger}_{s, \sigma}
c^{\phantom{\dagger}}_{r, \sigma}
{\cal P}_n^{\phantom{\dagger}}
&
\mbox{for $r \ne s$}
\\
c^{\phantom{\dagger}}_{r, \sigma}
{\cal P}_n^{\phantom{\dagger}}
c^{\dagger}_{r, \sigma}
=
{\cal P}_n
(1-n_{r, \uparrow}-n_{r, \downarrow})
{\cal P}_n
&
\mbox{for $r \in C_{n; U=\infty}$}
\\
c^{\phantom{\dagger}}_{r, \sigma}
{\cal P}_n^{\phantom{\dagger}}
c^{\dagger}_{r, \sigma}
=
{\cal P}_n
(1-n_{r, \sigma})
{\cal P}_n
&
\mbox{for $r \in C_{n; U=0}$},
\end{array}
\end{eqnarray}
where $C_{n; U=0}$ is the set of $U=0$ sites in $C_n$,
the cell Hamiltonian (\ref{eq:cellham}) is
\begin{eqnarray}
H_n
&=& \sum_{\sigma = \uparrow, \downarrow}
    \sum_{r,s \in C_n}
    \lambda_s^{(n)}
    \left( \lambda_r^{(n)} \right)^*
    c^{\phantom{\dagger}}_{s, \sigma}
    {\cal P}_n^{\phantom{\dagger}}
    c^{\dagger}_{r, \sigma}
\nonumber \\
&=& {\cal P}_n
    \left\{
    \sum_{r \in C_n} 2 \left\vert \lambda_r^{(n)} \right\vert^2
   -\left[
    \sum_{\sigma = \uparrow, \downarrow}
    \sum_{r\ne s(\in C_n)}\left(\lambda_r^{(n)}\right)^* \lambda_s^{(n)}
    c^{\dagger}_{r, \sigma}
    c^{\phantom{\dagger}}_{s, \sigma}
\right.\right.\nonumber \\
& &\hspace{15mm}\left.\left.
   +\sum_{\sigma = \uparrow, \downarrow}
    \sum_{r \in C_{n; U=\infty}}
    2 \left\vert \lambda_r^{(n)} \right\vert^2
    c^{\dagger}_{r, \sigma}
    c^{\phantom{\dagger}}_{r, \sigma}
   +\sum_{\sigma = \uparrow, \downarrow}
    \sum_{r \in C_{n; U=0}}
    \left\vert \lambda_r^{(n)} \right\vert^2
    c^{\dagger}_{r, \sigma}
    c^{\phantom{\dagger}}_{r, \sigma}
    \right]
    \right\}
    {\cal P}_n
\nonumber \\
&=& {\cal P}_n
    \left\{
    E_n
   -\sum_{\sigma = \uparrow, \downarrow}
    \sum_{r,s \in C_n}
    t_{r,s}^{(n)}
    c^{\dagger}_{r, \sigma}
    c^{\phantom{\dagger}}_{s, \sigma}
    \right\}
    {\cal P}_n,
\end{eqnarray}
where
\begin{eqnarray}
t_{r,s}^{(n)}
&=&\left \{ \begin{array}{@{\,}ll}
   \left( \lambda_r^{(n)} \right)^* \lambda_s^{(n)}
          & \mbox{for $r \ne s$}  \\
   2 \left\vert \lambda_r^{(n)} \right\vert^2
          & \mbox{for $r=s$ and $r,s \in C_{n; U=\infty}$} \\
     \left\vert \lambda_r^{(n)} \right\vert^2
          & \mbox{for $r=s$ and $r,s \in C_{n; U=0}$} \\
      \end{array} \right.\\
E_n &=& \sum_{r \in C_n} 2 \left\vert \lambda_r^{(n)} \right\vert^2.
\end{eqnarray}From (\ref{eq:ham1}), we have
\begin{equation}
H = - E_0 - {\cal P}
\sum_{\sigma = \uparrow, \downarrow}
\sum_{x,y \in \Lambda_N}
t_{x,y}^{\phantom{\dagger}}
c_{x,\sigma}^{\dagger}
c_{y,\sigma}^{\phantom{\dagger}}{\cal P},
\label{eq:hamstandard}
\end{equation}
where
\begin{eqnarray}
t_{x,y} &=& \sum_{n=1}^N t_{r,s}^{(n)}\ \ \ \
\mbox{for $x=f(n,r)$, $y=f(n,s)$}\\
E_0 &=& - \sum_{n=1}^N E_n,
\label{eq:staham}
\end{eqnarray}
and
\begin{eqnarray}
{\cal P} &=& \prod_{n=1}^N {\cal P}_n.
\label{eq:projec2}
\end{eqnarray}

It was shown in Refs.\ \cite{Ta93a,Ta93b}
that for the electron number, $2N$,
the unique ground state of the Hamiltonian (\ref{eq:ham1})
or (\ref{eq:hamstandard})
has zero energy and is given by
\begin{equation}
\Big\vert \Phi_{G.S.} \Big\rangle =
{\cal P}
\prod_{\sigma = \uparrow, \downarrow} \prod_{n=1}^N
\alpha_{n, \sigma}^{\dagger}
\Big\vert 0 \Big\rangle,
\label{eq:gs}
\end{equation}
where $\vert 0 \rangle$ is the vacuum state.
As we will see in (\ref{eq:gs2}),
we have a single valence bond in each cell.
Therefore, the filling of the ground state is
$1/N_a$, where $N_a$ is the number of sites in the unit cell.
In Secs.\ \ref{sec:corr} and \ref{sec:persistent},
we use the second term in (\ref{eq:hamstandard})
as the Hamiltonian and denote it by $H_S$.
The unique ground state of the Hamiltonian $H_S$
is given by (\ref{eq:gs}) with the energy $E_0$ in (\ref{eq:staham}).

\newpage
\section{Geometric Representation of the Correlation Functions}
\label{sec:georep}

We describe the geometric representation
of the norm of the ground state and the correlation functions
which was formulated by Tasaki \cite{Ta94} in arbitrary dimensions.
The geometric representation of the norm is described
in Refs. \cite{Ta93a,TK92,Taup1} for the lattice composed of $d$-sites
only.
Here, we have the lattice with both $p$- and $d$-sites.
The ground state (\ref{eq:gs}) can be written
\begin{equation}
\Big\vert \Phi_{G.S.} \Big\rangle
={\cal P}
 \prod_{n=1}^N
 \sum_{r,s \in C_n}
 \lambda_r^{\ast} \lambda_s^{\ast}
 c_{r,\uparrow}^{\dagger} c_{s,\downarrow}^{\dagger}
 \Big\vert 0 \Big\rangle
={\cal P}
 \prod_{n=1}^N
 \sum_{r \leq s \in C_n}
 \lambda_{rs}^{\ast} b_{r,s}^{\dagger}
 \Big\vert 0 \Big\rangle,
\label{eq:gs1}
\end{equation}
where
\begin{eqnarray*}
b_{r,s}^{\dagger}
&=& \left \{ \begin{array}{@{\,}ll}
c_{r,\uparrow}^{\dagger} c_{s,\downarrow}^{\dagger}
+c_{s,\uparrow}^{\dagger} c_{r,\downarrow}^{\dagger}
          & \mbox{for $r \ne s$}  \\
c_{r,\uparrow}^{\dagger} c_{s,\downarrow}^{\dagger},
          & \mbox{for $r=s$, } \\
      \end{array} \right.
\end{eqnarray*}
and
\begin{eqnarray*}
\lambda_{rs}^{\ast}&=&\lambda_r^{\ast} \lambda_s^{\ast}.
\end{eqnarray*}
The operator $b_{r,s}^{\dagger}$ is the creation operator
of the valence bond (i.e. a singlet pair) between sites $r$ and $s$
if $r \ne s$.
It creates a doubly occupied site if $r=s$.
They obey the commutation relations
\begin{equation}
[ b_{r,s}, b_{t,u}]
=[b_{r,s}^{\dagger},b_{t,u}^{\dagger}]
=[b_{r,s}^{\dagger},b_{t,u}^{\phantom{\dagger}}]=0,
\label{eq:bose}
\end{equation}
where $r$, $s$, $t$, and $u$ are different sites one another.
This operator satisfies the relation $b_{r,s}^{\dagger}=b_{s,r}^{\dagger}$
and the ground state (\ref{eq:gs1})
is a hopping-dominated RVB state
according to the terminology of \cite{TK92}.
It is different from the tunneling-dominated RVB states
\cite{TK92,PWA87,ABZH87,IC87,BS88,LDA88,RK88,MK88}.

Now we rewrite the ground state (\ref{eq:gs1}) in a convenient form
for diagrammatic evaluations of the norm and the correlation functions.
The diagrammatic method was first introduced
by Rumer \cite{Ru32} and Pauling \cite{Pa33}.
We denote a valence bond by $\{x,y\}$, and a doubly occupied site by
$\{x,x\}$ which is regarded as a self-closed bond
(Fig.\ \ref{fig:valencebond}).
Since a self-closed bond is actually a doubly occupied site,
it is allowed only at $p$-sites.
Let a valence-bond configuration $V$ be a set of $N$ bonds constructed
by choosing a single bond from each cell.
We show examples in Figs.\ \ref{fig:bondconfig}(a) and (b).
The bonds do not share a $d$-site
since it can have at most one electron.
A $p$-site is shared by at most two bonds.
In this way, the projection ${\cal P}$ defined
by (\ref{eq:projec}) and (\ref{eq:projec2})
is automatically taken into account.
We denote by ${\cal V}$
the set of all the possible valence-bond configurations.
The ground state (\ref{eq:gs1}) is rearranged, and written as
\begin{equation}
\Big\vert \Phi_{G.S.} \Big\rangle
=\sum_{V \in {\cal V}}
 \prod_{\{x,y\}\in V}
 \lambda_{x,y}^{\ast} b_{x,y}^{\dagger}
 \Big\vert 0 \Big\rangle.
\label{eq:gs2}
\end{equation}

\subsection{Norm of the ground state}
\label{subsec:normgs}From (\ref{eq:gs2}), the norm of the ground state is
\begin{eqnarray}
\langle \Phi_{G.S.} \vert\  \Phi_{G.S.} \rangle
=\sum_{V \in {\cal V}} \
 \sum_{V' \in {\cal V}}
 \lambda(V')\lambda^{\ast}(V)
 \Bigg\langle 0 \Bigg\vert
 \prod_{\{x',y'\}\in{V'}}
        b_{x',y'}^{\phantom{\dagger}}
 \prod_{\{x,y\}\in{V}}
        b_{x,y}^{\dagger}
 \Bigg\vert 0 \Bigg\rangle,
\label{eq:norm}
\end{eqnarray}
where $\lambda(V)=\prod_{\{x,y\}\in{V}}\lambda_{x,y}$.
Let us consider a graph $V \cup V'$ in the expectation value.
We call a bond which belongs to $V'$ as a ``bra-bond''
and one which belongs to $V$
as a ``ket-bond'' (Fig.\ \ref{fig:valencebond}).
We only consider graphs $V \cup V'$
in which numbers of ``bra-bonds'' and ``ket-bonds''
are equal at every sites.
(An example is shown in Fig.\ \ref{fig:bondconfig}(c).)
Otherwise, the expectation value is vanishing,
since the numbers of the creation and annihilation operators
are different at the site.

The graph $V \cup V'$ can be decomposed into connected subgraphs,
since the operators $b_{x',y'}$ and $b_{x,y}^{\dagger}$
commute (see (\ref{eq:bose})).
An example is shown in Fig.\ \ref{fig:bondconfig}(d)-(g).
We denote the number of the subgraphs by
$n \left( V \cup V' \right)$.
This decomposition is written
\begin{eqnarray}
V \cup V' = \sum_{i=1}^{n \left( V \cup V' \right)}
U_i \cup U'_i,
\end{eqnarray}
where $U_i \subset V$ and $U'_i \subset V'$.

By noting that $b_{x',y'}$ and $b_{x,y}^{\dagger}$ commute
with each other for distinct $x$, $y$, $x'$, and $y'$
(see (\ref{eq:bose})),
we find that the expectation value in (\ref{eq:norm}) can be
factorized into parts corresponding to connected subgraphs.
Thus, we have
\begin{equation}
\langle \Phi_{G.S.} \vert\  \Phi_{G.S.} \rangle
=\sum_{V,V' \in {\cal V}} \
\left\{
\prod_{i=1}^{n \left( V \cup V' \right)} \
       \lambda \left( U'_i \right)
       \lambda^{\ast} \left( U_i \right)
\Bigg\langle 0 \Bigg\vert
       \prod_{\{x',y'\}\in U'_i}
              b_{x',y'}^{\phantom{\dagger}}
       \prod_{\{x,y\}\in U_i}
              b_{x,y}^{\dagger}
\Bigg\vert 0 \Bigg\rangle
\right\}.
\label{eq:normg2}
\end{equation}
We note that each $U_i$, $U'_i$ in (\ref{eq:normg2})
depends on the whole configurations
$V$, $V'$, and $V \cup V'$.

It sometimes happen that two ``bra-bonds'' and two ``ket-bonds''
are connected to a single $p$-site in a connected subgraph
$U_i \cup U'_i$.
For our calculations, it is convenient to eliminate such sites.
This is done by using the identity
$b_{x,y}^{\dagger}b_{y,z}^{\dagger}
=-b_{y,y}^{\dagger}b_{x,z}^{\dagger}$
(see Fig.\ \ref{fig:elimination1}).
Examples of eliminations of such sites
are shown in Figs.\ \ref{fig:elimination2} diagrammatically.
The procedure in Fig.\ \ref{fig:elimination2}(a)
generates a minus sign.
We assign the sign to the non-closed bond.
We shall always apply this procedure hereafter
and it should be understood implicitly.
After this procedure, the subgraph $U_i \cup U'_i$ may
be decomposed into several graphs.
We denote the number of the graphs by
$n \left( U_i \cup U'_i \right)$.
The decomposition is unique and is written
\begin{eqnarray}
U_i \cup U'_i \to
\sum_{j=1}^{n \left( U_i \cup U'_i \right)} W_j \cup W'_j.
\label{eq:normw}
\end{eqnarray}
The arrow indicates that $W_j$, $W'_j$ are not necessarily
subsets of $U_i \cup U'_i$.
We only have three kinds of graphs $W_j \cup W'_j$:
self-closed bonds (Fig.\ \ref{fig:elimination3}(a)),
degenerate loops which consist of a pair of bonds
(Fig.\ \ref{fig:elimination3}(b)),
and non-degenerate loops which consist of even number
of distinct bonds
(Fig.\ \ref{fig:elimination3}(c)).
We call the graph $W_j \cup W'_j$ loop. From
the decomposition (\ref{eq:normw}),
we have
\begin{equation}
\Bigg\langle 0 \Bigg\vert
       \prod_{\{x',y'\}\in U'_i}
              b_{x',y'}^{\phantom{\dagger}}
       \prod_{\{x,y\}\in U_i}
              b_{x,y}^{\dagger}
\Bigg\vert 0 \Bigg\rangle
=\prod_{j=1}^{n \left( U_i \cup U'_i \right)}
(-1)^{m_j}
 w_j,
\label{eq:www}
\end{equation}
where $m_j$ is the number of the procedures
shown in Fig.\ \ref{fig:elimination2}(a) and
\begin{equation}
w_j
=\Bigg\langle 0 \Bigg\vert
 \prod_{\{x',y'\}\in W'_j}
        b_{x',y'}^{\phantom{\dagger}}
 \prod_{\{x,y\}\in W_j}
        b_{x,y}^{\dagger}
 \Bigg\vert 0 \Bigg\rangle.
\label{eq:wji}
\end{equation}
We call it weight. The value is classified to
\begin{equation}
w_j = \left \{ \begin{array}{@{\,}ll}
            1
& \mbox{if $W_j \cup W'_j$ is a pair of the self-closed bonds
(Fig.\ \ref{fig:elimination3}(a))}\\
            2
& \mbox{if $W_j \cup W'_j$ is a degenerate loop
(Fig.\ \ref{fig:elimination3}(b))}\\
            2(-1)^{{l_j}/2-1}
& \mbox{if $W_j \cup W'_j$ is a non-degenerate loop
(Fig.\ \ref{fig:elimination3}(c)),}\\
      \end{array} \right.
\label{eq:proof1}
\end{equation}
where $l_j$ is the number of the bonds in the $j$-th
non-degenerate loop. These weights are derived in Appendix B.
Now the norm of the ground state (\ref{eq:normg2}) is written
\begin{eqnarray}
\langle \Phi_{G.S.} \vert\  \Phi_{G.S.} \rangle
&=&
\sum_{V,V' \in {\cal V}} \
\left\{
\prod_{i=1}^{n \left( V \cup V' \right)}
\left[
\prod_{j=1}^{n \left( U_i \cup U'_i \right)}
       \lambda \left( W'_j \right)
       \lambda^{\ast} \Big( W_j \Big)
(-1)^{m_j}
 w_j
\right]
\right\}.
\label{eq:norm3a}
\end{eqnarray}

\subsection{Spin correlation function}
\label{sec:spinspincorr}

The spin correlation function is given by
\begin{equation}
\langle{S}_{x}^{z}{S}_{y}^{z}\rangle
= \frac{\langle\Phi_{G.S.}\vert{S}_{x}^{z}
{S}_{y}^{z}\vert\Phi_{G.S.}\rangle}
       {\langle\Phi_{G.S.}\vert\Phi_{G.S.}\rangle},
\label{eq:spin1}
\end{equation}
where $S_{x}^{z}=(c_{x,\uparrow}^{\dagger}
c_{x,\uparrow}^{\phantom{\dagger}}
-c_{x,\downarrow}^{\dagger}c_{x,\downarrow}^{\phantom{\dagger}})/2$
is the $z$-component of the spin operator on site $x$. From
(\ref{eq:gs2}), the numerator is written
\begin{equation}
\langle\Phi_{G.S.}\vert
{S}_{x}^{z}{S}_{y}^{z}
\vert\Phi_{G.S.}\rangle
=\sum_{V \in {\cal V}}
 \sum_{V' \in {\cal V}}
\lambda(V')\lambda^{\ast}(V)
\Bigg\langle 0 \Bigg\vert
\Bigg(
\prod_{\{u',v'\}\in{V'}}
       b_{u',v'}
\Bigg)
       S_x^z S_y^z
\prod_{\{u,v\}\in{V}}
       b_{u,v}^{\dagger}
\Bigg\vert 0 \Bigg\rangle.
\label{eq:spin2}
\end{equation}From the commutation relations (\ref{eq:bose}) and
$[S_x^z, b_{u,v}]=[S_x^z,b_{u,v}^{\dagger}]=0$
where $x$, $u$, and $v$ are different sites one another,
we can decompose the graph $V \cup V'$ into connected subgraphs
$U_i \cup U'_i$ as we did in subsection\ \ref{subsec:normgs}.
We only need to consider the case where
sites $x$ and $y$ belong to a single subgraph
denoted as $U^{(x,y)} \cup {U'}^{(x,y)}$.
Otherwise, the expectation value in (\ref{eq:spin2})
is vanishing.
The decomposition is written
\begin{eqnarray}
V \cup V'
=
U^{(x,y)} \cup {U'}^{(x,y)}
+\sum_{i=1}^{n \left( V \cup V' \right)-1}
U_i \cup U'_i.
\label{eq:decomps1}
\end{eqnarray}
After the elimination of $p$-sites with four non-closed bonds,
we only need to consider the case where
sites $x$ and $y$ belong to a single loop
(Fig.\ \ref{fig:geospin})
denoted as $W^{(x,y)} \cup {W'}^{(x,y)}$.
Otherwise, the expectation value in (\ref{eq:spin2})
is vanishing.
We have
\begin{eqnarray}
\lefteqn{V \cup V'}
\nonumber \\
&\to&
W^{(x,y)} \cup {W'}^{(x,y)}
+
\sum_{j=1}^{n \left( U^{(x,y)} \cup {U'}^{(x,y)} \right)-1}
W_j \cup W'_j
+
\sum_{i=1}^{n \left( V \cup V' \right)-1}
\left(
\sum_{j=1}^{n \left( U_i \cup U'_i \right)}
W_j \cup W'_j
\right)
\label{eq:decomps21} \\
&=&
W^{(x,y)} \cup {W'}^{(x,y)}
+
\sum_{j=1}^{{\cal N}_W}
W_j \cup W'_j.
\label{eq:decomps2}
\end{eqnarray}
We note that each $W_j$, $W'_j$ in the second term
of (\ref{eq:decomps21})
depends on the graphs $U^{(x,y)}$, ${U'}^{(x,y)}$,
and $U^{(x,y)} \cup {U'}^{(x,y)}$,
and those in the third term
depends on the graphs $U_i$, $U'_i$, and $U_i \cup U'_i$.
We denote the total number of the loops $W_j \cup W'_j$
in the second and the third terms of (\ref{eq:decomps21})
by ${\cal N}_W$.
Their labelings in (\ref{eq:decomps2}) were rearranged.
The weight for the loop $W^{(x,y)} \cup {W'}^{(x,y)}$
is
\begin{eqnarray}
w(x,y)
&=&
\Bigg\langle 0 \Bigg\vert
\Bigg(
\prod_{\{z',w'\}\in {W'}^{(x,y)}}
       b_{z',w'}^{\phantom{\dagger}}
\Bigg)
       S_{x}^{z}S_{y}^{z}
\prod_{\{z,w\}\in W^{(x,y)}}
       b_{z,w}^{\dagger}
\Bigg\vert 0 \Bigg\rangle
\nonumber \\
&=&
(-1)^{d(x,y)}\frac{1}{4}
\Bigg\langle 0 \Bigg\vert
\prod_{\{z',w'\} \in {W'}^{(x,y)}}
       b_{z',w'}^{\phantom{\dagger}}
\prod_{\{z,w\}\in W^{(x,y)}}
       b_{z,w}^{\dagger}
\Bigg\vert 0 \Bigg\rangle
\nonumber \\
&=&
(-1)^{d(x,y)}\frac{1}{4}w
\label{eq:spinloop}
\end{eqnarray}
where $d(x,y)$ is the number of the bonds between $x$ and $y$
along the loop and $w$ is given by (\ref{eq:proof1}).
A derivation is shown in Appendix\ \ref{app:spin}. From
(\ref{eq:www}), (\ref{eq:wji}), (\ref{eq:decomps2}),
and (\ref{eq:spinloop}), we obtain
\begin{eqnarray}
\lefteqn{\langle\Phi_{G.S.}\vert
S_x^z S_y^z
\vert\Phi_{G.S.}\rangle}
\nonumber \\
&=&
\sum_{V,V' \in {\cal V}} \
\left\{
(-1)^{d(x,y)}\frac{1}{4}
\prod_{i=1}^{n \left( V \cup V' \right)} \
       \lambda \left( U'_i \right)
       \lambda^{\ast} \left( U_i \right)
\Bigg\langle 0 \Bigg\vert
       \prod_{\{x',y'\}\in U'_i}
              b_{x',y'}^{\phantom{\dagger}}
       \prod_{\{x,y\}\in U_i}
              b_{x,y}^{\dagger}
\Bigg\vert 0 \Bigg\rangle
\right\}
\label{eq:spingeometric}
\\
&=&
\sum_{V, V' \in {\cal V}}
\left\{
(-1)^{d(x,y)}
\frac{1}{4}
\prod_{i=1}^{n \left( V \cup V' \right)}
\left[
\prod_{j=1}^{n \left( U_j \cup U'_j \right)}
       \lambda \left( W'_j \right)
       \lambda^{\ast} \Big( W_j \Big)
(-1)^{m_j}
 w_j
\right]\right\},
\label{eq:spin9}
\end{eqnarray}
where the weight for the loop
$W^{(x,y)} \cup {W'}^{(x,y)}$ is included in the product.

\subsection{
Correlation function
$\langle c_{x,\sigma}^{\phantom{\dagger}}
         c_{y,\sigma}^{\dagger} \rangle$
}
\label{sec:green}

We evaluate the correlation function defined by
\begin{equation}
\langle c_{x,\sigma}^{\phantom{\dagger}}
        c_{y,\sigma}^{\dagger} \rangle
=
\frac{
\langle\Phi_{G.S.}\vert
c_{x,\sigma}^{\phantom{\dagger}}c_{y,\sigma}^{\dagger}
\vert\Phi_{G.S.}\rangle
}
{
\langle\Phi_{G.S.}\vert\Phi_{G.S.}\rangle
}.
\label{eq:green1}
\end{equation}From (\ref{eq:gs2}), the numerator is written
\begin{equation}
\langle\Phi_{G.S.}\vert
c_{x,\sigma}^{\phantom{\dagger}} c_{y,\sigma}^{\dagger}
\vert\Phi_{G.S.}\rangle
=\sum_{V \in {\cal V}}
 \sum_{V' \in {\cal V}}
\lambda(V')\lambda^{\ast}(V)
\Bigg\langle 0 \Bigg\vert
\Bigg(
\prod_{\{u',v'\}\in{V'}}
       b_{u',v'}^{\phantom{\dagger}}
\Bigg)
       c_{x,\sigma}^{\phantom{\dagger}} c_{y,\sigma}^{\dagger}
\prod_{\{u,v\}\in{V}}
       b_{u,v}^{\dagger}
\Bigg\vert 0 \Bigg\rangle.
\label{eq:greenpro}
\end{equation}
We decompose the graph $V \cup V'$ into
connected subgraphs using the commutation relations (\ref{eq:bose})
and
$[c_{z,\sigma},b_{x,y}]
=[c_{z,\sigma}^{\phantom{\dagger}},b_{x,y}^{\dagger}]
=[c_{z,\sigma}^{\dagger},b_{x,y}^{\dagger}]=0$
where $z\neq x$ and $ z\neq y$
and obtain (\ref{eq:decomps1}).
After the elimination of $p$-sites with four non-closed bonds,
we only need to consider the case where
sites $x$ and $y$ belong to a single graph
and they satisfy one of the following four conditions:
(i) site $x(y)$ with one non-closed ket(bra)bond;
(ii) site $x(y)$ with one non-closed bra(ket)bond
and one self-closed ket(bra) bond.
(See Figs.\ \ref{fig:geoprop}
where a site with a pentagon represents the operator
$c_{x,\sigma}^{\phantom{\dagger}}$ or $c_{y,\sigma}^{\dagger}$.)
Otherwise, the expectation value is vanishing.
It is convenient to eliminate type (ii) sites using the identity
$c_{x,\sigma}^{\dagger} b_{x,y}^{\dagger}
=-c_{y,\sigma}^{\dagger} b_{x,x}^{\dagger}$
which is represented diagrammatically
in Fig.\ \ref{fig:elimination4}.
The results are shown in Figs.\ \ref{fig:elimination5}(a) and (b).
After this procedure,
the graph $W^{(x',y')} \cup {W'}^{(x',y')}$
in (\ref{eq:decomps1}) is a line
where ``bra-bond'' and ``ket-bond''
are placed alternately
and $x'$ and $y'$ are always at the end of the line
(Fig.\ \ref{fig:elimination5}(c)).
The weight is
\begin{equation}
w(x',y')=
\Bigg\langle 0 \Bigg\vert
\Bigg(
\prod_{\{u',v'\}\in {W'}^{(x'y')}}
       b_{u',v'}^{\phantom{\dagger}}
\Bigg)
       c_{x',\sigma}^{\phantom{\dagger}}c_{y',\sigma}^{\dagger}
\prod_{\{u,v\}\in W^{(x'y')}}
       b_{u,v}^{\dagger}
\Bigg\vert 0 \Bigg\rangle
=(-1)^{l(x',y')/2},
\label{eq:proof2}
\end{equation}
where $l(x',y')$ is the number of the bonds
(see Appendix\ \ref{app:green}).
Thus we have
\begin{eqnarray}
\lefteqn{\langle\Phi_{G.S.}\vert
c_{x,\sigma}^{\phantom{\dagger}} c_{y,\sigma}^{\dagger}
\vert\Phi_{G.S.}\rangle}
\nonumber \\
&=&
\sum_{V,V' \in {\cal V}}
\Bigg\{
 \lambda \left({U'}^{(x,y)}\right)
 \lambda^{\ast} \left(U^{(x,y)}\right)
\Bigg\langle 0 \Bigg\vert
\Bigg(
\prod_{\{z',w'\}\in {U'}^{(x,y)}}
       b_{z',w'}^{\phantom{\dagger}}
\Bigg)
c_{x,\sigma}^{\phantom{\dagger}} c_{y,\sigma}^{\dagger}
\prod_{\{z,w\}\in U^{(x,y)}}
       b_{z,w}^{\dagger}
\Bigg\vert 0 \Bigg\rangle
\nonumber \\
& & \hspace{32mm} \times
\left.
 \prod_{i=1}^{n \left( V \cup V' \right)-1}
\Bigg[
 \lambda \Big( U'_i \Big)
 \lambda^{\ast} \Big( U_i \Big)
\Bigg\langle 0 \Bigg\vert
 \prod_{\{u',v'\}\in{U'_i}}
        b_{u',v'}^{\phantom{\dagger}}
 \prod_{\{u,v\}\in{U_i}}
        b_{u,v}^{\dagger}
\Bigg\vert 0 \Bigg\rangle
\Bigg]\right\}
\label{eq:greengeometric}
\\
&=&
\sum_{V,V' \in {\cal V}}
\left\{
 \lambda \left({W'}^{(x',y')}\right)
 \lambda^{\ast} \left(W^{(x',y')}\right)
(-1)^{l(x',y')/2+m}
\prod_{j=1}^{{\cal N}_W} \
\lambda \left( W'_j \right) \lambda^{\ast}\Big(W_j \Big)
(-1)^{m_j} w_j
\right\},
\label{eq:propz}
\end{eqnarray}
where $m$ is the number of the procedures
shown in Fig.\ \ref{fig:elimination2}(a)
for the subgraph $U^{(x,y)} \cup {U'}^{(x,y)}$.

\subsection{Density correlation function}
\label{subsec:number}

We first evaluate the expectation value of the number operator
\begin{equation}
\langle n_{x,\sigma} \rangle
=
\frac{\langle \Phi_{G.S.} \vert n_{x,\sigma} \vert \Phi_{G.S.} \rangle}
     {\langle \Phi_{G.S.} \vert \Phi_{G.S.} \rangle}.
\label{eq:no}
\end{equation}From (\ref{eq:gs2}), the numerator is written
\begin{equation}
\langle\Phi_{G.S.}\vert
n_{x,\sigma}
\vert\Phi_{G.S.}\rangle
=\sum_{V \in {\cal V}}
 \sum_{V' \in {\cal V}}
\lambda(V')\lambda^{\ast}(V)
\Bigg\langle 0 \Bigg\vert
\Bigg(
\prod_{\{u',v'\}\in{V'}}
       b_{u',v'}^{\phantom{\dagger}}
\Bigg)
       n_{x,\sigma}^{\phantom{\dagger}}
\prod_{\{u,v\}\in{V}}
       b_{u,v}^{\dagger}
\Bigg\vert 0 \Bigg\rangle.
\label{eq:numberpro}
\end{equation}
We only need to consider the graph $V \cup V'$
which contains the site $x$.
Otherwise, the expectation value is vanishing.
The similar calculation to that
in Sec.\ \ref{sec:spinspincorr} leads
\begin{eqnarray}
V \cup V'
&=&
U^{(x)}\cup {U'}^{(x)}
+
\sum_{i=1}^{n \left( V \cup V' \right)-1}
U_i \cup U'_i
\nonumber \\
&\to&
W^{(x)} \cup {W'}^{(x)}
+
\sum_{j=1}^{n \left( U^{(x)} \cup {U'}^{(x)} \right)-1}
W_j \cup W'_j
+
\sum_{i=1}^{n \left( V \cup V' \right)-1}
\left(
\sum_{j=1}^{n \left( U_i \cup U'_i \right)}
W_j \cup W'_j
\right)
\nonumber \\
&=&
W^{(x)} \cup {W'}^{(x)}
+
\sum_{j=1}^{{\cal N}_W}
W_j \cup W'_j,
\label{eq:decomps3}
\end{eqnarray}
where $U^{(x)} \cup {U'}^{(x)}$
is a subgraph with the site $x$,
and $W^{(x)} \cup {W'}^{(x)}$ is a loop with the site $x$
after the elimination of $p$-sites with four non-closed bonds.
We distinguish two kinds of loops:
(i) site $x$ with self-closed bonds;
(ii) site $x$ with one non-closed ``bra-bond''
and one non-closed ``ket-bond''.
(See Figs.\ \ref{fig:geonumber},
where a site with a circle represents the number operator.)
(We note that a subgraph $U^{(x)} \cup {U'}^{(x)}$
with four non-closed bonds at the site $x$
is decomposed into type (ii) graph and loops.)
The weight is
\begin{eqnarray}
w(x)
&=&
\Bigg\langle 0 \Bigg\vert
\Bigg(
\prod_{\{z',w'\}\in {W'}^{(x)}}
       b_{z',w'}^{\phantom{\dagger}}
\Bigg)
       n_{x,\sigma}^{\phantom{\dagger}}
\prod_{\{z,w\}  \in W^{(x)}}
       b_{z,w}^{\dagger}
\Bigg\vert 0 \Bigg\rangle
\nonumber \\
&=&
 \left \{ \begin{array}{@{\,}ll}
       1                   & \mbox{for (i)}   \\
       (-1)^{{l(x)}/2-1} & \mbox{for (ii),} \\
          \end{array} \right.
\label{eq:proof3}
\end{eqnarray}
where $l(x)$ is the number of the bonds in the graph.
A derivation is given in Appendix\ \ref{app:number}. From
(\ref{eq:www}), (\ref{eq:wji}), (\ref{eq:decomps3}),
and (\ref{eq:proof3}), we obtain
\begin{eqnarray}
\lefteqn{
\langle\Phi_{G.S.}\vert
n_{x,\sigma}
\vert \Phi_{G.S.} \rangle }
\nonumber \\
& &=
\sum_{V,V' \in {\cal V}}
\left\{
 \lambda \left({U'}^{(x)}\right)
 \lambda^{\ast} \left(U^{(x)}\right)
\Bigg\langle 0 \Bigg\vert
\Bigg(
\prod_{\{z',w'\}\in {U'}^{(x)}}
       b_{z',w'}^{\phantom{\dagger}}
\Bigg)
n_{x,\sigma}^{\phantom{\dagger}}
\prod_{\{z,w\}\in U^{(x)}}
       b_{z,w}^{\dagger}
\Bigg\vert 0 \Bigg\rangle
\right.
\nonumber \\
& & \hspace{30mm} \left. \times
 \prod_{i=1}^{n \left( V \cup V' \right)-1}
\left[
 \lambda \Big( U'_i \Big)
 \lambda^{\ast} \Big( U_i \Big)
\Bigg\langle 0 \Bigg\vert
 \prod_{\{u',v'\}\in{U'_i}}
        b_{u',v'}^{\phantom{\dagger}}
 \prod_{\{u,v\}\in{U_i}}
        b_{u,v}^{\dagger}
\Bigg\vert 0 \Bigg\rangle
\right]\right\}
\label{eq:numbergeometric}
\\
& &=
\sum_{V,V' \in {\cal V}}
\left\{
\lambda \left( {W'}_j^{(x)} \right) \lambda^{\ast}\left(W_j^{(x)} \right)
(-1)^{m} w(x)
\prod_{j=1}^{{\cal N}_W} \
\lambda \left( W'_j \right) \lambda^{\ast}\Big(W_j \Big)
(-1)^{m_j} w_j
\right\}.
\label{eq:numa}
\end{eqnarray}

The density correlation function is
\begin{eqnarray}
\left\langle \delta n_x \delta n_y \right\rangle
&=&
\left\langle \left( n_x - \langle n_x \rangle \right)
             \left( n_y - \langle n_y \rangle \right)
\right\rangle \nonumber \\
&=&
\langle n_x n_y \rangle
-\langle n_x \rangle \langle n_y \rangle \nonumber \\
&=&\frac{\langle \Phi_{G.S.} \vert
          n_x n_y
         \vert \Phi_{G.S.} \rangle}
        {\langle \Phi_{G.S.} \vert \Phi_{G.S.} \rangle}
-\frac{\langle \Phi_{G.S.} \vert n_x \vert \Phi_{G.S.} \rangle
       \langle \Phi_{G.S.} \vert n_y \vert \Phi_{G.S.} \rangle}
      {\left( \langle \Phi_{G.S.} \vert \Phi_{G.S.} \rangle \right)^2}
      \nonumber \\
&=&
4 \left[
\frac{D(x,y;\sigma)}{\langle \Phi_{G.S.} \vert \Phi_{G.S.} \rangle}
-\langle n_{x,\sigma} \rangle
 \langle n_{y,\sigma} \rangle
\right],
\label{eq:dendendef}
\end{eqnarray}
where
\begin{eqnarray}
D(x,y; \sigma)=
\frac{1}{2}
\Big\langle \Phi_{G.S.} \Big\vert
 c_{x,\sigma}^{\phantom{\dagger}}c_{y,\sigma}^{\phantom{\dagger}}
 c_{y,\sigma}^{\dagger}c_{x,\sigma}^{\dagger}
 +
 c_{x,\sigma}^{\phantom{\dagger}}c_{y,-\sigma}^{\phantom{\dagger}}
 c_{y,-\sigma}^{\dagger}c_{x,\sigma}^{\dagger}
\Big\vert \Phi_{G.S.}\Big\rangle.
\label{eq:dendenD}
\end{eqnarray}From (\ref{eq:gs2}), we have
\begin{eqnarray}
D(x,y; \sigma)
&=&
\sum_{V \in {\cal V}}
\sum_{V' \in {\cal V}}
\lambda(V')\lambda^{\ast}(V)
\nonumber \\
& & \hspace{5mm}\times
\Bigg\langle 0 \Bigg\vert
\Bigg(
\prod_{\{u',v'\}\in{V'}}
       b_{u',v'}^{\phantom{\dagger}}
\Bigg)
\frac{1}{2}
\Bigg(
      c_{x,\sigma}^{\phantom{\dagger}}c_{y,\sigma}^{\phantom{\dagger}}
      c_{y,\sigma}^{\dagger}c_{x,\sigma}^{\dagger}
      +
      c_{x,\sigma}^{\phantom{\dagger}}c_{y,-\sigma}^{\phantom{\dagger}}
      c_{y,-\sigma}^{\dagger}c_{x,\sigma}^{\dagger}
\Bigg)
\prod_{\{u,v\}\in{V}}
       b_{u,v}^{\dagger}
\Bigg\vert 0 \Bigg\rangle.
\nonumber \\
\end{eqnarray}
We only need to consider the graph $V \cup V'$
which contains the sites $x$ and $y$.
Otherwise, the expectation value is vanishing.
The similar calculation to that
in Sec.\ \ref{sec:spinspincorr} leads (\ref{eq:decomps2})
where $W^{(x,y)} \cup {W'}^{(x,y)}$ is a loop (or two loops)
which contain(s) the sites $x$ and $y$.
We classify the loops $W^{(x,y)} \cup {W'}^{(x,y)}$ as
(i) sites $x$ and $y$ each belongs
to two distinct self-closed bonds,
(ii) site $x$ ($y$) belongs to the loop and site $y$ ($x$)
belongs to the self-closed bonds,
(iii) sites $x$ and $y$ each belongs to two distinct loops,
and (iv) sites $x$ and $y$ belong to a single graph.
(See Figs.\ \ref{fig:geoden}.)
Otherwise, the expectation value is vanishing.
The weight is
\begin{eqnarray}
\lefteqn{w(x,y;\sigma)}
\nonumber \\
&=&
\Bigg\langle 0 \Bigg\vert
\Bigg(
\prod_{\{u',v'\}\in{W'}^{(x,y)}}
       b_{u',v'}^{\phantom{\dagger}}
\Bigg)
\frac{1}{2}
\Bigg(
      c_{x,\sigma}^{\phantom{\dagger}}c_{y,\sigma}^{\phantom{\dagger}}
      c_{y,\sigma}^{\dagger}c_{x,\sigma}^{\dagger}
      +
      c_{x,\sigma}^{\phantom{\dagger}}c_{y,-\sigma}^{\phantom{\dagger}}
      c_{y,-\sigma}^{\dagger}c_{x,\sigma}^{\dagger}
\Bigg)
\prod_{\{u,v\}\in W^{(x,y)}}
       b_{u,v}^{\dagger}
\Bigg\vert 0 \Bigg\rangle
\nonumber \\
\label{eq:dendenw} \\
&=& \left \{ \begin{array}{@{\,}ll}
          1                                          & \mbox{for (i)} \\
          (-1)^{l(x)/2-1}                           & \mbox{for (ii)
                               when the site $x$ belongs to the loop} \\
          (-1)^{{l(x)}/2-1}\times(-1)^{{l_i(y)}/2-1} & \mbox{for (iii)}\\
          (-1)^{{l(x,y)}/2-1}/2                    & \mbox{for (iv)}, \\
      \end{array} \right.
\label{eq:proof4}
\end{eqnarray}
where $l(x)$ and $l(x,y)$ are the numbers of the bonds.
A derivation is given in Appendix\ \ref{app:den}. From
(\ref{eq:www}), (\ref{eq:wji}), and (\ref{eq:proof4}), we obtain
\begin{eqnarray}
\lefteqn{D(x,y; \sigma)}
\nonumber \\
& &=
\sum_{V,V' \in {\cal V}}
\Bigg\{
 \lambda \left({U'}^{(x,y)}\right)
 \lambda^{\ast} \left(U^{(x,y)}\right)
\nonumber \\
& & \hspace{10mm} \times
\Bigg\langle 0 \Bigg\vert
\Bigg(
\prod_{\{z',w'\}\in {U'}^{(x,y)}}
       b_{z',w'}^{\phantom{\dagger}}
\Bigg)
\frac{1}{2}
\Bigg(
      c_{x,\sigma}^{\phantom{\dagger}}c_{y,\sigma}^{\phantom{\dagger}}
      c_{y,\sigma}^{\dagger}c_{x,\sigma}^{\dagger}
      +
      c_{x,\sigma}^{\phantom{\dagger}}c_{y,-\sigma}^{\phantom{\dagger}}
      c_{y,-\sigma}^{\dagger}c_{x,\sigma}^{\dagger}
\Bigg)
\prod_{\{z,w\}\in U^{(x,y)}}
       b_{z,w}^{\dagger}
\Bigg\vert 0 \Bigg\rangle
\nonumber \\
& & \hspace{35mm} \left. \times
 \prod_{i=1}^{n \left( V \cup V' \right)-1}
\Bigg[
 \lambda \Big( U'_i \Big)
 \lambda^{\ast} \Big( U_i \Big)
\Bigg\langle 0 \Bigg\vert
 \prod_{\{u',v'\}\in{W'_i}}
        b_{u',v'}^{\phantom{\dagger}}
 \prod_{\{u,v\}\in{W_i}}
        b_{u,v}^{\dagger}
\Bigg\vert 0 \Bigg\rangle
\Bigg]\right\}
\label{eq:dengeometric}
\\
& &=
\sum_{V,V' \in {\cal V}}
\left\{
\lambda \left( W'^{(x,y)} \right) \lambda^{\ast}\left(W^{(x,y)} \right)
(-1)^m w(x,y;\sigma)
\prod_{j=1}^{{\cal N}_W} \
\lambda \left( W'_j \right) \lambda^{\ast}\Big(W_j \Big)
(-1)^{m_j} w_j
\right\}.
\label{eq:dendenfirst}
\end{eqnarray}

\subsection{Singlet-pair correlation function}

The singlet-pair correlation function is given by
\begin{equation}
\langle b_{x,y}^{\dagger} b_{u,v}^{\phantom{\dagger}}\rangle=
\frac{\langle\Phi_{G.S.}\vert
      b_{x,y}^{\dagger} b_{u,v}^{\phantom{\dagger}}
      \vert\Phi_{G.S.}\rangle}
     {\langle\Phi_{G.S.}\vert\Phi_{G.S.}\rangle}.
\label{eq:genebb}
\end{equation}From (\ref{eq:gs2}), we obtain
\begin{eqnarray}
\lefteqn{
\langle \Phi_{G.S.} \vert
b_{x,y}^{\dagger} b_{u,v}^{\phantom{\dagger}}
\vert \Phi_{G.S.} \rangle
}
\nonumber \\
&=&\sum_{V \in {\cal V}} \
 \sum_{V' \in {\cal V}}
 \lambda(V')\lambda^{\ast}(V)
 \Bigg\langle 0 \Bigg\vert
\left(
 \prod_{\{x',y'\}\in{V'}}
        b_{x',y'}^{\phantom{\dagger}}
\right)
      b_{x,y}^{\dagger} b_{u,v}^{\phantom{\dagger}}
 \prod_{\{x,y\}\in{V}}
        b_{x,y}^{\dagger}
 \Bigg\vert 0 \Bigg\rangle
\nonumber \\
&=&
\sum_{V,V' \in {\cal V}}
\left\{
\prod_{i=1}^{n \Big(V\cup V' \cup \{x,y\}\cup \{u,v\} \Big)}
 \lambda \Big(U'_j \Big)
 \lambda^{\ast}\Big(U_j \Big)
 \Bigg\langle 0 \Bigg\vert
\left(
 \prod_{\{x',y'\}\in{U'}}
        b_{x',y'}^{\phantom{\dagger}}
\right)
      b_{x,y}^{\dagger} b_{u,v}^{\phantom{\dagger}}
 \prod_{\{x,y\}\in{U}}
        b_{x,y}^{\dagger}
 \Bigg\vert 0 \Bigg\rangle
\right\}
\nonumber \\& &
\label{eq:singetgeometric} \\
&=&
\sum_{V,V' \in {\cal V}}
\left\{
\prod_{i=1}^{n \Big(V\cup V' \cup \{x,y\}\cup \{u,v\} \Big)}
\left[
\prod_{j=1}^{n \left( U_i \cup U'_i \right) } \
 \lambda \Big(W'_j \Big)
 \lambda^{\ast}\Big(W_j \Big)
(-1)^{m_j} w_j
\right]
\right\},
\label{eq:singletgeo}
\end{eqnarray}
where $V$ and $V'$ satisfy the condition
that the graph $V \cup V' \cup \{x,y\} \cup \{u,v\}$
consists of connected subgraphs.

\newpage
\section{Explicit calculations of the correlation functions}
\label{sec:corr}

The results of Secs.\ \ref{sec:ham} and \ref{sec:georep}
are valid in any dimensions.
At the moment, however, a practical use of them is limited to models
in one dimension in which the transfer matrix method can be applied.
In this section,
the equal-time correlation functions are evaluated exactly
for one-dimensional models.
We shall show the analytical procedures for obtaining them
for Model A defined below in subsection\ \ref{sec:modela},
illustrating the method in details.
The results are shown for a system size $N$
and in the thermodynamic limit.
For Models B and C,
we shall only show the results in the thermodynamic limit.

\subsection{Model A}
\label{sec:modela}

\subsubsection{Hamiltonian}
\label{sec:modelahami}

Let us consider a lattice constructed from cells with three sites.
We have two lattices
which satisfy the uniqueness condition of Ref.\ \cite{Ta93b}.
One of them, which we call Model A
(the other is called Model B, see subsection\ \ref{sec:modelb})
is constructed by a cell with two $d$-sites and one $p$-site
(Fig.\ \ref{fig:modelAcell}(a)).
Note that a cell is not a unit cell.
A unit cell is composed of a $d$-site and a $p$-site.
In the models constructed by the cell construction,
Model A is the simplest one for the following two reasons.
The structure of the lattice is the simplest.
(We can construct lattices from cells with two sites.
The exact ground state, however, contains
two electrons per site and is fully-filled.)
The calculation of the correlation function is easier than
that of Model B.

The cell Hamiltonian (\ref{eq:cellham}) is obtained by choosing
$\alpha_{n, \sigma}^{(A)}
= \sum_{r=1}^3 \lambda_r c_{r,\sigma}
\equiv \lambda_1^{\phantom{P}} c_{1,\sigma}^d
+\lambda_2^{\phantom{P}} c_{2,\sigma}^d
+\lambda_3^{\phantom{P}} c_{3,\sigma}^p $ in (\ref{eq:alpha})
and setting $\lambda_3=1$ without loss of generality
(see Fig.\ \ref{fig:modelAcell}(a) for intra-cell index).
Here $c_{r,\sigma}^d$ ($c_{r,\sigma}^p$) is
the annihilation operator on a $d$($p$)-site.
The full Hamiltonian is obtained by identifying the site $1$
in the $(n-1)$-th cell with the site $2$ in the $n$-th cell
(Fig.\ \ref{fig:modelAcell}(b)).
The Hamiltonian is
\begin{eqnarray}
H_S &=&  {\cal P}
\sum_{\sigma=\uparrow,\downarrow}
\Big\{ \sum_{n=1}^{N}
\left[ \ (-\lambda_1 \lambda_2
{c_{n+1,\sigma}^d}^{\dagger} c_{n,\sigma}^d
  -\lambda_1   {c_{n+1,\sigma}^d}^{\dagger} c_{n,\sigma}^p
  -\lambda_2   {c_{n,\sigma}^d}^{\dagger} c_{n,\sigma}^p  + h.c.)
\right.
\nonumber\\
& &\hspace{40mm}
\left.\left.
  +\epsilon_n^d {c_{n,\sigma}^d}^{\dagger} c_{n,\sigma}^d
  +\epsilon^p {c_{n,\sigma}^p}^{\dagger} c_{n,\sigma}^p \ \right]
+\epsilon_{N+1}^d {c_{N+1,\sigma}^d}^{\dagger} c_{N+1,\sigma}^d \right\}
 {\cal P},
\label{eq:hama}
\end{eqnarray}
where the on-site potentials are
$\epsilon_1^d = -2 \lambda_2^2$,
$\epsilon_n^d = -2( \lambda_1^2 + \lambda_2^2 )$ ($2 \le n \le N$),
$\epsilon_{N+1}^d = -2 \lambda_1^2$,
and $\epsilon^p=-1$.
A unit cell is labeled by $n$.
The ground state is
\begin{eqnarray}
\Big\vert \Phi_{G.S.}^A \Big\rangle &=&
{\cal P}
\prod_{n=1}^N
\prod_{\sigma = \uparrow, \downarrow}
{\alpha_{n, \sigma}^{(A)}}^{\dagger}
\Big\vert 0 \Big\rangle \\
&=&
{\cal P}
\prod_{n=1}^N
\prod_{\sigma = \uparrow, \downarrow}
\left(
 \lambda_1 c_{n,\sigma}^{d \ \dagger}
+\lambda_2 c_{n+1,\sigma}^{d \ \dagger}
+          c_{n,\sigma}^{p \ \dagger}
\right)
\Big\vert 0 \Big\rangle
\label{eq:gsmodela}
\end{eqnarray}
which is a half-filled state.

\subsubsection{Band structure in the single-electron problem}
\label{subsec:banda}

Before studying the ground state (\ref{eq:gsmodela}),
we investigate the corresponding non-interacting system.
We consider the system with an even number of unit cells
under the periodic boundary condition
$c_{N+1}^d=c_1^d$.
A one-particle state can be written
\begin{equation}
\Big\vert \Phi_F \Big\rangle
= \sum_n \sum_{\alpha =p,d}
\varphi_n^{\alpha} {c^{\alpha}_{n,\uparrow}}^{\dagger}
\Big\vert 0 \Big\rangle,
\label{eq:oneparticle}
\end{equation}
where $\varphi_n^{\alpha}$ are complex coefficients.
The single-electron Schr\"odinger equation
$H \vert \Phi_F \rangle = E \vert \Phi_F \rangle$,
where $E$ is the energy eigenvalue,
corresponding to the Hamiltonian (\ref{eq:hama}) is
\begin{eqnarray}
\left\{
\begin{array}{@{\,}l}
E \varphi_n^d =
-\lambda_1 \lambda_2 ( \varphi_{n-1}^d + \varphi_{n+1}^d )
-2 ( \lambda_1^2 + \lambda_2^2 ) \varphi_n^d
-\lambda_1 \varphi_{n-1}^p
-\lambda_2 \varphi_{n+1}^p  \\
E \varphi_n^p =
-\lambda_1 \varphi_{n-1}^d
-\lambda_2 \varphi_{n+1}^d
-\varphi_n^p
\end{array}\right..
\end{eqnarray}From the Fourier transformation
$\varphi_n^{\alpha}
= \frac{1}{\sqrt{N}} \sum_k e^{ikn} \varphi_k^{\alpha}$
where
\begin{equation}
k=0,
\pm\frac{2\pi}{N},
\pm\frac{4\pi}{N},
\cdots,
\pm2\pi\frac{\frac{N}{2}-1}{N},
\pi,
\label{eq:hasuu}
\end{equation}
the Schr\"odinger equation in the momentum space is
\begin{eqnarray}
\left\{
\begin{array}{@{\,}l}
E \varphi_k^d =
-2 ( \lambda_1 \lambda_2 \cos{k}
+ \lambda_1^2 + \lambda_2^2 ) \varphi_k^d
-  ( \lambda_1 + \lambda_2 e^{-ik}) \varphi_k^p \\
E \varphi_k^p =
-  ( \lambda_1 + \lambda_2 e^{ik} ) \varphi_k^d
-  \varphi_k^p  \\
\end{array}\right..
\end{eqnarray}
The eigen energies are
\begin{eqnarray}
E_{\pm} =
-\frac{1}{2}
\left[
2 \lambda_1 \lambda_2 \cos{k} + 2 \lambda_1^2 + \lambda_2^2 +1
\mp
\sqrt{\left(2 \lambda_1 \lambda_2 \cos{k}
+2\lambda_1^2 +2\lambda_2^2\right)^2
-4\left(\lambda_1^2 + \lambda_2^2\right)
}
\right],
\end{eqnarray}
where $-$, $+$ are the band index with $-$ ($resp.$ $+$)
corresponding to the $+$ ($resp.$ $-$) sign.
The energy gap between two bands is
$\Delta$ $=\sqrt{\left(2 \lambda_1 \lambda_2 -1\right)
\left(1+4\lambda_1^2  -2\lambda_1\lambda_2  +4\lambda_2^2\right)}$.
When $2 \lambda_1 \lambda_2 -1=0$,
the gap closes at $k=\pi$. (See Fig.\ \ref{fig:modelaband}.)

In the ground state (\ref{eq:gsmodela}),
there are $2N$ electrons.
Since there are $2N$ sites in the lattice,
the electron number corresponds to full-filling of the lower band.
Therefore,
the ground state of the non-interactiong system is insulating
for $2 \lambda_1 \lambda_2 -1 \ne 0$.
It is metallic when $2 \lambda_1 \lambda_2 -1 = 0$.

\subsubsection{Norm of the ground state}
\label{subsec:norm}

Before calculating the correlation functions,
we evaluate the norm of the ground state (\ref{eq:gsmodela}),
since the state is not normalized.
For the sake of convenience,
we draw the lattice shown in Fig.\ \ref{fig:modelAcell}(b)
as Fig.\ \ref{fig:modelAcell}(c).
The ground state admits the geometric representation (\ref{eq:gs2}),
where an example of the valence-bond configuration $V$
is shown in Fig.\ \ref{fig:modelAgeorep}(a).
The geometric representation of the norm is (\ref{eq:normg2}),
where an example of the graph $V \cup V'$ is shown
in Fig.\ \ref{fig:modelAgeorep}(b).
We first evaluate the contribution from a graph $V \cup V'$.
It can be decomposed into the subgraphs
$U_i \cup U'_i$ $(i=1$, $2$, $\cdots$, $n(V \cup V'))$.
No loop extends over more than two cells
and there is no $p$-site with four bonds,
since the sites which is identified in the cell construction
are $d$-sites.
We do not need the procedures shown in Figs.\ \ref{fig:elimination2}.
Therefore, the graph $U_i \cup U'_i$ cannot be decomposed further
and we find $n \left( U_i \cup U'_i \right) =1$
in (\ref{eq:normw}) and $m_j=0$ in (\ref{eq:norm3a}).
The cells with the graph are classified to four kinds
(Figs.\ \ref{fig:modelAnorm}(a)-(d)).
Consider the graph shown in Fig.\ \ref{fig:modelAnorm}(a). From
(\ref{eq:proof1}), the weight for the degenerate loop is 2
and the contribution from $\lambda \left( U'_i \right)
\lambda^{\ast} \left(U_i \right)$ in (\ref{eq:norm3a}) is
$\lambda_2^2$.
Therefore, the contribution from the graph is $2 \lambda_2^2$.
For other graphs see Figs.\ \ref{fig:modelAnorm}.

The sum over the graph $V \cup V'$ in (\ref{eq:normg2})
is equivalent to that over all the combination
of above four kinds of cells
under the restriction that a $d$-site has at most two valence-bonds.
(The restriction means, for example, that the identification
of the right $d$-site in Fig.\ \ref{fig:modelAnorm}(a)
with the left $d$-site in Fig.\ \ref{fig:modelAnorm}(c)
is forbidden.)
Hereafter we shall always take into account the restriction
and it should be understood implicitly.
To evaluate the sum we use the transfer matrix method.
We have to distinguish two cases due to the restriction.
Let $A_n$ and $B_n$ be the quantity defined by the right-hand side
of (\ref{eq:normg2}) on the lattice $\Lambda_n$.
For $A_n$, the sum is taken over all the combination of the cells
shown in Fig.\ \ref{fig:modelAnorm} with the restriction
that the $n$-th cell is represented
by Fig.\ \ref{fig:modelAnorm}(a) or (b).
For $B_n$, the sum is taken as was done for $A_n$ with the restriction
that the $n$-th cell is represented by (c) or (d).
They are represented diagrammatically
\setlength{\unitlength}{1mm}
\begin{eqnarray}
A_n &=& \\
B_n &=&.
\end{eqnarray}
We note that the norm in the system size $N$ is
\begin{eqnarray}
\langle \Phi_{G.S.} \vert \Phi_{G.S.} \rangle
=A_N + B_N.
\label{eq:normnote}
\end{eqnarray}
Given $A_{n-1}$ and $B_{n-1}$,
we can form $A_n$ and $B_n$ by attaching them to the $n$-th cell.
Let us consider $A_n$ first.
When the $n$-th cell is represented by Fig.\ \ref{fig:modelAnorm}(a),
we can attach $A_{n-1}$ and cannot attach $B_{n-1}$.
When the $n$-th cell is represented by Fig.\ \ref{fig:modelAnorm}(b),
we can attach $A_{n-1}$ or $B_{n-1}$ to it.
Thus we have the recursion relation
\begin{eqnarray}
A_n &=& \nonumber\\
    &=& 2 \lambda_2^2 A_{n-1} +A_{n-1} +B_{n-1}.
\label{eq:an}
\end{eqnarray}
For $B_n$ the similar calculation leads
\begin{eqnarray}
B_n &=& \nonumber \\
    &=& 2\lambda_1^2 A_{n-1} +2\lambda_1^2 B_{n-1}
       +2\lambda_1^2 \lambda_2^2 A_{n-1}.
\label{eq:bn}
\end{eqnarray}
They are conveniently written in a matrix form as
\begin{equation}
\left( \begin{array} {c}
       A_n \\
       B_n
       \end{array} \right)
=
{\bf T}_n
\left( \begin{array} {c}
       A_{n-1} \\
       B_{n-1}
       \end{array} \right),\
{\bf T}_n
=\left( \begin{array} {cc}
       1 +2\lambda_2^2                        & 1 \\
       2\lambda_1^2 + 2\lambda_1^2\lambda_2^2  & 2 \lambda_1^2 \\
       \end{array} \right).
\label{eq:transfmodelA}
\end{equation}
The initial vector is $\vec{I} \equiv (A_0, B_0)^T=(1,0)^T$,
since any cell in Fig.\ \ref{fig:modelAnorm}
is allowed as the first cell.
To obtain the quantity (\ref{eq:normnote}),
we choose the final vector $\vec{F} \equiv (A_N, B_N)^T=(1,1)^T$.
Since the transfer matrix is not symmetric,
it is convenient to diagonalize it using the right and left eigenvectors.
The matrix can be diagonalized as
\begin{eqnarray}
{\bf T}_n = {\bf R}{\bf D}{\bf R}^{-1} = {\bf L}^{-1}{\bf D}{\bf L},
\end{eqnarray}
where
\begin{equation}
{\bf D}
=\left( \begin{array} {cc}
       e_1 & 0   \\
       0   & e_2 \\
       \end{array} \right), \
{\bf R}
=\left( \begin{array} {cc}
       R_{11} & R_{12} \\
       R_{21} & R_{22} \\
       \end{array} \right), \
{\bf L}
=\left( \begin{array} {cc}
       L_{11} & L_{12} \\
       L_{21} & L_{22} \\
       \end{array} \right),
\label{eq:transfsmodelA}
\end{equation}
where $e_i$ are the eigenvalues of ${\bf T}_n$,
$e_1 =(2\lambda_1^2 +2\lambda_2^2 +1 +\omega_1)/2$ and
$e_2 =(2\lambda_1^2 +2\lambda_2^2 +1 -\omega_1)/2$
with $\omega_1
=\sqrt{4\lambda_1^4+4\lambda_2^4+4\lambda_1^2+4\lambda_2^2+1}$.
They satisfy $e_1 > e_2  > 0 $ for $\lambda_1 \neq 0$
and $\lambda_2 \neq 0$.
We choose the left eigenvectors
$\vec{L}_{1}$ $=(L_{11}, L_{12})^T$ and
$\vec{L}_{2}$ $=(L_{21}, L_{22})^T$ corresponding to the eigenvalues
$e_1$ and $e_2$, respectively,
and the right eigenvectors
$\vec{R}_{1}$ $=(R_{11}, R_{21})^T$ and
$\vec{R}_{2}$ $=(R_{12}, R_{22})^T$.
Using the diagonal matrix ${\bf C}$ $={\bf L}{\bf R}$,
we obtain
\begin{equation}
{\bf T}_n= {\bf R} {\bf D} {\bf C}^{-1} {\bf L},
\label{eq:diagonalization}
\end{equation}
where
\begin{equation}
{\bf C}
=\left( \begin{array} {cc}
       c_1 & 0 \\
       0   & c_2
       \end{array} \right).
\label{eq:matc}
\end{equation}From the quantities, the norm of the ground state
in the system size $N$ is
\begin{eqnarray}
\Big\langle \Phi_{G.S.}^A \Big\vert \Phi_{G.S.}^A \Big\rangle
&=& \vec{F}^T
    {\bf T}_N
    \cdots
    {\bf T}_2
    {\bf T}_1
    \vec{I}  \nonumber \\
&=& \vec{F}^T
    {\bf R}
    {\bf D}^{N}
    {\bf C}^{-1}
    {\bf L}
    \vec{I}
\label{eq:normmodelA}\\
&=&
\frac{1 +2\lambda_1^2 +2\lambda_2^2 +4\lambda_1^2 \lambda_2^2 +\omega_1}
     {2 \omega_1}
e_1^N
-
\frac{1 +2\lambda_1^2 +2\lambda_2^2 +4\lambda_1^2 \lambda_2^2 -\omega_1}
     {2 \omega_1}
e_2^N,
\nonumber
\end{eqnarray}
where we used the relation
${\bf C}^{-1} {\bf L} {\bf R} ={\bf I}$,
Here ${\bf I}$ is the identity matrix.
In the thermodynamic limit,
the eigenvalue $e_1$ dominates
and we have
\begin{eqnarray}
\Big\langle \Phi_{G.S.}^A \Big\vert \Phi_{G.S.}^A \Big\rangle
=
\frac{1 +2\lambda_1^2 +2\lambda_2^2 +4\lambda_1^2 \lambda_2^2 + \omega_1}
     {2 \omega_1}.
\end{eqnarray}

\subsubsection{Expectation value of the number operator}
\label{subsec:onepoint}

We calculate the occupation on a $p$-site
$\langle n_{i,\sigma}^p \rangle$.
The geometric representation of the expectation value
is (\ref{eq:numbergeometric}),
where an example of the graph $V \cup V'$ is shown
in Fig.\ \ref{fig:modelAgeorep}(c).
We do not need the procedures shown in Figs.\ \ref{fig:elimination2},
because there is no site with four bonds.
Therefore, we find $n \left( U_i \cup U'_i \right) =1$
in (\ref{eq:decomps3}) and $m_j=0$ in (\ref{eq:numa}).
In the representation, the operator $n_{i,\sigma}^p$ modifies
the weight associated with the graph which contains the $i$-th cell.
Therefore, we replace the transfer matrix ${\bf T}_i$
by ${\bf N}_i^{(p)}$ which is a matrix associated with
the operator $n_{i,\sigma}^p$.
The expectation value can be written
\begin{equation}
\Big\langle \Phi_{G.S.}^A \Big\vert
n_{i,\sigma}^p
\Big\vert \Phi_{G.S.}^A \Big\rangle
=\vec{F}^T
{\bf T}_N
\cdots
{\bf T}_{i+1}
{\bf N}_i^{(p)}
{\bf T}_{i-1}
\cdots
{\bf T}_1
\vec{I}
\end{equation}
We derive the matrix ${\bf N}_i^{(p)}$.
We have three kinds of graphs on the $i$-th cell
(Fig.\ \ref{fig:modelAoccp}(a)-(c)).
Let $A^{(p)}_i$ and $B^{(p)}_i$ be the quantity defined
by the right-hand side of (\ref{eq:numbergeometric})
on the lattice $\Lambda_i$.
The restriction for the sum is that
the $i$-th cell is represented either
by Fig.\ \ref{fig:modelAoccp}(a) or (b) for $A^{(p)}_i$
and by (c) for $B^{(p)}_i$.
They are represented diagrammatically
\begin{eqnarray}
A^{(p)}_i &=& \\
B^{(p)}_i &=&,
\end{eqnarray}
and the recursion relations are
\begin{eqnarray}
A^{(p)}_i &=& \nonumber\\
          &=& \lambda_2^2 A_{i-1} + A_{i-1} + B_{i-1} \\
B^{(p)}_i &=& \nonumber \\
          &=& \lambda_1^2 A_{i-1} +\lambda_1^2 B_{i-1}.
\label{eq:apn}
\end{eqnarray}From (\ref{eq:proof3}), the loop with operator
$n_{i,\sigma}$ has weight 1.
It is written
\begin{equation}
\left(
 \begin{array}{c}
    A^{(p)}_i \\
    B^{(p)}_i
 \end{array}
\right)
=
{\bf N}_i^{(p)}
\left(
 \begin{array}{c}
    A_{i-1} \\
    B_{i-1}
 \end{array}
\right), \ \
{\bf N}_i^{(p)}
=
\left(
 \begin{array}{cc}
    1+ \lambda_2^2 & 1          \\
    \lambda_1^2    & \lambda_1^2
 \end{array}
\right).
\label{eq:ntildep}
\end{equation}

 From (\ref{eq:onepoint}),
(\ref{eq:onepointnum}), (\ref{eq:normmodelA}), and (\ref{eq:ntildep}),
we have
\begin{eqnarray}
\langle n_{i,\sigma}^p \rangle
&=&\frac{\vec{F}^T
         {\bf T}^{N-i}
         {\bf N}_{i}^{(p)}
         {\bf T}^{i-1}
         \vec{I}}
        {\vec{F}^T
         {\bf T}^{N}
         \vec{I}} \nonumber \\
&=&\frac{\vec{F}^T
         {\bf R}
         {\bf D}^{N-i}
         {\bf C}^{-1}
         {\bf L}
         {\bf N}_{i}^{(p)}
         {\bf R}
         {\bf D}^{i-1}
         {\bf C}^{-1}
         {\bf L}
         \vec{I}}
        {\vec{F}^T
         {\bf R}
         {\bf D}^{N}
         {\bf C}^{-1}
         {\bf L}
         \vec{I}}
\label{eq:occmodelA}
\\
&=&
\frac{
C_1
-C_2 \left( \frac{e_2}{e_1} \right)^i
-C_3 \left( \frac{e_2}{e_1} \right)^{N-i}
+C_4 \left( \frac{e_2}{e_1} \right)^N
}{
\frac{\omega_1}{2}
\left\{
\left[
\left(1+2\lambda_1^2\right)\left(1+2\lambda_2^2\right)+\omega_1
\right]
-
\left[
\left(1+2\lambda_1^2\right)\left(1+2\lambda_2^2\right)-\omega_1
\right] \left( \frac{e_2}{e_1} \right)^N
\right\}
},
\nonumber
\end{eqnarray}
where
\begin{eqnarray}
C_1
&=&
\left( e_1-2\lambda_1^2 \right)
\left( e_1-2\lambda_2^2 \right)
\left( e_1-\lambda_1^2 -\lambda_2^2 \right)
\nonumber \\
C_2
&=&
\left( e_1-2\lambda_2^2 \right)
\left( e_2-2\lambda_1^2 \right)
\left( e_1 e_2 -\lambda_2^2 e_1 -\lambda_1^2 e_2 \right)/e_2
\nonumber \\
C_3
&=&
\left( e_1-2\lambda_1^2 \right)
\left( e_2-2\lambda_2^2 \right)
\left( e_1 e_2 -\lambda_1^2 e_1 -\lambda_2^2 e_2 \right)/e_1
\nonumber \\
C_4
&=&
\left( e_2-2\lambda_1^2 \right)
\left( e_2-2\lambda_2^2 \right)
\left( e_2-2\lambda_1^2 -2\lambda_2^2 \right),
\nonumber
\end{eqnarray}
in the system size $N$.
In the thermodynamic limit, $N$, $i$, $N-i$ $\to \infty$,
we obtain
\begin{eqnarray}
\langle n_{i,\sigma}^p \rangle
=
\frac{1}{2} \left(1 + \frac{1}{\omega_1} \right).
\label{eq:npx2}
\end{eqnarray}
Since there are two electrons in a unit cell, from
(\ref{eq:npx2}) we have the occupation on a $d$-site
\begin{eqnarray}
\langle n_{i,\sigma}^d \rangle
= \frac{1}{2} \left( 1 -\frac{1}{\omega_1} \right).
\end{eqnarray}
We can also obtain the same quantity from the geometric representation
of the expectation value. From the graphs shown
in Fig.\ \ref{fig:modelAoccp}(d)-(f),
we have the matrix associated with the operator $n_{i, \sigma}^d$
\begin{eqnarray}
{\bf N}_i^{(d)}
=
\left(
 \begin{array}{cc}
    \lambda_2^2            & 0            \\
    \lambda_1^2\lambda_2^2 & \lambda_1^2
 \end{array}
\right),
\label{eq:And}
\end{eqnarray}
which will be used to obtain the density correlation function.

The results are shown in Figs.\ \ref{fig:modelA:occp:res}(a) and (b)
for $\alpha$ $=p$ and $d$, respectively.
We consider the following cases:
(i) $\lambda_1$, $\lambda_2$ $\gg 1$ ($\lambda_1=\lambda_2$);
(ii) $\vert \lambda_1 \vert$, $\vert \lambda_2 \vert$ $\ll 1$
($\lambda_1=\lambda_2$);
(iii) $\vert \lambda_1 \vert \ll 1$, $\lambda_2$ $\gg 1$.
For (i),
on-site potentials satisfy the relation $\epsilon_d \ll \epsilon_p$.
There is almost one electron per a site.
For (ii),
on-site potentials satisfy the relation $\epsilon_p < \epsilon_d$,
$\epsilon_d - \epsilon_p \gg \lambda_1 \lambda_2$.
The $d$-sites are almost empty. The $p$-sites are almost doubly occupied.
For (iii), the system decouples to a collection of the pairs
of $p$- and $d$-sites.

\subsubsection{Two-point correlation functions}
\label{subsec:twopointcorr}

We calculate the density correlation function for the $p$-site.
We first evaluate the first term in (\ref{eq:dendendef}).
The geometric representation of the expectation value
is (\ref{eq:dengeometric}),
where an example of the graph $V \cup V'$ is shown
in Fig.\ \ref{fig:modelAgeorep}(d).
We do not need the procedures shown in Figs.\ \ref{fig:elimination2},
because there is no site with four bonds.
Therefore, we find $n \left( U_i \cup U'_i \right) =1$
in (\ref{eq:decomps21}) and $m_j=0$ in (\ref{eq:dendenfirst}).
In the representation, the operators modify the weight
associated with the graph which contains cells
between the cells $i$ and $j$. From the derivation
of the matrix ${\bf N}_i^{(p)}$,
the expectation value is
\begin{equation}
D(x,y;\sigma)
=\vec{F}^T
{\bf T}_N
\cdots
{\bf T}_{j+1}
{\bf N}_j^{(p)}
{\bf T}_{j-1}
\cdots
{\bf T}_{i+1}
{\bf N}_i^{(p)}
{\bf T}_{i-1}
\cdots
{\bf T}_1
\vec{I}.
\label{eq:modeladendenfir}
\end{equation}

 From (\ref{eq:ntildep}) and (\ref{eq:modeladendenfir}),
the first term in the right hand side of (\ref{eq:dendendef}) is
\begin{eqnarray}
\lefteqn{\frac{D(x,y;\sigma)}{\langle \Phi_{G.S.}
\vert \Phi_{G.S.} \rangle}}
\nonumber \\
& &=
\frac{\vec{F}^T
     {\bf T}_{N}
     {\bf T}_{N-1}
     \cdots
     {\bf T}_{j+1}
     {\bf N}_{j}^{(p)}
     {\bf T}_{j-1}
     \cdots
     {\bf T}_{i+1}
     {\bf N}_{i}^{(p)}
     {\bf T}_{i-1}
     \cdots
     {\bf T}_{1}
     \vec{I}}
     {\vec{F}^T
     {\bf T}^{N}
     \vec{I}}
\label{eq:density1st} \\
& &=
\frac{
C_1
-C_2 \left( \frac{e_2}{e_1} \right)^i
+C_3 \left( \frac{e_2}{e_1} \right)^j
-C_4 \left( \frac{e_2}{e_1} \right)^{j-i}
+C_5 \left( \frac{e_2}{e_1} \right)^{N-j+i}
-C_6 \left( \frac{e_2}{e_1} \right)^{N-j}
+C_7 \left( \frac{e_2}{e_1} \right)^{N-i}
-C_8 \left( \frac{e_2}{e_1} \right)^N
}{
\frac{\omega_1^2}{2}
\left\{
\left[
\left(1+2\lambda_1^2\right)\left(1+2\lambda_2^2\right)+\omega_1
\right]
-
\left[
\left(1+2\lambda_1^2\right)\left(1+2\lambda_2^2\right)-\omega_1
\right] \left( \frac{e_2}{e_1} \right)^N
\right\}
},
\nonumber
\end{eqnarray}
where
\begin{eqnarray}
C_1
&=&
\left( e_1-2\lambda_1^2 \right)
\left( e_1-2\lambda_2^2 \right)
\left( \lambda_1^2 +\lambda_2^2 -e_1\right)
\nonumber \\
C_2
&=&
\left( e_1-2\lambda_2^2 \right)
\left( e_2-2\lambda_1^2 \right)
\left( \lambda_1^2 +\lambda_2^2 -e_1\right)
\left( \lambda_2^2 e_1 +\lambda_1^2 e_2 -e_1 e_2 \right)/e_2
\nonumber \\
C_3
&=&
\left( e_1-2\lambda_1^2 \right)
\left( e_2-2\lambda_2^2 \right)
\left( \lambda_1^2 +\lambda_2^2 -e_2 \right)
\left( \lambda_2^2 e_1 +\lambda_1^2 e_2 -e_1 e_2 \right)/e_2
\nonumber \\
C_4
&=&
\left( e_1-2\lambda_1^2 \right)
\left( e_1-2\lambda_2^2 \right)
\left( \lambda_1^2 e_1 +\lambda_2^2 e_2 -e_1 e_2 \right)
\left( e_1 e_2 -\lambda_2^2 e_1 -\lambda_1^2 e_2 \right)/(e_1 e_2)
\nonumber \\
C_5
&=&
\left( e_2-2\lambda_1^2 \right)
\left( e_2-2\lambda_2^2 \right)
\left( \lambda_2^2 e_1 +\lambda_1^2 e_2 -e_1 e_2 \right)
\left( e_1 e_2 -\lambda_1^2 e_1 -\lambda_2^2 e_2 \right)/(e_1 e_2)
\nonumber \\
C_6
&=&
\left( e_1-2\lambda_1^2 \right)
\left( e_2-2\lambda_2^2 \right)
\left( \lambda_1^2 +\lambda_2^2 -e_1 \right)
\left( \lambda_1^2 e_1 +\lambda_2^2 e_2 -e_1 e_2 \right)/e_1
\nonumber \\
C_7
&=&
\left( e_1-2\lambda_1^2 \right)
\left( e_2-2\lambda_2^2 \right)
\left( \lambda_1^2 +\lambda_2^2 -e_2\right)
\left( \lambda_1^2 e_1 +\lambda_2^2 e_2 -e_1 e_2 \right)/e_1
\nonumber \\
C_8
&=&
\left( e_2-2\lambda_1^2 \right)
\left( e_2-2\lambda_2^2 \right)
\left( \lambda_1^2 +\lambda_2^2 -e_2\right),
\nonumber
\end{eqnarray}
in the system size $N$.
In the thermodynamic limit
$N-j$, $i \to \infty$
keeping $\vert j-i \vert$ finite, we have
\begin{eqnarray}
\frac{D(x,y;\sigma)}{\langle \Phi_{G.S.} \vert \Phi_{G.S.} \rangle}
=
\left[
\frac{1}{2}\left(\frac{1}{\omega_1}+1\right)
\right]^2
-
\left(\frac{e_2}{e_1} \right)^{- \vert i-j \vert}
\frac{ \lambda_1^2 \lambda_2^2
\left[ 1+ 2(\lambda_1^2 +\lambda_2^2
+\lambda_1^4 +\lambda_2^4 ) \right]}
      {e_1^2 \omega_1^2}.
\end{eqnarray}From (\ref{eq:npx2}),
the first term in the right hand side cancels out with
the last term in (\ref{eq:dendendef}).
We obtain
\begin{eqnarray}
\langle n_i^p n_j^p \rangle
-\langle n_i^p \rangle \langle n_j^p \rangle
=
-\left(\frac{e_2}{e_1} \right)^{- \vert i-j \vert}
 \frac{4 \lambda_1^2 \lambda_2^2
 \left[ 1+ 2(\lambda_1^2 +\lambda_2^2
+\lambda_1^4 +\lambda_2^4 ) \right]}
      {e_1^2 \omega_1^2}.
\label{eq:denden}
\end{eqnarray}
For the $d$-site, the similar calculation leads
\begin{eqnarray}
\langle n_i^d n_j^d \rangle
-\langle n_i^d \rangle \langle n_j^d \rangle
=
 -\left( \frac{e_2}{e_1} \right)^{- \vert i-j \vert}
  \frac{4 \lambda_1^2 \lambda_2^2}{ (e_1 +e_2) \omega_1}.
\end{eqnarray}
The density correlation functions take negative value
and decay exponentially with distance.
We show the results
in Figs.\ \ref{fig:modelA:denp:res}
and \ \ref{fig:modelA:dend:res}.
For the parameter region (i) identified at the end
of subsection\ \ref{subsec:onepoint},
the density correlation between $p$-sites is enhanced
and that between $d$-sites is suppressed.
For (ii) and (iii), they are suppressed.

Since we have no non-degenerate loop,
the spin correlation functions
$\langle S_i^z S_j^z \rangle$ are vanishing
for $\vert j-i \vert \ge 2$.
Since we have no self-closed bond at the sites
where the adjacent cells are identified,
$\langle b_{i,j}^{\dagger} b_{k,l} \rangle$ is
vanishing
for $\vert k-i \vert \ge 1$.

We evaluate the correlation function
$\langle c_{i,\sigma}^p c_{j,\sigma}^{p \dagger} \rangle$.
The geometric representation of the expectation value is
(\ref{eq:greengeometric}),
where an example of the graph $V \cup V'$ is shown
in Fig.\ \ref{fig:modelAgeorep}(e).
We do not need the procedures shown in Figs.\ \ref{fig:elimination2},
because there is no site with four bonds.
Therefore, we find $n \left( U_i \cup U'_i \right) =1$
in (\ref{eq:decomps21}) and $m_j=0$ in (\ref{eq:propz}).
In the representation, the operators modify
the weight associated with the graph(s) which contains the cells
between $i$ and $j$.
Let the transfer matrix associated with the operator
$c_{i,\sigma}^{p}$ ($c_{j,\sigma}^{p \dagger}$) be
${\bf G}_{i}^{R,(p)}$ (${\bf G}_{j}^{L,(p)}$).
We need a new matrix ${\bf G}_n$ for the $n$-th cell
($i+1 \leq n \leq j-1$). From these matrices the expectation value
can be written
\begin{equation}
\langle c_{i,\sigma}^p c_{j,\sigma}^{p \dagger} \rangle
=\vec{F}^T
 {\bf T}_N
 \cdots
 {\bf T}_{j+1}
 {\bf G}_{j}^{L,(p)}
 {\bf G}_{j-1}
 \cdots
 {\bf G}_{i+1}
 {\bf G}_{i}^{R,(p)}
 {\bf T}_{i-1}
 \cdots
 {\bf T}_1
 \vec{I}.
\label{eq:modelagreen}
\end{equation}
We derive the matrices.
We first consider the matrix ${\bf G}_n$.
It is reduced to a number,
because we have only one kind of graph (a line) on the $n$-th cell
($i+1 \leq n \leq j-1$)
(Fig.\ \ref{fig:modelAprop}(a)).
Let $G_n$ be the quantity defined by a sum.
The sum is taken over $V \cup V'$ on the lattice $\Lambda_n$
such that the graph consists of loops shown
in Fig.\ \ref{fig:modelAnorm} on the $k$-th cell ($1 \le k \le i-1$)
and the line shown in Fig.\ \ref{fig:modelAprop}(a)
on the $n$-th cell ($i \le n \le j$).
A line with $2n$ bonds is on $n$ cells,
since a cell has two valence bonds.
The weight for the line is $(-1)^n$.
We assign $-1$ to each $n$ cells.
In this way, the weight (\ref{eq:proof2})
is automatically taken into account.
They are represented diagrammatically
\begin{eqnarray}
G_n &=& \nonumber\\
&=& \nonumber\\
&=& -\lambda_1\lambda_2G_{n-1}.
\label{eq:gtpp}
\end{eqnarray}
We have
\begin{eqnarray}
{\bf G}_n = - \lambda_1 \lambda_2.
\label{eq:modelAgpp}
\end{eqnarray}
For the matrix ${\bf G}_{i}^{(p)}$,
let $G_i^{R,(p)}$ be the quantity defined by a sum.
The sum is taken over $V \cup V'$ on the lattice $\Lambda_i$
such that the graph consists of loops shown
in Fig.\ \ref{fig:modelAnorm} on the $k$-th cell ($1 \le k \le i-1$)
and the graph shown in Fig.\ \ref{fig:modelAprop}(b) or (c)
on the $i$-th cell.
The recursion relation is
\begin{eqnarray}
G_i^{(p)} &=& \nonumber\\
&=& \nonumber\\
&=& \lambda_1 A_{i-1}
    +\lambda_1 B_{i-1}
    +\lambda_1 \lambda_2^2 A_{i-1},
\end{eqnarray}
and is written
\begin{eqnarray}
G_i^{(p)} = {\bf G}_i^{R,(p)}
  \left( \begin{array} {c}
         A_{j-1} \\
         B_{j-1}
         \end{array} \right), \ \
{\bf G}_i^{R,(p)} = \left( \begin{array} {cc}
         \lambda_1 + \lambda_1 \lambda_2^2, & \lambda_1
         \end{array} \right),
\label{eq:gypp}
\end{eqnarray}
For ${\bf G}_j^{L,(p)}$,
let $A^{(p)}_j$ and $B^{(p)}_j$ be the quantity defined
by the right-hand side of (\ref{eq:greengeometric})
on the lattice $\Lambda_j$.
The sum is taken over the graph $V \cup V'$
such that the graph consists of loops shown
in Fig.\ \ref{fig:modelAnorm}
on the $k$-th cell $(1 \le k \le i-1)$
and the graph shown in Fig.\ \ref{fig:modelAprop}(a)
on the $l$-th cell $(i \le l \le j)$.
The restriction for the sum is that the $j$-th cell
is represented by either
Fig.\ \ref{fig:modelAprop}(d) for $A^{(p)}_j$ and (e) for $B^{(p)}_j$.
The recursion relations are
\begin{eqnarray}
A^{(p)}_j &=& \nonumber \\
        &=& \lambda_2 G_{j-1} \\
B^{(p)}_j &=& \nonumber \\
        &=& \lambda_1^2 \lambda_2 G_{j-1},
\end{eqnarray}
and are written
\begin{eqnarray}
\left(
 \begin{array}{c}
     A^{(p)}_j \\
     B^{(p)}_j
 \end{array}
\right)
= {\bf G}_j^{L,(p)} G_{j-1}, \ \
{\bf G}_j^{L,(p)}
=
\left(
 \begin{array} {c}
    \lambda_2 \\
    \lambda_1^2 \lambda_2
 \end{array}
\right),
\label{eq:gxpp}
\end{eqnarray}From (\ref{eq:normmodelA}), (\ref{eq:modelagreen}),
(\ref{eq:modelAgpp}), (\ref{eq:gypp}), and (\ref{eq:gxpp}),
we obtain
\begin{eqnarray}
\langle c_{i,\sigma}^{p} c_{j,\sigma}^{p \dagger} \rangle
&=&
\frac{
   \vec{F}^T
   {\bf T}^{N-j}
   {\bf G}_{j}^{L,(p)}
   {\bf G}^{j-i-1}
   {\bf G}_{i}^{R,(p)}
   {\bf T}^{i-1}
   \vec{I}
}{ \vec{F}^T
   {\bf T}^{N}
   \vec{I} }
\label{eq:greentrans} \\
&=&
-(-\lambda_1 \lambda_2)^{j-i}
\frac{
C_1 \left( \frac{1}{e_1} \right)^{j-i-1}
-C_2 \left( \frac{e_2}{e_1} \right)^{i-1}
-C_3 \left( \frac{e_2}{e_1} \right)^{N-j}
+C_4 \left( \frac{e_2}{e_1} \right)^N
}{
\omega_1
\left\{
\left[
\left(1+2\lambda_1^2\right)\left(1+2\lambda_2^2\right)+\omega_1
\right]
-
\left[
\left(1+2\lambda_1^2\right)\left(1+2\lambda_2^2\right)-\omega_1
\right] \left( \frac{e_2}{e_1} \right)^N
\right\}
},
\nonumber
\end{eqnarray}
where
\begin{eqnarray}
C_1
&=&
\left( e_1-2\lambda_1^2 \right)
\left( e_1- \lambda_1^2 \right)
\left( e_1-2\lambda_2^2 \right)
\left( e_1- \lambda_2^2 \right)/e_1
\nonumber \\
C_2
&=&
\left( e_1- \lambda_1^2 \right)
\left( e_1-2\lambda_2^2 \right)
\left( e_2-2\lambda_1^2 \right)
\left( e_2- \lambda_2^2 \right)/e_1^{j-i+1}
\nonumber \\
C_3
&=&
\left( e_1-2\lambda_1^2 \right)
\left( e_1- \lambda_2^2 \right)
\left( e_2- \lambda_1^2 \right)
\left( e_2-2\lambda_2^2 \right)/e_1^{j-i+1}
\nonumber \\
C_4
&=&
\left( e_1-2\lambda_1^2 \right)
\left( e_1- \lambda_1^2 \right)
\left( e_1-2\lambda_2^2 \right)
\left( e_1- \lambda_2^2 \right)/e_2^{j-i+1},
\nonumber
\end{eqnarray}
in the system size $N$.
In the thermodynamic limit
$N-j$, $i \to \infty$ keeping $\vert j-i \vert$ finite,
we obtain
\begin{eqnarray}
\langle c_{i,\sigma}^{p} c_{j,\sigma}^{p \dagger} \rangle
&=&
-\left( -\frac{\lambda_1 \lambda_2 }{e_1} \right)^{- \vert i-j \vert}
 \frac{(2\lambda_1^2+1+\omega_1) (2\lambda_2^2+1+\omega_1)}
      {4 e_1 \omega_1}.
\label{eq:pp}
\end{eqnarray}

Next, we evaluate the correlation function
$\langle c_{i,\sigma}^d c_{j,\sigma}^{d \dagger} \rangle$.
We show an example of the graph $V \cup V'$
in Fig.\ \ref{fig:modelAgeorep}(f).
(We have only one kind of the line.)
Using the graph of the ends of the line
(Figs.\ \ref{fig:modelAprop}(f) and (g))
the matrices in (\ref{eq:modelagreen}) are
\begin{eqnarray}
{\bf G}_i^{R,(d)}
=
\left(
 \begin{array}{cc}
    -\lambda_1\lambda_2, & 0
 \end{array}
\right), \ \
{\bf G}_{j}^{L,(d)}
&=&
\left(
 \begin{array}{c}
     1  \\
     2 \lambda_1^2
 \end{array}
\right),
\label{eq:modelAgdd}
\end{eqnarray}From these matrices we obtain
\begin{eqnarray}
\langle c_{i,\sigma}^{d} c_{j,\sigma}^{d \dagger} \rangle
&=&
\frac{1}{ \omega_1}
\left( -\frac{\lambda_1 \lambda_2 }{e_1} \right)^{- \vert i-j \vert}.
\end{eqnarray}
The correlation functions decay exponentially with the oscillating sign.

For a finite lattice under open boundary condition,
the system is not translational invariant.
In the thermodynamic limit, however, by the Fourier transformation
of the correlation function
$\langle c_{i,\sigma}^{\phantom{\dagger}}c_{j,\sigma}^{\dagger} \rangle$,
we obtain the momentum distribution function
\cite{BL93} for $\alpha=p$, $d$
\begin{eqnarray}
\langle n_{k,\sigma}^{\alpha} \rangle
=f^{(\alpha)} F(k,r) + f_0^{(\alpha)},
\end{eqnarray}
where $f_0^{(\alpha)} = \langle n_{i,\sigma}^{\alpha} \rangle$
and
\begin{eqnarray}
F(k,r)=\frac{2 r[\cos{k}-r]}{1+r^2-2 r \cos{k}},
\end{eqnarray}
where
$f^{(p)} = -(2\lambda_1^2+1+\omega_1)
(2\lambda_2^2+1+\omega_1)/4 e_1 \omega_1$,
$f^{(d)} =1/\omega_1$,
and $r= - \lambda_1 \lambda_2 /e_1$.
The results are shown
in Figs.\ \ref{fig:modelA:res:mom1} and \ref{fig:modelA:res:mom2}.
There is no singularity in $\langle n_{k,\sigma}^{\alpha} \rangle$.
For the parameter range (i) in subsection\ \ref{subsec:onepoint},
the momentum distribution for the $d$-site is completely flat,
while that for the $p$-site has a broad peak around $k=0$.
For (ii), the momentum distribution for the $p$-site
is almost unity for every $k$ and is completely flat,
while that for the $d$-site is almost zero.
For (iii), both of them are completely flat.

\subsubsection{Discussion}
\label{subsec:modeladis}

All the correlation functions under consideration
decay exponentially with distance.
These results suggest the existence of a finite excitation gap.
Therefore, it is expected that
the state does not exhibit a metallic state
but rather an insulating one.
The correlation lengths are given by
$\xi_{nn} =\left[ \ln \left(\frac{e_2}{e_1} \right) \right]^{-1}$
for the density correlation functions and
$\xi_{cc}
=\left[ \ln \left(\frac{\lambda_1 \lambda_2}{e_1} \right) \right]^{-1}$
for the correlation functions
$\langle c_{i,\sigma}^{\alpha} c_{j,\sigma}^{\beta \dagger} \rangle$
(Fig.\ \ref{fig:modelA:corrleng}).
The correlation lengths of the correlation between $p$-sites
and that between $d$-sites are the same.
The spin correlation functions vanish for $\vert j-i \vert \ge 2$.
The singlet-pair correlation functions vanish for any $\vert j-i \vert$.

We consider the limit
$\vert \lambda_1 \vert$, $\vert \lambda_2 \vert$ $\ll 1$
($\lambda_1=\lambda_2$).
We obtain $\xi_{nn}$, $\xi_{cc}$ $= 0$.
The density correlation functions for the nearest neighbor
sites vanish.
The ground state is described by a collection of
the decoupled $p$-sites which are doubly occupied.

We consider the limit $\lambda_1$, $\lambda_2$ $\gg 1$
($\lambda_1=\lambda_2$).
The correlation length converges to a finite value:
$\xi_{nn}$ $=$
$\left[ \ln \left( \frac{2-\sqrt{2}}{2+\sqrt{2}} \right) \right]^{-1}$
and
$\xi_{cc}$ $=$
$\left[ \ln \left( \frac{1}{4+2\sqrt{2}} \right) \right]^{-1}$.
The density correlation function for the nearest neighbor $p$-sites
remains finite,
while that for the nearest neighbor $d$-sites vanish,
Since there is almost one electron per a site,
the correlation between $d$-sites is suppressed
and that between $p$-sites is enhanced.

For $\vert \lambda_1 \vert \ll 1$, $\lambda_2$ $\gg 1$,
the correlations are suppressed.
This is due to that
the system decouples to a collection of the pairs
of $p$- and $d$-sites.

The ground state (\ref{eq:gsmodela}) is a half-filling state.
In the non-interacting system,
the filling factor corresponds to that of a metallic state
at $2 \lambda_1 \lambda_2 -1 = 0$
and that of a band insulator
for $2 \lambda_1 \lambda_2 -1 \ne 0$.
Therefore, we have a metal-insulator transition
when $\lambda_1$, $\lambda_2$ are fixed to satisfy
$2 \lambda_1 \lambda_2 -1 = 0$,
and the on-site Coulomb interaction $U$ on $d$-sites is
varied from $0$ to $\infty$.
In the non-interacting system, the correlation length,
which is proportional to the inverse of the energy gap,
takes a finite value for
$\vert \lambda_1 \vert$, $\vert \lambda_2 \vert$ $\ll 1$,
it diverges when $2 \lambda_1 \lambda_2 -1 = 0$ hold,
and it goes to zero for $\lambda_1$, $\lambda_2$ $\gg 1$.
These properties are completely different from that
of the ground state (\ref{eq:gsmodela}).

\subsection{Notation of the transfer matrices}
\label{sec:transferm}

In order to present the calculations in latter models efficiently
we fix some notations. From the derivation
in subsection (\ref{subsec:norm}),
the norm of the ground state can be generally written
\begin{eqnarray}
\langle \Phi_{G.S.} \vert \Phi_{G.S.} \rangle
&=& \vec{F}^T
    {\bf T}_N
    \cdots
    {\bf T}_2
    {\bf T}_1
    \vec{I},
\label{eq:gneralnorm}
\end{eqnarray}
where the matrices depend on the model under consideration.

We describe the expectation value of a local operator ${\cal O}_i$
by using the transfer matrices.
In the geometric representation,
when there is an operator ${\cal O}_i$,
the weight associated with the graph which contains the $i$-th cell
is modified.
Therefore, we replace the transfer matrix ${\bf T}_i$
by ${\bf O}_i$ which is a matrix associated with
the operator ${\cal O}_i$.
The expectation value is written
\begin{eqnarray}
\langle \Phi_{G.S.} \vert {\cal O}_i \vert \Phi_{G.S.} \rangle
&=& \vec{F}^T
    {\bf T}_N
    \cdots
    {\bf T}_{i+1}
    {\bf O}_i
    {\bf T}_{i-1}
    \cdots
    {\bf T}_1
    \vec{I}.
\label{eq:onepoint}
\end{eqnarray}
When ${\cal O}_i$ is the number operator
$n_{i,\sigma}^{\alpha}$
where $\alpha=d$$(p)$ for a $p$($d$)-site,
let the corresponding matrix be
\begin{equation}
{\bf O}_i = {\bf N}_i^{(\alpha)}.
\label{eq:onepointnum}
\end{equation}

For the two-point correlation function
$\langle {\cal O}_i^R {\cal O}_j^L \rangle$,
the weight associated with the graph which contains
the cells between $i$ and $j$ is modified.
Let ${\bf P}_{k}$ be the transfer matrix
between the sites $i+1$ and $j-1$,
and ${\bf O}_i^R$ (${\bf O}_j^L$) be the matrix associated
with the operator ${\cal O}_i^R$ (${\cal O}_j^L$).
We have
\begin{eqnarray}
\langle \Phi_{G.S.} \vert {\cal O}_i^R {\cal O}_j^L
\vert \Phi_{G.S.} \rangle
&=& \vec{F}^T
    {\bf T}_{N}
    {\bf T}_{N-1}
    \cdots
    {\bf T}_{j+1}
    {\bf O}_{j}^L
    {\bf P}_{j-1}
    \cdots
    {\bf P}_{i+1}
    {\bf O}_{i}^R
    {\bf T}_{i-1}
    \cdots
    {\bf T}_{1}
    \vec{I}.
\label{eq:corr1}
\end{eqnarray}
We use the following notations:
\begin{eqnarray}
\begin{array}{@{\,}llll}
{\bf P}_k = {\bf T}_k,
&
{\bf O}_i^R = {\bf N}_i^{(\alpha)},
&
{\bf O}_j^L = {\bf N}_j^{(\alpha)}
&
\mbox{for the quantity $D(i,j;\sigma)$ in (\ref{eq:dendenD})}
\\
{\bf P}_k = {\bf S}_k,
&
{\bf O}_i^R = {\bf S}_i^{R,(\alpha)},
&
{\bf O}_j^L = {\bf S}_j^{L,(\alpha)}
&
\mbox{for the spin correlation function}
\\
{\bf P}_k = {\bf H}_k,
&
{\bf O}_i^R = {\bf H}_i^{R,(\gamma)},
&
{\bf O}_j^L = {\bf H}_j^{L,(\gamma)}
&
\mbox{for the singlet-pair correlation function}
\\
{\bf P}_k = {\bf G}_k,
&
{\bf O}_i^R = {\bf G}_i^{R,(\alpha)},
&
{\bf O}_j^L = {\bf G}_j^{L,(\alpha)}
&
\mbox{for the correlation function
$\langle c_{i,\sigma}^{\phantom{\dagger}}
c_{j,\sigma}^{\dagger} \rangle$.}
\end{array}
\label{eq:twopointgne}
\end{eqnarray}
For the singlet-pair correlation function
$\langle b_{i,j}^{\dagger} b_{k,l}^{\phantom{\dagger}} \rangle$,
we distinguish the following four cases by $\gamma$:
\begin{eqnarray}
\left \{ \begin{array}{@{\,}lll}
\mbox{(i)}   & \mbox{$\gamma$=$p$}\
& \mbox{sites $i$ and $j$ are $p$-sites and $i=j$;}\\
\mbox{(ii)}  & \mbox{$\gamma$=$pp$}\
& \mbox{sites $i$ and $j$ are $p$-sites and $i \ne j$;}\\
\mbox{(iii)} & \mbox{$\gamma$=$pd$}\
& \mbox{site $i$ is $p$-site and site $j$ is $d$-site;}\\
\mbox{(iv)}  & \mbox{$\gamma$=$dd$}\
& \mbox{sites $i$ and $j$ are $d$-sites}.
\end{array} \right.
\label{eq:classifysinglet}
\end{eqnarray}

We can evaluate multi-point correlation functions
for operators which are constructed from fermion operators.
The numerator of the correlation function is obtained by
an insertion of the transfer matrices
which are associated with the operators.

\subsection{Model B}
\label{sec:modelb}

\subsubsection{Hamiltonian}
\label{sec:modelbhami}

Model B is constructed from a cell
with two $p$-sites and one $d$-site
(Fig.\ \ref{fig:modelBcell}(a)).
This is one of the models of Strack \cite{St93}
which was studied by Bares and Lee \cite{BL93}.
The cell Hamiltonian (\ref{eq:cellham}) is obtained by choosing
$\alpha_{n,\sigma}^{(B)}
= \sum_{r=1}^3 \lambda_r c_{r,\sigma}
\equiv \lambda_1^{\phantom{P}} c_{1,\sigma}^p
 +\lambda_2^{\phantom{P}} c_{2,\sigma}^p
+\lambda_3^{\phantom{P}} c_{3,\sigma}^d $ in (\ref{eq:alpha})
and setting $\lambda_3=1$ without loss of generality
(see Fig.\ \ref{fig:modelBcell}(a) for intra-cell index).
The full Hamiltonian is obtained by identifying site $1$
in the $(n-1)$-th cell with site $2$ in the $n$-th cell
(Fig.\ \ref{fig:modelBcell}(b)).
The Hamiltonian is
\begin{eqnarray}
H_S &=& {\cal P}
 \sum_{\sigma = \uparrow, \downarrow}
\Big\{ \sum_{n=1}^{N}
\Big[\ (-\lambda_1 \lambda_2 {c_{n+1,\sigma}^p}^{\dagger} c_{n,\sigma}^p
     -\lambda_1          {c_{n+1,\sigma}^p}^{\dagger} c_{n,\sigma}^d
     -\lambda_2          {c_{n,\sigma}^p}^{\dagger} c_{n,\sigma}^d +h.c.)
\nonumber \\
& &\hspace{40mm}  +\epsilon_n^p {c_{n,\sigma}^p}^{\dagger} c_{n,\sigma}^p
 +\epsilon^d {c_{n,\sigma}^d}^{\dagger} c_{n,\sigma}^d \ \Big]
 +\epsilon_{N+1}^p {c_{N+1,\sigma}^p}^{\dagger} c_{N+1,\sigma}^p \Big\}
 {\cal P},
\label{eq:hamb}
\end{eqnarray}
where the on-site potentials are
$\epsilon_1^p = - \lambda_2^2$,
$\epsilon_n^p =-( \lambda_1^2 + \lambda_2^2 )$ ($2 \le n \le N$),
$\epsilon_{N+1}^p = - \lambda_1^2$,
and $\epsilon^d =-2$.
A unit cell is labeled by $n$.
The ground state is
\begin{eqnarray}
\Big\vert \Phi_{G.S.}^B \Big\rangle &=&
{\cal P}
\prod_{n=1}^N
\prod_{\sigma = \uparrow, \downarrow}
{\alpha_{n,\sigma}^{(B)}}^{\dagger}
\Big\vert 0 \Big\rangle
\nonumber \\
&=&
{\cal P}
\prod_{n=1}^N
\prod_{\sigma = \uparrow, \downarrow}
\left(
 \lambda_1 c_{n,\sigma}^{p \ \dagger}
+\lambda_2 c_{n+1,\sigma}^{p \ \dagger}
+          c_{n,\sigma}^{d \ \dagger}
\right)
\Big\vert 0 \Big\rangle,
\label{eq:gsmodb}
\end{eqnarray}
which is a half-filled state.
In the parameter space, $\lambda_2=-\lambda_1$,
this model reduces one of the models in Ref.\ \cite{St93}.
The model in Ref.\ \cite{BL93} is recovered by setting
$\lambda_1=-\lambda_2=\widetilde{V}^{-1}$.

\subsubsection{Band structure in the single-electron problem}
\label{subsec:bandb}

We investigate the single-electron problem for the Hamiltonian
(\ref{eq:hamb}).
We consider the system with an even number of unit cells
under the periodic boundary condition.
The similar calculation to that in Sec.\ \ref{subsec:banda} leads to
the dispersion relations
\begin{eqnarray}
E_{\pm} =
-\frac{1}{2}
\left[
2 \lambda_1 \lambda_2 \cos{k} + \lambda_1^2 + \lambda_2^2 + 2
\mp
\sqrt{
\left(2 \lambda_1 \lambda_2 \cos{k}
+\lambda_1^2 +\lambda_2^2\right)^2+4
}
\right],
\end{eqnarray}
where $-$, $+$ are the band index with $-$ ($resp.$ $+$)
corresponding to the $+$ ($resp.$ $-$) sign,
and $k$ is the wave vector with (\ref{eq:hasuu}).
The energy gap between two bands is
$\Delta$ $=\frac{1}{2} \sqrt{(\lambda_1 + \lambda_2)^4+4}
+\frac{1}{2} \sqrt{(\lambda_1 - \lambda_2)^4+4}
-2 \lambda_1 \lambda_2$,
which is nonvanishing for any finite $\lambda_1$, $\lambda_2$.

The electron number in the ground state (\ref{eq:gsmodb})
corresponds to full-filling of the lower band.
Therefore, the ground state of the non-interacting system is insulating.

\subsubsection{Correlation functions}
\label{subsec:corrfunmodb}From (\ref{eq:gneralnorm}),
the norm of the ground state is
\begin{eqnarray}
\Big\langle \Phi_{G.S.}^B \Big\vert \Phi_{G.S.}^B \Big\rangle
&=& \frac{e_1^{N}}{c_1}
    (R_{11}L_{11} +2R_{21}L_{21} +R_{31}L_{31}),
\label{eq:norm3b}
\end{eqnarray}
where the corresponding matrices in (\ref{eq:transfmodelA}),
(\ref{eq:transfsmodelA}), and (\ref{eq:matc})
are
\begin{eqnarray}
{\bf T}_n
&=&\left(
 \begin{array}{ccc}
   2\lambda_2^2 +\lambda_2^4  & 2\lambda_2^2   & 0   \\
   \lambda_1^2  +\lambda_1^2\lambda_2^2
& 2\lambda_1^2 +\lambda_1^2\lambda_2^2 & \lambda_1^2 \\
   \lambda_1^4     & 2\lambda_1^4      & \lambda_1^4 \\
 \end{array}
 \right), \ \
\vec{I}
=\left(
 \begin{array}{c}
  1 \\
  0 \\
  0 \\
 \end{array}
 \right), \ \
\vec{F}
=\left(
 \begin{array}{c}
  1 \\
  2 \\
  1 \\
 \end{array}
 \right),
\nonumber \\
{\bf D}
&=&\left(
 \begin{array}{ccc}
  e_1 & 0   & 0   \\
  0   & e_2 & 0   \\
  0   & 0   & e_3 \\
 \end{array}
 \right), \ \
{\rm and} \ \
{\bf C}
=\left(
 \begin{array} {ccc}
   c_1 & 0 & 0 \\
   0 & c_2 & 0 \\
   0 & 0 & c_3 \\
 \end{array}
\right).
\label{eq:mat:transfb}
\end{eqnarray}
The matrix ${\bf T}_n$ and the initial and the final vectors are
derived in Appendix\ \ref{app:transfb}.
Here $e_i$ ($i=1$, $2$, and $3$) are the eigenvalues of ${\bf T}_n$
\begin{eqnarray*}
e_1 &=& \frac{s}{3}
       +\frac{1}{3}\left( p^{\frac{1}{3}} -tp^{\frac{1}{3}} \right) \\
e_2 &=& \frac{s}{3}
       -\frac{1}{6}
        \left[ \left( p^{\frac{1}{3}} -tp^{\frac{1}{3}} \right)
              +i\sqrt{3} \left( p^{\frac{1}{3}} +tp^{\frac{1}{3}} \right)
        \right] \\
e_3 &=& \frac{s}{3}
       -\frac{1}{6}
        \left[ \left( p^{\frac{1}{3}} -tp^{\frac{1}{3}} \right)
              -i\sqrt{3} \left( p^{\frac{1}{3}} +tp^{\frac{1}{3}} \right)
        \right],
\end{eqnarray*}
where
$s= 2\lambda_1 +2\lambda_2 +\lambda_1^2 +\lambda_1\lambda_2 +\lambda_2^2$,
$t= -4\lambda_1^2 -2\lambda_1\lambda_2 -4\lambda_2^2 -4\lambda_1^3
    -2\lambda_1^2 \lambda_2 -2\lambda_1 \lambda_2^2 -4\lambda_2^3
    -\lambda_1^4 +\lambda_1^3\lambda_2 +\lambda_1\lambda_2^3 -\lambda_2^4$,
and $p=(p_1+ 3^{\frac{3}{2}} \sqrt{p_2})/2$.
Here
$p_1 =16\lambda_1^3 +12\lambda_1^2\lambda_2 +12\lambda_1\lambda_2^2
     +16\lambda_2^3
     +24\lambda_1^4 +18\lambda_1^3 \lambda_2 +6\lambda_1^2\lambda_2^2
     +18\lambda_1\lambda_2^3 +24\lambda_2^4
     +12\lambda_1^5 -12\lambda_1^3\lambda_2^2 -12\lambda_1^2\lambda_2^3
     +12\lambda_2^5
     +2\lambda_1^6 -3\lambda_1^5\lambda_2 +14\lambda_1^3\lambda_2^3
     -3\lambda_1\lambda_2^5 +2\lambda_2^6$
and
$p_2= \lambda_1^2 \lambda_2^2 (
     -16\lambda_1^2 -16\lambda_2^2
     -48\lambda_1^3 -32\lambda_1^2\lambda_2 -32\lambda_1\lambda_2^2
     -48\lambda_2^3
     -68\lambda_1^4 -72\lambda_1^3\lambda_2 -60\lambda_1^2\lambda_2^2
     -72\lambda_1\lambda_2^3 -68\lambda_2^4
     -56\lambda_1^5 -40\lambda_1^4\lambda_2 -32\lambda_1^3\lambda_2^2
     -32\lambda_1^2\lambda_2^3 -40\lambda_1\lambda_2^4 -56\lambda_2^5
     -28\lambda_1^6 +12\lambda_1^5\lambda_2 +12\lambda_1^4\lambda_2^2
     +8\lambda_1^3\lambda_2^3 +12\lambda_1^2\lambda_2^4
     +12\lambda_1\lambda_2^5 -28\lambda_2^6
     -8\lambda_1^7 +16\lambda_1^6\lambda_2 -8\lambda_1^4\lambda_2^3
     -8\lambda_1^3\lambda_2^4 +16\lambda_1\lambda_2^6 -8\lambda_2^7
     -\lambda_1^8 +4\lambda_1^7\lambda_2 -4\lambda_1^6\lambda_2^2
     -4\lambda_1^5\lambda_2^3 +10\lambda_1^4\lambda_2^4
     -4\lambda_1^3\lambda_2^5 -4\lambda_1^2\lambda_2^6
     +4\lambda_1\lambda_2^7 -\lambda_2^8 )$.
They satisfy $e_1>e_2>e_3>0$ for $\lambda_1 \neq 0$
and $\lambda _2 \neq 0$.
The matrix ${\bf L}=(L_{ij})$ (${\bf R}=(R_{ij})$) is
constructed from the left(right) eigenvectors.
We choose the left eigenvectors
$\vec{L}_1$ $=(L_{11}, L_{12}, L_{13})^T$,
$\vec{L}_2$ $=(L_{21}, L_{22}, L_{23})^T$, and
$\vec{L}_3$ $=(L_{31}, L_{32}, L_{33})^T$
corresponding to the eigenvalues $e_1$, $e_2$, and $e_3$, respectively,
where
$L_{j1}=\lambda_1^2(\lambda_1^4-e_j)(1+\lambda_2^2)-\lambda_1^6$,
$L_{j2}=(\lambda_1^4-e_j)(2\lambda_2^2-\lambda_2^4-e_j)$,
and
$L_{j3}=\lambda_1^2(2\lambda_2^2-\lambda_2^4-e_j)$.
We choose the right eigenvectors
$\vec{R}_1$ $=(R_{11}, R_{21}, R_{31})^T$,
$\vec{R}_2$ $=(R_{12}, R_{22}, R_{32})^T$, and
$\vec{R}_3$ $=(R_{13}, R_{23}, R_{33})^T$,
where
$R_{1j}=2\lambda_2^2(\lambda_1^4-e_j)$,
$R_{2j}=(\lambda_1^4-e_j)(e_j-2\lambda_2^2-\lambda_2^4)$, and
$R_{3j}=2\lambda_1^4(\lambda_2^2+\lambda_2^4-e_j)$.

We evaluate the expectation value of the number operator
$\langle n_{i,\sigma}^{\alpha} \rangle$.
The transfer matrices associated with $n_{i,\sigma}^{\alpha}$ are
\begin{eqnarray}
{\bf N}_i^{(p)}
=
\left(
 \begin{array}{ccc}
    \lambda_2^2 +\lambda_2^4
  & 2\lambda_2^2
  & 0           \\
   \frac{1}{2}\lambda_1^2\lambda_2^2
  & \lambda_1^2 +\lambda_1^2 \lambda_2^2
  & \lambda_1^2 \\
    0
  & \lambda_1^4
  & 0           \\
 \end{array}
\right), \ \
{\bf N}_i^{(d)}
=
\left(
 \begin{array}{ccc}
    \lambda_2^2
  & \lambda_2^2
  & 0                      \\
    \frac{1}{2}\lambda_1^2
  & \lambda_1^2
  & \frac{1}{2}\lambda_1^2 \\
    0
  & 0
  & 0                      \\
 \end{array}
\right).
\label{eq:Bn}
\end{eqnarray}
They are derived in Appendix\ \ref{app:matoccb}. From
(\ref{eq:occmodelA}), (\ref{eq:norm3b}), (\ref{eq:mat:transfb}),
and (\ref{eq:Bn}),
we obtain
\begin{eqnarray}
\langle n_{i,\sigma}^{\alpha} \rangle
&=&
\left \{
\begin{array}{@{\,}ll}
       \left( L_{11}-\frac{1}{2} \lambda_1L_{12} \right)\lambda_2 R_{11}
       +(L_{12}-L_{13})R_{21}
       +L_{13} R_{31}
     & {\rm for} \ \ \alpha=p; \\
       (\lambda_2L_{11} +\frac{1}{2} \lambda_1L_{12})R_{11}
       +(\lambda_2L_{11} +\lambda_1L_{12}       )R_{21}
       +\frac{1}{2} \lambda_1L_{12} R_{31}
     & {\rm for} \ \ \alpha=d,
     \end{array} \right.
\end{eqnarray}
in the thermodynamic limit, $N$, $i$, $N-i$ $\to \infty$.
It can be verified that there are two electrons per a unit cell as
$\langle n_i^p \rangle
+\langle n_i^d \rangle =2$.
The results are shown in Figs.\ \ref{fig:modelB:occ:res}.
The case $\lambda_2=\vert \lambda_1 \vert$ was disscussed
in Ref.\ \cite{BL93}.
We consider the case
$\vert \lambda_1 \vert \ll 1$, $\lambda_2$ $\gg 1$.
The occupation on $p$- ($d$-) site have a minimum (maximum) at
a intermediate value of $\lambda_1$.

 From (\ref{eq:dendendef}), (\ref{eq:occmodelA}),
(\ref{eq:density1st}), (\ref{eq:norm3b}), (\ref{eq:mat:transfb}),
and (\ref{eq:Bn}),
the density correlation functions are
\begin{eqnarray}
\langle n_i^{\alpha} n_j^{\alpha} \rangle
-\langle n_i^{\alpha} \rangle \langle n_j^{\alpha} \rangle
&=&
  \left(\frac{e_2}{e_1} \right)^{- \vert i-j \vert}
  \frac{1}{e_1 e_2 c_1 c_2} R_2^{(\alpha)} L_2^{(\alpha)}
+ \left(\frac{e_3}{e_1} \right)^{- \vert i-j \vert}
  \frac{1}{e_1 e_3 c_1 c_3} R_3^{(\alpha)} L_3^{(\alpha)},
\end{eqnarray}
where
\begin{eqnarray*}
R_j^{(p)}
&=&
\left( L_{11} -L_{12}/2 \right) R_{1j}
+( L_{12} -L_{13} ) R_{2j}
+ L_{13}R_{3j} \\
L_j^{(p)}
&=&
R_{11}L_{j1}
+\left( -R_{11}/2 +R_{21}\right) L_{j2}
+( -R_{21} +R_{31} ) L_{j3} \\
R_j^{(d)}
&=&
\left( \lambda_2L_{11} +\lambda_1L_{12}/2 \right) R_{1j}
+\left( \lambda_2L_{11} +\lambda_1L_{12}           \right) R_{2j}
+\lambda_1L_{12}R_{3j}/2 \\
L_j^{(d)}
&=&
\lambda_2 (R_{11} +R_{21}) L_{21}
+\lambda_1 \left( R_{11}/2 +R_{21} +/2R_{31} \right) L_{22},
\end{eqnarray*}
in the thermodynamic limit
$N-j$, $i \to \infty$ keeping $\vert j-i \vert$ finite.
The density correlation functions take negative value
and decay exponentially with distance.
We show the results
in Figs.\ \ref{fig:modelB:den:res1} and \ref{fig:modelB:den:res2}.
For $\vert \lambda_1 \vert \ll 1$, $\lambda_2$ $\gg 1$,
the density correlations are suppressed.
They have a minimum at a intermediate value of $\lambda_1$.

We evaluate the spin correlation function.
The transfer matrices are
\begin{eqnarray}
{\bf S}_n &=& \lambda_1^2 \lambda_2^2, \nonumber \\
{\bf S}_{i}^{R,(p)}
= \left( \begin{array} {ccc}
              \frac{1}{2}\lambda_1^2 \lambda_2^2,  & 0,  & 0 \\
         \end{array} \right),
& &
{\bf S}_{j}^{L,(p)}=
\left(
 \begin{array}{c}
    0 \\
    - \lambda_1^2 \\
    - \lambda_1^4 \\
 \end{array}
\right),
\label{eq:spintrb1} \\
{\bf S}_{i}^{R,(d)}
= \left( \begin{array} {ccc}
    \frac{1}{2}\lambda_1^2, & \lambda_1^2, & \frac{1}{2}\lambda_1^2 \\
    \end{array} \right),
& &
{\bf S}_j^{L,(d)}
=
\left( \begin{array}{c}
        - \lambda_2^2 \\
        0 \\
        0 \\
       \end{array} \right). \nonumber
\end{eqnarray}
They are derived in Appendix\ \ref{app:matspinb}. From
(\ref{eq:corr1}), (\ref{eq:twopointgne}), (\ref{eq:norm3b}),
and (\ref{eq:spintrb1}), we have
\begin{eqnarray}
\lefteqn{\langle S_i^z S_j^z \rangle }
\nonumber \\
&=&
\frac{
      \vec{F}^T
      {\bf T}^{N-j}
      {\bf S}_{j}^{L,(\alpha)}
      {\bf S}^{j-i-1}
      {\bf S}_{i}^{R,(\beta)}
      {\bf T}^{i-1}
      \vec{I}
    }{
      \vec{F}^T
      {\bf T}^{N}
      \vec{I}
    }
\label{eq:spincorrB}    \\
&=&
\left(
\frac{\lambda_1^2 \lambda_2^2}{e_1}
\right)^{- \vert i-j \vert}
\times
\left \{
 \begin{array}{@{\,}ll}
    -\frac{\lambda_1^2}{2 e_1 c_1}
     \left( L_{12} + \lambda_1^2 L_{13} \right) R_{11}
& \mbox{for $\beta=p$ and $\alpha=p$} \\
    -\frac{1  }{2 e_1 c_1}
     L_{11} \left( R_{11} +2 R_{21} +R_{31} \right)
& \mbox{for $\beta=d$ and $\alpha=d$} \\
    -\frac{\lambda_1^2}{2\lambda_2^2 e_1c_1}
     \left( L_{12} +\lambda_1^2 L_{13} \right)
     \left( R_{11} +2 R_{21} +R_{31} \right)
& \mbox{for $\beta=p$ and $\alpha=d$} \\
    -\frac{\lambda_2^2}{2 e_1 c_1}
     \left( L_{11} R_{11} \right)
& \mbox{for $\beta=d$ and $\alpha=p$.} \\
 \end{array}
\right.
\hspace{5mm}
\end{eqnarray}
We note that $\langle S_{x+1}^{z,p} S_{x}^{z,d} \rangle
\neq \langle S_x^{z,p} S_{x}^{z,d} \rangle$,
because the Hamiltonian is not invariant
under the reflection of the lattice.
All the spin correlation functions take negative value
and decay exponentially with distance.
We show the results
in Figs.\ \ref{fig:modelB:spin:res}.
For $\vert \lambda_1 \vert \ll 1$, $\lambda_2$ $\gg 1$,
the spin correlations are suppressed.

We evaluate the singlet-pair correlation function (\ref{eq:genebb})
where $i$ and $j$ ($k$ and $l$) are in the same cell.
For (\ref{eq:classifysinglet})-(ii),
the correlation function is vanishing.
We evaluate them for (\ref{eq:classifysinglet})-(i) and (iii).
(The position of the operators are shown
in Fig.\ \ref{fig:modelBdemo}(g) and (h)
in Appendix\ \ref{app:matsingleb},
where a double solid (broken) line represents the operator
$b_{i,j}^{\dagger}$ $(b_{k,l})$.)
The transfer matrices are
\begin{eqnarray}
{\bf H}_n &=& 2 \nonumber \\
{\bf H}_i^{R,(pp)} =
 \lambda_1^2 \lambda_2^2 \left( \begin{array} {ccc}
              1,  & 0,  & 0 \\
           \end{array} \right),
& &
{\bf H}_k^{L,(pp)}
=\left( \begin{array}{c}
        0 \\
        2 \lambda_1^2 \\
        2 \lambda_1^4 \\
       \end{array} \right)
\label{eq:singletrb}
\\
{\bf H}_i^{R,(dp)}
=2\lambda_1^2\lambda_2 \left( \begin{array} {ccc}
              1,  & 1,  & 0 \\
           \end{array} \right),
& &
{\bf H}_k^{L,(dp)}
=\left( \begin{array}{c}
        0 \\
        2 \lambda_1 \lambda_2^2 \\
        0 \\
       \end{array} \right). \nonumber
\end{eqnarray}
They are derived in Appendix\ \ref{app:matsingleb}. From
(\ref{eq:corr1}), (\ref{eq:twopointgne}), (\ref{eq:norm3b}),
and (\ref{eq:singletrb}),
we obtain
\begin{eqnarray}
\langle  b_{i,j}^{\dagger} b_{k,l} \rangle
&=&
\frac{
      \vec{F}^T
      {\bf T}^{N-k}
      {\bf H}_{k}^{L,(\gamma)}
      {\bf H}^{k-i-1}
      {\bf H}_{i}^{R,(\gamma)}
      {\bf T}^{i-1}
      \vec{I}
    }{
      \vec{F}^T
      {\bf T}^{N}
      \vec{I}
    }
\label{eq:singletpairB}    \\
&=&
\left( \frac{\lambda_1^2 \lambda_2^2}{e_1}
\right)^{- \vert i-k \vert} \times
\left \{
\begin{array}{@{\,}ll}
       L_{13} R_{11}
     & {\rm for} \ \ \gamma=p \\
       \frac{4\lambda_1\lambda_2}{e_1c_1}L_{12}(R_{11}+R_{21})
     & {\rm for} \ \ \gamma=dp.
\end{array} \right.
\end{eqnarray}
The singlet-pair correlation function
decay exponentially with distance.
The results are shown in Fig.\ \ref{fig:modelB:single:res}
for $\gamma=p$.
For $\vert \lambda_1 \vert \ll 1$, $\lambda_2$ $\gg 1$,
the singlet-pair correlations are suppressed.

We evaluate the correlation function
$\langle c_{i,\sigma}^{\alpha} c_{j,\sigma}^{\alpha \dagger} \rangle$.
The transfer matrices are
\begin{eqnarray}
{\bf G}_n
= -\lambda_1 \lambda_2 & &
\left(
 \begin{array}{cc}
  1 + \lambda_2^2  & 1     \\
  \lambda_1^2      & \lambda_1^2
 \end{array}
\right) \nonumber \\
{\bf G}_{i}^{R,(p)}
=
 -\lambda_1 \lambda_2
\left(
 \begin{array}{ccc}
   1 +\lambda_2^2 & 1           & 0  \\
   \lambda_1^2    & \lambda_1^2 & 0
 \end{array}
\right),
& & \ \
{\bf G}_{j}^{L,(p)}
=
\left(
 \begin{array}{cc}
   \lambda_2^2                                    & 0           \\
   \lambda_1^2 +\frac{1}{2}\lambda_1^2\lambda_2^2 & \lambda_1^2 \\
   \lambda_1^4                                    & \lambda_1^4
 \end{array}
\right)
\label{eq:bgreen1} \\
{\bf G}_{i}^{R,(d)}
=
 -\lambda_1
\left(
 \begin{array}{ccc}
    \lambda_2^2 & \lambda_2^2  & 0  \\
    \lambda_1^2 & 2\lambda_1^2 & \lambda_1^2
 \end{array}
\right),
& & \ \
{\bf G}_{j}^{L,(d)}
= - \lambda_2
\left(
 \begin{array}{cc}
    \lambda_2^2            & 0                       \\
    \frac{1}{2}\lambda_1^2 & \frac{1}{2} \lambda_1^2 \\
    0                      & 0
 \end{array}
\right). \nonumber
\end{eqnarray}
They are derived in Appendix\ \ref{app:propb}.
The matrix ${\bf G}_n$ is diagonalized
\begin{equation}
{\bf G}_n= -\lambda_1 \lambda_2
\widetilde{\bf R} \widetilde{\bf D}
\widetilde{\bf C}^{-1} \widetilde{\bf L},
\label{eq:diagonalization2}
\end{equation}
where
\begin{equation}
\widetilde{\bf D}
=\left( \begin{array} {cc}
       g_1 & 0   \\
       0   & g_2 \\
       \end{array} \right), \
\widetilde{\bf R}
=\left( \begin{array} {cc}
       \widetilde{R}_{11} & \widetilde{R}_{12} \\
       \widetilde{R}_{21} & \widetilde{R}_{22} \\
       \end{array} \right), \
\widetilde{\bf L}
=\left( \begin{array} {cc}
       \widetilde{L}_{11} & \widetilde{L}_{12} \\
       \widetilde{L}_{21} & \widetilde{L}_{22} \\
       \end{array} \right), \
\widetilde{\bf C}
=\left( \begin{array} {cc}
       \widetilde{c_1} & 0 \\
       0               & \widetilde{c_2}
       \end{array} \right).
\label{eq:greentrmodelB}
\end{equation}
Here $g_k$ are the eigenvalues of ${\bf G}_n$,
$g_1 =(1+\lambda_1+\lambda_2+\omega_2)/2$
and
$g_2 =(1+\lambda_1+\lambda_2-\omega_2)/2$
with
$\omega_2 =\sqrt{1+2(\lambda_1+\lambda_2) +(\lambda_1-\lambda_2)^2}$.
They satisfy $g_1 > g_2  > 0 $ for $\lambda_1 \neq 0$
and $\lambda_2 \neq 0$.
We choose the left eigenvectors
$\widetilde{L}_1$ $=(\widetilde{L}_{11}, \widetilde{L}_{12})^T$ and
$\widetilde{L}_2$ $=(\widetilde{L}_{21}, \widetilde{L}_{22})^T$
corresponding to the eigenvalues $g_1$ and $g_2$, respectively,
where
$\widetilde{L}_{j1}=g_j-\lambda_1^2$,
$\widetilde{L}_{j2}=1$,
and the right eigenvectors
$\widetilde{R}_1$ $=(\widetilde{R}_{11}, \widetilde{R}_{21})^T$ and
$\widetilde{R}_2$ $=(\widetilde{R}_{12}, \widetilde{R}_{22})^T$
where
$\widetilde{R}_{1j}=g_j -\lambda_1^2$,
$\widetilde{R}_{2j}=\lambda_1^2$.
Here $\widetilde{\bf C}$ $= \widetilde{\bf L} \widetilde{\bf R}$.

 From these matrices and (\ref{eq:greentrans}), we obtain
\begin{eqnarray}
\langle c_{i,\sigma} c_{j,\sigma}^{\dagger} \rangle
&=&
 \left( -\lambda_1 \lambda_2
        \frac{g_1}{e_1}
 \right)^{- \vert i-j \vert} f_1^{(\alpha)}
+\left( -\lambda_1 \lambda_2
        \frac{g_2}{e_1}
 \right)^{- \vert i-j \vert} f_2^{(\alpha)},
\end{eqnarray}
where
\begin{eqnarray}
f_m^{(p)}
&=&
 \frac{1}{c_1 \widetilde{c}_m}
 \left[ \frac{1}{2}(g_m-\lambda_1^2)L_{12} +\lambda_1^2L_{13} \right]
 \left[ ( g_m-\lambda_1^2 )R_{11} +R_{21} \right], \\
f_m^{(d)} &=&
 \frac{1}{e_1 g_m c_1 \widetilde{c}_m}
 \left[ \lambda_2^2(g_m-\lambda_1^2)L_{11}
        +\frac{1}{2}g_m\lambda_1^2L_{12} \right]
\nonumber \\
& & \hspace{25mm}\times
 \left[ \lambda_2^2 (g_m-\lambda_1^2) \left( R_{11} +R_{21} \right)
        +\lambda_1^2( R_{11} +2R_{21} +R_{31})\right].
\end{eqnarray}
The correlation functions decay exponentially.

By the Fourier transformation of the correlation function
$\langle c_{i,\sigma}^{\phantom{\dagger}}c_{j,\sigma}^{\dagger} \rangle$,
we obtain the momentum distribution function for
$\alpha=p$ and $d$
\begin{eqnarray}
\langle n_{k,\sigma}^{\alpha} \rangle
=f_1^{(\alpha)} F_1(k,r_1) + f_2^{(\alpha)} F_2(k,r_2) + f_0^{(\alpha)},
\end{eqnarray}
where $f_0^{(\alpha)} = \langle n_{i,\sigma}^{\alpha} \rangle$
and
\begin{eqnarray}
F_i(k,r_i)=\frac{2 r_i[\cos{k}-r_i]}{1+r_i^2-2 r_i \cos{k}}.
\label{eq:fourierF}
\end{eqnarray}
where
$r_i = -\lambda_1 \lambda_2 g_i/e_1$.
The results are shown in Figs. \ref{fig:modelB:mom:res1}
and \ref{fig:modelB:mom:res2}.
There is no singularity in $\langle n_{k,\sigma}^{\alpha} \rangle$.
For $\vert \lambda_1 \vert \ll 1$, $\lambda_2$ $\gg 1$,
the momentum distributions for the $p$- and $d$-sites are flat.

\subsubsection{Discussion}
\label{subsec:modelbdis}

All the correlation functions under consideration
decay exponentially with distance.
These resluts suggest the existence of a finite excitation gap.
The existence of the energy gap is numerically confirmed
in Ref.\ \cite{Kimura}.
Therefore, it is expected that the state is insulating.
The correlation lengths are given by
$\xi_{nn}=\left[ \ln \left( \frac{e_2}{e_1} \right) \right]^{-1}$,
$\xi_{ss}=\xi_{bb}=
\left[ \ln \left( \frac{\lambda_1^2 \lambda_2^2}
                       {e_1} \right) \right]^{-1}$,
for the spin and singlet-pair correlation functions, respectively,
and
$\xi_{cc}
=\left[ \ln \left( \frac{\lambda_1^2 \lambda_2^2 g_1}{e_1}
                         \right) \right]^{-1}$
(Fig.\ \ref{fig:modelB:corrleng}).
(They satisfy the relation
$\xi_{cc}$ $> \xi_{nn}$ $> \xi_{ss}$ $= \xi_{bb}$.)
The correlation lengths of the correlation between
$p$-sites and that between $d$-sites are same.
We note that the spin correlation is ferromagnetic.

We consider the region $\lambda_1$, $\lambda_2$ $\gg 1$.
The $d$-sites are almost empty and the $p$-sites
are almost doubly occupied.
The correlation lengths behave
$\xi_{cc} \sim \lambda_1$ and
$\xi_{nn}$, $\xi_{ss}$, $\xi_{bb} \sim \frac{\lambda_1}{2}$
(for $\lambda_1=\lambda_2$).
The correlation functions for the nearest neighbor sites
are suppressed.

For $\vert \lambda_1 \vert$, $\vert \lambda_2 \vert$ $\ll 1$,
there is almost one electron per a site.
The density correlation function for the nearest neighbor
$p$-sites are enhanced,
while that for $d$-sites are suppressed.
The spin correlation function for the nearest neighbor
$p$-sites are suppressed, and that for $d$-sites are enhanced.
Therefore, the electrons on the $d$-sites have a tendency
to behave like a localized spin.
It corresponds to a kind of Kondo lattice regime \cite{BL93}
in the sense that there are one localized electron
and one conduction electron per a unit cell.
The ground state is described by a collection
of local singlets between them.
The effective exchange coupling between the $p$- and $d$-sites
$J \sim \frac{\lambda_1 \lambda_2}{\epsilon^p - \epsilon^d}$
is comparable to the hopping amplitude between $p$-sites.

For $\vert \lambda_1 \vert \ll 1$, $\lambda_2$ $\gg 1$,
the correlations are suppressed.
This is due to that
the system decouples to a collection of the pairs
of $p$- and $d$-sites.

The ground state (\ref{eq:gsmodb}) is a half-filling state.
The filling factor corresponds to that of a band insulator
in the non-interacting system
where the excitation gap satisfies the relation $1 < \Delta < 2$.
Therefore, the correlation length is finite and is almost independent
on the parameters $\lambda_1$ and $\lambda_2$,
which is different from that of the ground state (\ref{eq:gsmodb}).
The properties of the ground state (\ref{eq:gsmodb})
are completely different from that of the non-interacting system.

\subsection{Model C}
\label{sec:modelc}

\subsubsection{Hamiltonian}
\label{sec:modelchami}

The lattice of Model C is constructed from a cell
with four $d$-sites and one $p$-site (Fig.\ \ref{fig:modelCcell}(a)).
The model has four free parameters.
We investigate the simplest model with one parameter.
The cell Hamiltonian is obtained by choosing
$\alpha_{n,\sigma}^{(C)}$
$=$
$\sum_{r=1}^5 \lambda_r c_{r,\sigma}$
$\equiv$
$ \lambda_1^{\phantom{P}} c_{1,\sigma}^{d}$
$+\lambda_2^{\phantom{P}} c_{2,\sigma}^{d}$
$+\lambda_3^{\phantom{P}} c_{3,\sigma}^d$
$+\lambda_4^{\phantom{P}} c_{4,\sigma}^d$
$+\lambda_5^{\phantom{P}} c_{5,\sigma}^p$
and setting $\lambda_r=1$ $(r=1,2,3,4)$ and $\lambda_5=\lambda$
(see Fig.\ \ref{fig:modelCcell}(a) for intra-cell index).
The full Hamiltonian is obtained by identifying sites $1$ and $2$
in the $(n-1)$-th cell with sites $3$ and $4$ in the $n$-th cell,
respectively
(Fig.\ \ref{fig:modelCcell}(b)).
The Hamiltonian is
\begin{eqnarray}
H_S = {\cal P}
 \sum_{\sigma = \uparrow, \downarrow} \Big\{
 \sum_{n=1}^{N}
 \Big[ & \Big( &
           -\lambda c_{n,\sigma}^{d1 \dagger} c_{n,\sigma}^p
           -\lambda c_{n,\sigma}^{d2 \dagger} c_{n,\sigma}^p
           -\lambda c_{n+1,\sigma}^{d1 \dagger} c_{n,\sigma}^p
           -\lambda c_{n+1,\sigma}^{d2 \dagger} c_{n,\sigma}^p
\nonumber \\
       & & -       c_{n,\sigma}^{d1 \dagger} c_{n,\sigma}^{d2}
           -        c_{n+1,\sigma}^{d1 \dagger} c_{n,\sigma}^{d1}
           -        c_{n+1,\sigma}^{d2 \dagger} c_{n,\sigma}^{d2}
           -        c_{n+1,\sigma}^{d1 \dagger} c_{n,\sigma}^{d2}
           -        c_{n+1,\sigma}^{d2 \dagger} c_{n,\sigma}^{d1} + h.c.
         \Big) \nonumber \\
       &+& \epsilon^p c_{n,\sigma}^{p \dagger} c_{n,\sigma}^p
          +\epsilon_n^d c_{n,\sigma}^{d1 \dagger} c_{n,\sigma}^{d1}
          +\epsilon_n^d c_{n,\sigma}^{d2 \dagger} c_{n,\sigma}^{d2}
        \Big] \nonumber \\
     &+&\epsilon_{N+1}^d c_{N+1,\sigma}^{d1 \dagger} c_{N+1,\sigma}^{d1}
      + \epsilon_{N+1}^d c_{N+1,\sigma}^{d2 \dagger} c_{N+1,\sigma}^{d2}
 \Big\} {\cal P},
\label{eq:hamc}
\end{eqnarray}
where the on-site potentials are
$\epsilon^p =-\lambda^2$,
$\epsilon_n^d =-4$ ($2 \le n \le N$),
and
$\epsilon_1^d$ $=\epsilon_{N+1}^d =-2$.
A unit cell is labeled by $n$.
Here $c_{n,\sigma}^{d1}$ ($c_{n,\sigma}^{d2}$) is
the annihilation operator on a $d$-site for $r=3$ ($4$)
in the $n$-th cell.
The ground state is
\begin{eqnarray}
\Big\vert \Phi_{G.S.}^C \Big\rangle &=&
{\cal P}
\prod_{i=1}^N
\prod_{\sigma = \uparrow, \downarrow}
{\alpha_{n,\sigma}^{(C)}}^{\dagger}
\Big\vert 0 \Big\rangle
\nonumber \\
&=&
{\cal P}
\prod_{n=1}^N
\prod_{\sigma = \uparrow, \downarrow}
\left(
 c_{n,\sigma}^{d1 \ \dagger}
+c_{n,\sigma}^{d2 \ \dagger}
+c_{n+1,\sigma}^{d1 \ \dagger}
+c_{n+1,\sigma}^{d2 \ \dagger}
+\lambda c_{n,\sigma}^{p \ \dagger}
\right)
\Big\vert 0 \Big\rangle,
\label{eq:gsmodc}
\end{eqnarray}
which is a $1/3$-filling state.

\subsubsection{Band structure in the single-electron problem}
\label{subsec:bandc}

We investigate the single-electron problem for the Hamiltonian
(\ref{eq:hamc}).
We consider the system with an even number of the cells
under the periodic boundary conditions.
The similar calculation to that in Sec.\ \ref{subsec:banda} leads
the dispersion relations
\begin{eqnarray}
E_1 &=&
-\frac{1}{2}
\left[
6 + \lambda^2 + 4 \cos{k}
+\sqrt{
\left( 6 + \lambda^2 + 4 \cos{k} \right)^2 -8 \lambda^2
}
\right]
\nonumber \\
E_2 &=& -2
\\
E_3 &=&
-\frac{1}{2}
\left[
6 + \lambda^2 + 4 \cos{k}
-\sqrt{
\left( 6 + \lambda^2 + 4 \cos{k} \right)^2 -8 \lambda^2
}
\right],
\nonumber
\end{eqnarray}
where $1$, $2$, and $3$ are the band index,
and $k$ is the wave vector with (\ref{eq:hasuu}).
The energy gap between the lowest two bands is
$\Delta=0$ for $0 < \lambda < \sqrt{2}$
and $\Delta=\lambda^2$ for $\lambda > \sqrt{2}$.

The electron number in the ground state (\ref{eq:gsmodc})
corresponds to full-filling of the lowest band.
Therefore, the ground state of the non-interacting system is metallic
for $0 < \lambda < \sqrt{2}$
and is insulating
for $\lambda > \sqrt{2}$.

\subsubsection{Correlation functions}
\label{subsec:corrfunmodc}From (\ref{eq:gneralnorm}),
the norm of the ground state is
\begin{eqnarray}
\Big\langle \Phi_{G.S.}^C \Big\vert \Phi_{G.S.}^C \Big\rangle
&=&\frac{e_1^{N}}{c_1}
\left( R_{11}L_{11}+4R_{21}L_{21}+R_{31}L_{31} \right),
\label{eq:norm(c)}
\end{eqnarray}
where the corresponding matrices in (\ref{eq:mat:transfb}) are
\begin{eqnarray}
{\bf T}_n
&=&\left(
 \begin{array}{ccc}
  2 +4\lambda^2 +\lambda^4 & 4\lambda^2 +4\lambda^4 & \lambda^4 \\
  2 +\lambda^2             & 2 +4\lambda^2          & \lambda^2 \\
  2                        & 8                      & 2         \\
 \end{array}
 \right), \ \
\vec{I}
=\left(
 \begin{array}{c}
  1 \\
  0 \\
  0 \\
 \end{array}
 \right), \ \
\vec{F}
=\left(
 \begin{array}{c}
  1 \\
  4 \\
  1 \\
 \end{array}
 \right) \nonumber \\
{\bf D}
&=&\left(
 \begin{array}{ccc}
  e_1 & 0   & 0   \\
  0   & e_2 & 0   \\
  0   & 0   & e_3 \\
 \end{array}
 \right), \ \
{\rm and} \ \
{\bf C}
=\left(
 \begin{array} {ccc}
   c_1 & 0 & 0 \\
   0 & c_2 & 0 \\
   0 & 0 & c_3 \\
 \end{array}
\right).
\label{eq:mat:transf}
\end{eqnarray}
The matrix ${\bf T}_n$ and the initial and the final vectors are
derived in Appendix\ \ref{app:transfc}.
Here $e_i$ ($i=1$, $2$, and $3$) are the eigenvalues of ${\bf T}_n$
\begin{eqnarray}
e_1 &=& \frac{1}{3} \left( s + \frac{t}{q} \right) \nonumber \\
e_2 &=& \frac{s}{3}
        +\frac{i}{12} \left[ \left(1+2\sqrt{3} \right) \frac{t}{q}
                            +\left(1-2\sqrt{3} \right) q
                      \right] \nonumber \\
e_3 &=& \frac{s}{3}
        -\frac{i}{12} \left[ \left(1+2\sqrt{3} \right) \frac{t}{q}
                            +\left(1-2\sqrt{3} \right) q
                      \right],\nonumber
\end{eqnarray}
where
$s=6+8\lambda^2+\lambda^4$,
$t=\lambda^2(48 +58\lambda^2 +16\lambda^4 +\lambda^6)$,
and
$q=\left(p_1+ 3^{\frac{3}{2}} p_2 i \right)^{\frac{1}{3}}$
with
$p_1=468\lambda^4 +512\lambda^6 +183\lambda^8
+24\lambda^{10} +\lambda^{12}$
and
$p_2=\lambda^6(4096 +6736\lambda^2 +4288\lambda^4 $
$+1328\lambda^6 +192\lambda^8 +9\lambda^{10})$.
They satisfy $e_1>e_2>e_3>0$ for $\lambda \neq 0$.
The matrix ${\bf L}=(L_{ij})$ (${\bf R}=(R_{ij})$) is
constructed from the left(right) eigenvectors.
We choose them as we did in Sec.\ \ref{sec:modelb},
where
$L_{j1}=(e_j-2)^2 -4e_j\lambda^2$,
$L_{j2}=4\lambda^2(e_j\lambda^2+e_j-2)$,
$L_{j3}=\lambda^4(e_j+2)$,
$R_{1j}=(e_j -2)^2 -4e_j\lambda^2$,
$R_{2j}=(2+\lambda^2) e_j -4$, and
$R_{3j}=2(e_j+6)$.

We evaluate the expectation value of the number operator
$\langle n_{i,\sigma}^{\alpha} \rangle$.
The transfer matrices associated with $n_{i,\sigma}^{\alpha}$ are
\begin{eqnarray}
{\bf N}_i^{(p)}
=
\left(
\begin{array}{ccc}
  2\lambda^2 +\lambda^4 & 2\lambda^2 +4\lambda^4 & \lambda^4 \\
 \frac{1}{2}\lambda^2  & 2\lambda^2   & \frac{1}{2}\lambda^2 \\
  0                     & 0                      & 0         \\
\end{array}
\right), \ \
{\bf N}_i^{(d)}
=
\left(
\begin{array}{ccc}
  1 +\lambda^2 & 2\lambda^2 +\lambda^4 & \frac{1}{2}\lambda^4 \\
  \frac{1}{2}  & 1 +\lambda^2          & \frac{1}{2}\lambda^2 \\
  0            & 2                     & 1         \\
\end{array}
\right).
\label{eq:Cn}
\end{eqnarray}
They are derived in Appendix\ \ref{app:matocc}. From
(\ref{eq:occmodelA}), (\ref{eq:norm(c)}), (\ref{eq:mat:transf}),
and (\ref{eq:Cn}),
we obtain
\begin{eqnarray}
\langle n_{i,\sigma}^{\alpha} \rangle
=
\left \{
\begin{array}{@{\,}ll}
  \frac{\lambda^2}{4 c_1}
  \left( 4 L_{11}R_{21} + L_{12}R_{31}  \right)
     & {\rm for} \ \ \alpha=p \\
  \frac{1}{4 c_1}
\left[ \left(2L_{11} -L_{12} +2L_{13}\right) R_{11}
      +\left(2L_{12} -4L_{13}\right) R_{21}
      +2L_{13} R_{31}
\right]
     & {\rm for} \ \ \alpha=d.
     \end{array} \right.
\end{eqnarray}
It can be verified that there are two electrons per a unit cell as
$\langle n_i^p \rangle
 + 2\langle n_i^d \rangle
 =2$.
The results are shown in Fig.\ \ref{fig:modelC:occ:res}.
For $\vert \lambda \vert \ll 1$,
on-site potentials satisfy the relations $\epsilon_d < \epsilon_p$
and $|\epsilon_p - \epsilon_d|$ $\gg$ $\vert \lambda \vert$
(the hybridization).
There is almost one electron per $d$-site.
The $p$-sites are almost empty.
For $\lambda \gg 1$,
on-site potentials satisfy the relations $\epsilon_p \ll \epsilon_d$
and $\epsilon_d - \epsilon_p \gg \lambda$.
The $p$-sites are almost doubly occupied.
The $d$-sites are almost empty.

 From (\ref{eq:dendendef}), (\ref{eq:density1st}), (\ref{eq:mat:transf}),
(\ref{eq:norm(c)}), and (\ref{eq:Cn}),
the density correlation functions are
\begin{eqnarray}
\langle n_i^{\alpha} n_j^{\alpha} \rangle
-\langle n_i^{\alpha} \rangle \langle n_j^{\alpha} \rangle
=
\left \{
\begin{array}{@{\,}ll}
 \left(\frac{e_2}{e_1} \right)^{- \vert i-j \vert}
 \frac{\lambda^4}{4c_1 c_2} R_2^{(p)} L_2^{(p)}
+\left(\frac{e_3}{e_1} \right)^{- \vert i-j \vert}
 \frac{\lambda^4}{4c_1 c_3} R_3^{(p)} L_3^{(p)}
     & {\rm for} \ \ \alpha=p \\
 \left(\frac{e_2}{e_1} \right)^{- \vert i-j \vert}
 \frac{1}{4 c_1 c_2} R_2^{(d)} L_2^{(d)}
+\left(\frac{e_3}{e_1} \right)^{- \vert i-j \vert}
 \frac{1}{4 c_1 c_3} R_3^{(d)} L_3^{(d)}
     & {\rm for} \ \ \alpha=d,
\end{array} \right.
\end{eqnarray}
where
\begin{eqnarray*}
R_j^{(p)} &=& 4L_{11}R_{1j} +L_{12}R_{3j} \\
L_j^{(p)} &=& 4L_{j1}R_{21} +L_{j2}R_{31} \\
R_j^{(d)} &=&  \left( 2L_{11}-L_{12}+2L_{13} \right) R_{1j}
         +2\left( L_{12}-2L_{13}         \right) R_{2j}
         +2L_{13}                                R_{3j} \\
L_j^{(d)} &=&  L_{j1} \left( 2R_{11}                 \right)
         + L_{j2} \left( -R_{11} +2R_{21}        \right)
         +2L_{j3} \left( R_{11} -2R_{21} +R_{31} \right).
\end{eqnarray*}
The density correlation functions take negative value
and decay exponentially with distance.
We show the results.
in Fig.\ \ref{fig:modelC:den:res}.
For $\vert \lambda \vert \ll 1$ and $\lambda \gg 1$,
the density correlations are suppressed.

We evaluate the spin correlation function.
The transfer matrices are
\begin{eqnarray}
{\bf S}_n &=& -2 \nonumber \\
{\bf S}_i^{R,(p)}
= \left( \begin{array} {ccc}
         \frac{1}{2}\lambda^2,  & \frac{5}{2}\lambda^2,  & 0 \\
         \end{array} \right),
\ & & \
{\bf S}_j^{L,(p)}=
 \left(
 \begin{array}{c}
    2\lambda^2 \\
    0 \\
    0
 \end{array}
 \right)
\label{eq:spintrc1} \\
{\bf S}_i^{R,(d)}
= \left( \begin{array} {ccc}
          \frac{1}{2}\lambda^2,  & 0,  & 0 \\
         \end{array} \right),
\ & & \
{\bf S}_j^{L,(d)}=
 \left( \begin{array} {c}
             -\lambda^4 \\
             -\lambda^2 \\
             -2
             \end{array} \right).
\nonumber
\end{eqnarray}
They are derived in Appendix\ \ref{app:matspinc}.  From
(\ref{eq:spincorrB}), (\ref{eq:norm(c)}),
and (\ref{eq:spintrc1}), we have
\begin{eqnarray}
\langle S_i^z S_j^z \rangle
=
\left(
-\frac{2}{e_1}
\right)^{- \vert i-j \vert}
\times
\left \{ \begin{array}{@{\,}ll}
 -\frac{\lambda^4}{2 e_1 c_1} \left( L_{11}R_{11} +5L_{12}R_{21} \right)
    & \mbox{for $\beta=p$ and $\alpha=p$} \\
 -\frac{\lambda^2}{2 e_1 c_1} L_{11} R_{11}
    & \mbox{for $\beta=p$ and $\alpha=d$} \\
 -\frac{\lambda^2}{2 e_1 c_1} L_{13} R_{11}
    & \mbox{for $\beta=d$ and $\alpha=d$.} \\
         \end{array} \right.
\end{eqnarray}
The spin correlation functions decay exponentially
with the oscillating sign.
We show the results
in Fig.\ \ref{fig:modelC:spin:res}.
For $\vert \lambda \vert \ll 1$ and $\lambda \gg 1$,
the spin correlations are suppressed.

We evaluate the singlet-pair correlation function.
For (\ref{eq:classifysinglet})-(iv),
the correlation functions are vanishing for $\vert k-i \vert \ge 2$.
We evaluate them for (\ref{eq:classifysinglet})-(i) and (iii).
(The position of the operators are shown
in Figs.\ \ref{fig:modelCdemo}(e) and (f)
in Appendix\ \ref{app:matsinglec}.)
The transfer matrices are
\begin{equation}
{\bf H}_n = 2, \ \
{\bf H}_{i}^{R,(\gamma)}
= \lambda^2\left( \begin{array} {ccc}
              1,  & 5,  & 0 \\
         \end{array} \right),
\ \
{\bf H}_{j}^{L,(\gamma)}=
 \left(
 \begin{array} {c}
   -4\lambda^2 \\
    0 \\
    0
 \end{array}
 \right),
\label{eq:singletrc}
\end{equation}
for $\gamma=p$ and $dp$.
They are derived in Appendix\ \ref{app:matsinglec}. From
(\ref{eq:singletpairB}), (\ref{eq:norm(c)}), and (\ref{eq:singletrc}),
we obtain
\begin{eqnarray}
\langle b_{i,j}^{\dagger} b_{k,l} \rangle
=
-\left( \frac{2}{e_1} \right)^{- \vert i-k \vert}
\frac{2\lambda^4}{e_1 c_1}
\left( L_{11} R_{11} +5L_{12}R_{21} \right)
\ \ \mbox{for $\alpha$, $\beta=p$, $dp$.}
\end{eqnarray}
The singlet-pair correlation functions decay exponentially
with distance.
We show the results
in Fig.\ \ref{fig:modelC:single:res}.
For $\vert \lambda \vert \ll 1$ and $\lambda \gg 1$,
the correlations are suppressed.

We evaluate the correlation function
$\langle c_{i,\sigma}^{\phantom{\dagger}}c_{j,\sigma}^{\dagger} \rangle$.
The transfer matrices are
\begin{eqnarray}
{\bf G}_{n}
= -2 & &
\left(
 \begin{array}{cc}
   1 + \lambda^2  & \lambda^2     \\
   1              & 1
 \end{array}
\right)
\nonumber \\
{\bf G}_i^{R,(p)}
= -\lambda
\left(
 \begin{array}{ccc}
   2 & 2 & 0  \\
   1 & 4 & 1
 \end{array}
\right),
\ & &\
{\bf G}_j^{L,(p)}
= -\lambda
\left(
 \begin{array}{cc}
   2 & 0  \\
   1 & 1  \\
   0 & 0
 \end{array}
\right)
\label{eq:cgreen1} \\
{\bf G}_i^{R,(d)}
= -
\left(
 \begin{array}{ccc}
   1 + \lambda^2  & \lambda^2 & 0    \\
   1              & 1         & 0
 \end{array}
\right),
\  & & \
{\bf G}_j^{L,(d)}
= -
\left(
 \begin{array}{cc}
   \lambda^2 +\lambda^4   & \lambda^4  \\
   \frac{1}{2} +\lambda^2 & \lambda^2  \\
   2                      & 2
 \end{array}
\right). \nonumber
\end{eqnarray}
They are derived in Appendix\ \ref{app:prpc1}.
The matrix ${\bf G}_n$ is diagonalized as
\begin{equation}
{\bf G}_n= -2
\widetilde{\bf R} \widetilde{\bf D}
\widetilde{\bf C}^{-1} \widetilde{\bf L},
\end{equation}
where the matrices are shown in (\ref{eq:diagonalization2}).
The corresponding quantities are
$g_1 =(2+\lambda^2+\lambda \omega_3)/2$
and
$g_2 =(2-\lambda^2+\lambda \omega_3)/2$
with $\omega_3 = \sqrt{4+\lambda^2}$.
They satisfy $g_1 > g_2  > 0 $ for $\lambda \neq 0$.
We choose the matrices
${\bf \widetilde{L}}=(\widetilde{L}_{ij})$ and
${\bf \widetilde{R}}=(\widetilde{R}_{ij})$
as we did in Sec.\ \ref{sec:modelb},
where
$\widetilde{L}_{j1}=(g_j-1)/\lambda^2$,
$\widetilde{L}_{j2}=1$,
$\widetilde{R}_{1j}=g_j -1$, and
$\widetilde{R}_{2j}=1$.
(See also (\ref{eq:greentrmodelB}).) From
these matrices and (\ref{eq:greentrans}), we obtain
\begin{eqnarray}
\langle c_{i,\sigma} c_{j,\sigma}^{\dagger} \rangle
&=&
 \left( -2 \frac{g_1}{e_1} \right)^{- \vert i-j \vert} f_1^{(\alpha)}
+\left( -2 \frac{g_2}{e_1} \right)^{- \vert i-j \vert} f_2^{(\alpha)}
,
\end{eqnarray}
where
\begin{eqnarray}
f_m^{(p)}
&=&
-\frac{1}{2 c_1 \widetilde{c_m} e_1 g_m}
  \left[ 2(g_m-1) L_{11} +g_m L_{12} \right]
\nonumber \\
& &
\hspace{35mm}\times
 \left[ 2(g_m-1)(R_{11} +R_{21})
+\lambda^2(R_{11}+4R_{21}+R_{31}) \right] \\
f_m^{(d)}
&=&
\frac{2}{\lambda^2 c_1 \widetilde{c_m}}
 \left[ (g_m-1)L_{12} +4L_{13} \right]
 \left[ (g_m-1)R_{11} +\lambda^2R_{21} \right].
\end{eqnarray}
The correlation functions decay exponentially
with the oscillating sign.

By the Fourier transformation of the correlation function
$\langle c_{i,\sigma}^{\phantom{\dagger}}c_{j,\sigma}^{\dagger} \rangle$,
we obtain the momentum distribution functions for $\alpha=p$ and $d$
\begin{eqnarray}
\langle n_{k,\sigma}^{\alpha} \rangle
=f_1^{(\alpha)} F_1(k,r_1) + f_2^{(\alpha)} F_2(k,r_2) + f_0^{(\alpha)},
\end{eqnarray}
where $F_i(k,r_i)$ is defined by (\ref{eq:fourierF}),
$r_i = -g_i/e_1$,
and $f_0^{(\alpha)} = \langle n_{i,\sigma}^{\alpha} \rangle$.
The results are shown in Fig.\ \ref{fig:modelC:mom:res1}.
There is no singularity in the momentum distribution functions.
For the momentum distribution for $p$-site,
it has a sharp peak at $k=\pi$ for $\vert \lambda \vert \ll 1$.
The peak is, however, expected to vanish in the complete limit.
It is almost flat for $\lambda \gg 1$.
The momentum distribution for $d$-site
is expected to be flat for the complete limit
$\vert \lambda \vert \ll 1$.

\subsubsection{Discussion}
\label{subsec:modelcdis}

All the correlation functions under consideration
decay exponentially with distance.
These results suggest the existence of a finite excitation gap.
Therefore, it is expected that the state is insulating.
Their correlation lengths are
$\xi_{nn} =\left[ \ln \left( \frac{e_2}{e_1}\right) \right]^{-1}$,
$\xi_{ss}$ $=\xi_{bb}$,
$=\left[ \ln \left( \frac{2}{e_1}\right) \right]^{-1}$,
and
$\xi_{cc} =\left[ \ln \left( \frac{2g_1}{e_1}\right) \right]^{-1}$
(Fig.\ \ref{fig:modelC:corrleng}).
(They satisfy the relation
$\xi_{cc}$ $> \xi_{nn}$ $> \xi_{ss}$ $= \xi_{bb}$.)
We note that the spin correlation is antiferromagnetic.

We consider the region $\vert \lambda \vert \ll 1$.
There is almost one electron per a $d$-site.
The $p$-sites are almost empty.
The correlation lengths are large.
However, the correlation functions for the nearest neighbor sites vanish.

For $\lambda \gg 1$,
There is almost two electrons per a $p$-site.
The $d$-sites are almost empty.
The correlation lengths
and the correlation functions for the nearest neighbor sites vanish.

The ground state (\ref{eq:gsmodc}) is a $1/3$-filling state.
The filling factor corresponds to that of the band insulator
in the non-interacting system for $\lambda > \sqrt{2}$.
In the non-interacting system,
we have the metal-insulator transition at $\lambda=\sqrt{2}$
by the variation of $\lambda$.
However, the ground state (\ref{eq:gsmodc}) is insulating
for any $\lambda$.
The properties of these states are completely different.

\newpage
\section{Absence of the persistent current}
\label{sec:persistent}

In Sec.\ \ref{sec:corr}, we have assumed the parameter
$\lambda_r$ to be real.
Relaxing the condition,
effects of a magnetic field is included by taking
hopping matrix elements to be complex.
Thus the effects of the magnetic field
is investigated exactly
for the systems of strongly correlated electrons described in this paper.
Let us calculate the persistent current.
Considering the system in a ring geometry
and putting a flux through the ring,
we can measure
the Aharonov-Bohm effect and the persistent current
\cite{BY61,BIR,LDDB,CWBKGK,MCB}.
In the ring geometry,
we include the effect of the flux $\Phi$ by changing the hopping matrix
elements of the $N$-th cell.
We first classify sites in the $N$-th cell into two classes as:
(i) sites which belong to the $N$-th unit cell
and (ii) sites which are identified with sites in the cell $C_1$.
We denote the sets of the sites (i) and (ii) by $C_{N;u}$
and $C_{N;e}$, respectively.
The cell Hamiltonian (\ref{eq:cellham}) associated
with the $N$-th cell is obtained by choosing
\begin{eqnarray}
\alpha_{N,\sigma}(\Phi)
=
\sum_{r=1}^{\vert C_N \vert}
\widetilde{\lambda}_r^{(N)} (\Phi) \ c_{r,\sigma},
\label{eq:peralp}
\end{eqnarray}
where
\begin{eqnarray}
\widetilde{\lambda}_r^{(N)} (\Phi)
&=&
\left \{
\begin{array}{@{\,}ll}
\lambda_r^{(N)} e^{i \Phi} & \mbox{for $r$ $\in C_{N;e}$}\\
\lambda_r^{(N)}            & \mbox{for $r$ $\in C_{N;u}$},
\end{array} \right.
\label{eq:percell}
\end{eqnarray}
with real $\lambda_r^{(N)}$. From
(\ref{eq:peralp}) and (\ref{eq:alpha}) for $1 \leq n \leq N-1$,
the Hamiltonian is
\begin{equation}
H_S (\Phi) = - {\cal P}
\sum_{\sigma = \uparrow, \downarrow}
\sum_{x,y \in \Lambda_N}
t_{x,y} (\Phi) \ c_{x,\sigma}^{\dagger} c_{y,\sigma}{\cal P},
\end{equation}
where
\begin{eqnarray}
t_{x,y}(\Phi) = \sum_{n=1}^{N-1} t_{r,s}^{(n)}+t_{r,s}^{(N)}(\Phi),\ \
\mbox{for $x=f(n,r)$ and $y=f(n,s)$,}
\end{eqnarray}
with
\begin{eqnarray}
t_{r,s}^{(n)}
&= &
\left \{ \begin{array}{@{\,}ll}
  {\lambda_r^{(n)}} \lambda_s^{(m)} \delta_{n,m}
          & \mbox{for $r \ne s$}  \\
   2  \left( \lambda_r^{(n)} \right)^2
          & \mbox{for $r=s$ and $r \in C_{n; U=\infty}$} \\
      \left( \lambda_r^{(n)} \right)^2
          & \mbox{for $r=s$ and $r \in C_{n; U=0}$}, \\
      \end{array} \right.
\\
& & \hspace{40mm}
\mbox{($1 \leq m \leq N-1$, \ $1 \leq n \leq N-1$)}
\nonumber \\
t_{r,s}^{(N)}(\Phi)
&= &
\left \{ \begin{array}{@{\,}ll}
\left( \widetilde{\lambda}_r^{(N)}  ( \Phi ) \right)^{*} \lambda_s^{(N)}
        & \mbox{for $r \ne s$ and $r \in C_{N;e}$, $s \in C_{N;u}$}  \\
  {\lambda_r^{(N)}} \  \widetilde{\lambda}_s^{(N)}(\Phi)
        & \mbox{for $r \ne s$ and $r \in C_{N;u}$, $s \in C_{N;e}$}  \\
  {\lambda_r^{(N)}} \ \lambda_s^{(N)}
          & \mbox{for $r \ne s$ and $r,s \in C_{N;u}$}  \\
   2 \left( \lambda_r^{(N)} \right)^2
          & \mbox{for $r=s$ and $r \in C_{N; U=\infty}$} \\
     \left( \lambda_r^{(N)} \right)^2
          & \mbox{for $r=s$ and $r \in C_{N; U=0}$}. \\
      \end{array} \right.
\end{eqnarray}
The ground state of $H_S (\Phi)$ is
\begin{equation}
\Big\vert \Phi_{G.S.} \left( \Phi \right) \Big\rangle =
{\cal P}
\prod_{\sigma = \uparrow, \downarrow}
\left[
\left(
\prod_{n=1}^{N-1}
\alpha_{n,\sigma}^{\dagger}
\right)
\alpha_{N,\sigma}^{\dagger} (\Phi)
\right]
\Big\vert 0 \Big\rangle.
\label{eq:gsper}
\end{equation}
The ground state energy is given explicitly by
\begin{eqnarray}
E_0 (\Phi)
&=& -\sum_{n=1}^{N-1}\sum_{r \in C_n}
2\Big\vert \lambda_r^{(n)} \Big\vert^2
  -\sum_{r \in C_N} 2\Big\vert
\widetilde{\lambda}_r^{(N)}(\Phi) \Big\vert^2
\nonumber \\
&=& -\sum_{n=1}^{N}\sum_{r \in C_n}
2\Big\vert \lambda_r^{(n)} \Big\vert^2.
\label{eq:perene}
\end{eqnarray}
It is independent of the flux $\Phi$.
The persistent current $I$ is evaluated
by using the Byers-Yang relation \cite{BY61}.
We obtain
\begin{equation}
I \propto - \frac{\partial E_0(\Phi)}{\partial \Phi}=0.
\label{eq:persis}
\end{equation}
The persistent current is vanishing for any of the solvable models
discussed here.
This is consistent with our conclusion that the ground state
is insulating.

Extending the discussion here,
the absence of the persistent current can be shown
in any dimensions for the models discussed in this paper.

\section{Summary}
\label{sec:summary}

We investigated three models with strongly correlated electrons
which have the RVB state as an exact ground state.
The number of the electrons per unit cell is restricted to be 2.
The correlation functions are evaluated exactly
using the transfer matrix method for the geometric representations
of the valence-bond states \cite{Ta94}.
The two-point correlation functions for spin, density,
and singlet-Cooper-pairs are obtained for any distance.
All the correlation functions decay
exponentially with distance.
The momentum distribution functions are also evaluated
and there is no singularity.
The results suggest that the ground states of the models are
insulating.
The persistent currents are also considered
and turned out to be vanishing.

\acknowledgments

We are grateful to H. Tasaki for valuable discussions.
The method used in this work were developed by him.
We would like to thank P.-A. Bares and P.-A. Lee
for useful discussions and sending
their article prior to publication.
This work is, in part, supported
by a grant for Priority Area from Ministry of Education,
Science, and Culture Japan.

\newpage
\appendix
\section{}
\label{app:models}

We describe the models by several authors \cite{BG92,Mi92,St93,Ta93a}
and those investigated in this paper
using the cell construction of Tasaki
(see Table\ \ref{table:models}).
The lattice is constructed by the cell in the second column
in Table\ \ref{table:models}.
The cell for the line graph is constructed as follows.
We define a lattice
${\cal L}=(\Lambda, B)$ where $\Lambda$
is the set of the sites (vertices)
and $B$ is the set of the bonds (edges).
The line graph $L( {\cal L})=(\Lambda^L, B^L)$
constructed from a lattice $\Lambda$ has the bonds of ${\cal L}$
as sites $(\Lambda^L=B)$,
and two sites are connected by a bond in $B^L$
if the corresponding bonds in $B$ have a site in common.
The cell is defined by sites ($\in \Lambda^L$)
which are connected by a bond ($\in B$)
to a same site ($\in \Lambda$).

\section{}
\label{app:loop}

We first note the equalities
\begin{eqnarray}
c_{i,\sigma}^{\phantom{\dagger}}
b_{i,j}^{\dagger}
=
sgn(\sigma) \
c_{j,\sigma}^{\dagger}
\left(
c_{i,\sigma}^{\phantom{\dagger}}
c_{i,\sigma}^{\dagger}
\right),
\label{eq:identcb}
\end{eqnarray}
and
\begin{equation}
b_{i,j}^{\phantom{\dagger}}b_{i,k}^{\dagger}
=
\sum_{\sigma=\uparrow,\downarrow}
\left(
c_{i,\sigma}^{\phantom{\dagger}}
c_{i,\sigma}^{\dagger}
\right)
\left(
c_{j,-\sigma}^{\phantom{\dagger}}
c_{k,-\sigma}^{\dagger}
\right)
+
\Delta_{i},
\label{eq:identbb}
\end{equation}
where
$\Delta_{i}=
-\sum_{\sigma=\uparrow,\downarrow}
c_{i,\sigma}^{\dagger}
\left(
c_{k,-\sigma}^{\dagger}
c_{j,\sigma}^{\phantom{\dagger}}
\right)
c_{i,-\sigma}^{\phantom{\dagger}}$
which has the property
$\Delta_{i}\vert0\rangle= \langle0\vert\Delta_{i}=0$.

We show an equality.
Consider a connected graph $W \cup W'$ with $2n$ $(n \geq 1)$ bonds.
We set
$W=\{\{2,3\},\{4,5\},\cdots,\{2n,2n+1\}\}$ and
$W'=\{\{1,2\},\{3,4\},\cdots,\{2n-1,2n\}\}$.
We define the quantity
\begin{eqnarray}
L(n;\sigma,\rho)
&=&
\left\langle 0 \left\vert
\Bigg(
\prod_{k=1}^{n}b_{2k,2k+1}^{\phantom{\dagger}}
\Bigg)
               c_{1,\sigma}^{\phantom{\dagger}}
               c_{2n+1,\rho}^{\dagger}
\prod_{k=1}^{n}b_{2k-1,2k}^{\dagger}
\right\vert 0 \right\rangle.
\label{eq:appB1}
\end{eqnarray}
For $n=1$, from (\ref{eq:identcb}) we have
\begin{eqnarray}
L(1;\sigma,\rho)
&=&
\left\langle 0 \left\vert
b_{2,3}^{\phantom{\dagger}}
c_{1,\sigma}^{\phantom{\dagger}}
c_{3,\rho}^{\dagger}
b_{1,2}^{\dagger}
\right\vert 0 \right\rangle
\nonumber \\
&=&
-\left\langle 0 \left\vert
b_{2,3}^{\phantom{\dagger}}
c_{3,\rho}^{\dagger}
\left(
c_{1,\sigma}^{\phantom{\dagger}}
b_{1,2}^{\dagger}
\right)
\right\vert 0 \right\rangle
\nonumber \\
&=&
-sgn(\sigma)
\left\langle 0 \left\vert
b_{2,3}^{\phantom{\dagger}}
c_{3,\rho}^{\dagger}
c_{2,-\sigma}^{\phantom{\dagger}}
\right\vert 0 \right\rangle
\nonumber \\
&=&
sgn(\sigma)
sgn(-\rho)
\left\langle 0 \left\vert
c_{2,-\sigma}^{\phantom{\dagger}}
c_{2,-\sigma}^{\dagger}
c_{3,\rho}^{\phantom{\dagger}}
c_{3,\rho}^{\dagger}
\right\vert 0 \right\rangle
\nonumber \\
&=&
- \delta_{\sigma, \rho}
\end{eqnarray}From (\ref{eq:identcb}), we have
\begin{eqnarray}
L(n;\sigma,\rho)
&=&
\left\langle 0 \left\vert
\Bigg(
\prod_{k=1}^{n}b_{2k,2k+1}^{\phantom{\dagger}}
\Bigg)
               c_{1,\sigma}^{\phantom{\dagger}}
               c_{2n+1,\rho}^{\dagger}
               b_{1,2}^{\dagger}
\prod_{k=2}^{n}b_{2k-1,2k}^{\dagger}
\right\vert 0 \right\rangle
\nonumber \\
&=&
-sgn(\sigma)
\left\langle 0 \left\vert
\Bigg(
\prod_{k=1}^{n}b_{2k,2k+1}^{\phantom{\dagger}}
\Bigg)
               c_{2n+1,\rho}^{\dagger}
               c_{2,-\sigma}^{\dagger}
\prod_{k=2}^{n}b_{2k-1,2k}^{\dagger}
\right\vert 0 \right\rangle
\nonumber \\
&=&
-sgn(\sigma)
\left\langle 0 \left\vert
\Bigg(
\prod_{k=2}^{n}b_{2k,2k+1}^{\phantom{\dagger}}
\Bigg)
               b_{2,3}^{\phantom{\dagger}}
               c_{2n+1,\rho}^{\dagger}
               c_{2,-\sigma}^{\dagger}
\prod_{k=2}^{n}b_{2k-1,2k}^{\dagger}
\right\vert 0 \right\rangle
\nonumber \\
&=&
-sgn(\sigma)\
 sgn(-\sigma)
\left\langle 0 \left\vert
\Bigg(
\prod_{k=2}^{n}b_{2k,2k+1}^{\phantom{\dagger}}
\Bigg)
              (-c_{3,\sigma}^{\phantom{\dagger}})
               c_{2n+1,\rho}^{\dagger}
\prod_{k=2}^{n}b_{2k-1,2k}^{\dagger}
\right\vert 0 \right\rangle
\nonumber \\
&=&
-
\left\langle 0 \left\vert
\Bigg(
\prod_{k=2}^{n}b_{2k,2k+1}^{\phantom{\dagger}}
\Bigg)
               c_{3,\sigma}^{\phantom{\dagger}}
               c_{2n+1,\rho}^{\dagger}
\prod_{k=2}^{n}b_{2k-1,2k}^{\dagger}
\right\vert 0 \right\rangle.
\end{eqnarray}
We change the labeling of the lattice sites by the rule
$j \to j-2$ and obtain
\begin{eqnarray}
L(n;\sigma,\rho)
&=&
-
\left\langle 0 \left\vert
\Bigg(
\prod_{k=1}^{n-1}b_{2k,2k+1}^{\phantom{\dagger}}
\Bigg)
               c_{1,\sigma}^{\phantom{\dagger}}
               c_{2n-1,\rho}^{\dagger}
\prod_{k=1}^{n-1}b_{2k-1,2k}^{\dagger}
\right\vert 0 \right\rangle
\nonumber \\
&=& -L(n-1;\sigma,\rho) \nonumber \\
&=& \cdots              \nonumber \\
&=& (-1)^{n-1} L(1;\sigma,\rho) \nonumber \\
&=& (-1)^{n} \delta_{\sigma, \rho}.
\label{eq:identL}
\end{eqnarray}

We show (\ref{eq:proof1}).
For the self-closed bonds and degenerate graph
we have $w_j=1$ and $w_j=2$, respectively.
Consider the non-degenerate loop $W_j \cup W'_j$
with $l_j=2n$ $(n \geq 1)$ bonds.
We set
$W_j=\{\{2,3\},\{4,5\},\cdots,\{2n-2,2n-1\},\{2n,2n+1(=1)\}\}$ and
$W'_j=\{\{1,2\},\{3,4\},\cdots,\{2n-1,2n\}\}$. From
(\ref{eq:identbb}) and (\ref{eq:identL}) the weight is
\begin{eqnarray}
w_j
&=&
\left\langle 0 \left\vert
\prod_{k=1}^{n}b_{2k,2k+1}^{\phantom{\dagger}}
\prod_{k=1}^{n}b_{2k-1,2k}^{\dagger}
\right\vert 0 \right\rangle
\nonumber \\
&=&
\left\langle 0 \left\vert
\Bigg(
\prod_{k=1}^{n-1}b_{2k,2k+1}^{\phantom{\dagger}}
\Bigg)
b_{2n,2n+1}^{\phantom{\dagger}}
b_{2n-1,2n}^{\dagger}
\prod_{k=1}^{n-1}b_{2k-1,2k}^{\dagger}
\right\vert 0 \right\rangle
\nonumber \\
&=&
\left\langle 0 \Bigg\vert
\Bigg(
\prod_{k=1}^{n-1}b_{2k,2k+1}^{\phantom{\dagger}}
\Bigg)
\Bigg(
\sum_{\sigma=\uparrow,\downarrow}
c_{2n,\sigma}^{\phantom{\dagger}}
c_{2n,\sigma}^{\dagger}
c_{1,-\sigma}^{\phantom{\dagger}}
c_{2n-1,-\sigma}^{\dagger}
+\Delta_{2n}
\Bigg)
\prod_{k=1}^{n-1}b_{2k-1,2k}^{\dagger}
\Bigg\vert 0 \right\rangle
\nonumber \\
&=&
\sum_{\sigma=\uparrow,\downarrow}
\left\langle 0 \left\vert
\Bigg(
\prod_{k=1}^{n-1}b_{2k,2k+1}^{\phantom{\dagger}}
\Bigg)
c_{2n,\sigma}^{\phantom{\dagger}}
c_{2n,\sigma}^{\dagger}
c_{1,-\sigma}^{\phantom{\dagger}}
c_{2n-1,-\sigma}^{\dagger}
\prod_{k=1}^{n-1}b_{2k-1,2k}^{\dagger}
\right\vert 0 \right\rangle
\nonumber \\
&=&
\sum_{\sigma=\uparrow,\downarrow}
\left\langle 0 \left\vert
\Bigg(
\prod_{k=1}^{n-1}b_{2k,2k+1}^{\phantom{\dagger}}
\Bigg)
c_{1,-\sigma}^{\phantom{\dagger}}
c_{2(n-1)+1,-\sigma}^{\dagger}
\prod_{k=1}^{n-1}b_{2k-1,2k}^{\dagger}
\right\vert 0 \right\rangle
\nonumber \\
&=&
\sum_{\sigma=\uparrow,\downarrow}
L(n-1;-\sigma,-\sigma)
\nonumber \\
&=&
\sum_{\sigma=\uparrow,\downarrow}
(-1)^{n-1}\delta_{-\sigma,-\sigma}
\nonumber \\
&=&
2(-1)^{n-1}
\nonumber \\
&=&
2(-1)^{l_j/2-1},
\label{eq:a2}
\end{eqnarray}
where we used the relation $l=2n$ in the last line.

\section{}
\label{app:spin}

We first note the equalities
\begin{eqnarray}
S_i^z b_{i,j}^{\dagger}
&=& \frac{1}{2}
\tilde{b}_{i,j}^{\dagger}+\delta_{i,j}^{\phantom{\dagger}}
 \nonumber \\
S_i^z \tilde{b}_{i,j}^{\dagger}
&=& \frac{1}{2}b_{i,j}^{\dagger}
-\widetilde{\delta}_{ji}^{\phantom{\dagger}},
\label{eq:equality1}
\end{eqnarray}
where
\begin{eqnarray}
\widetilde{b}_{i,j}^{\dagger}
&=&
c_{i,\uparrow}^{\dagger}
c_{j,\downarrow}^{\dagger}
+
c_{i,\downarrow}^{\dagger}
c_{j,\uparrow}^{\dagger},
\end{eqnarray}
is the creation operator of the triplet-pair between sites $i$ and $j$
and satisfies the relation
$\widetilde{b}_{i,j}^{\dagger}=-\widetilde{b}_{ji}^{\dagger}$.
Here
\begin{eqnarray}
\delta_{i,j}
&=& -\frac{ c_{i,\uparrow}^{\dagger} c_{j,\downarrow}^{\dagger}
          n_{i,\downarrow}^{\phantom{\dagger}}
         -c_{i,\downarrow}^{\dagger} c_{j,\uparrow}^{\dagger}
          n_{i,\uparrow}^{\phantom{\dagger}}}
        {2} \nonumber \\
\widetilde{\delta}_{i,j}
&=& -\frac{ c_{i,\uparrow}^{\dagger} c_{j,\downarrow}^{\dagger}
          n_{i,\downarrow}^{\phantom{\dagger}}
         +c_{i,\downarrow}^{\dagger} c_{j,\uparrow}^{\dagger}
          n_{i, \uparrow}^{\phantom{\dagger}}}
        {2}
\end{eqnarray}
which have the properties
$\delta_i \vert 0 \rangle$
$= \langle 0 \vert \delta_i =0$
and
$\tilde{\delta}_i \vert 0 \rangle$
$= \langle 0 \vert \tilde{\delta}_i = 0$.

We show (\ref{eq:spinloop}).
Consider the connected graph $W^{(x,y)} \cup {W'}^{(x,y)}$
with $2n$ $(n \geq 1)$ bonds.
We set
$W^{(x,y)}=\{\{2,3\},\{4,5\},\cdots,\{2n,2n+1\}\}$ and
${W'}^{(x,y)}=\{\{1,2\},\{3,4\},\cdots,\{2n-1,2n\}\}$.
We denote the right hand side of (\ref{eq:spinloop}) by $S(m)$ as
\begin{eqnarray}
S(m)
&=&
\left\langle 0 \left\vert
\Bigg(
\prod_{k=1}^{n}b_{2k,2k+1}^{\phantom{\dagger}}
\Bigg)
               S_1^z
               S_{m}^z
\prod_{k=1}^{n}b_{2k-1,2k}^{\dagger}
\right\vert 0 \right\rangle,
\label{eq:appC1}
\end{eqnarray}
where we set $x=1$ and $y=m$. From (\ref{eq:equality1}), we have
\begin{eqnarray}
S(2m+1)
&=&\left\langle 0  \Bigg\vert
   \Bigg(
   \prod_{k=1}^{n}b_{2k,2k+1}^{\phantom{\dagger}}
   \Bigg)
                  S_1^z S_{2m+1}^z
   \prod_{k=1}^{n}b_{2k-1,2k}^{\dagger}
   \Bigg\vert 0 \right\rangle   \nonumber \\
&=&\left\langle 0  \Bigg\vert
   \Bigg(
   \prod_{k=1}^{m-1}b_{2k,2k+1}^{\phantom{\dagger}}
   \Bigg)
   \Bigg(
   \prod_{k=m+1}^{n}b_{2k,2k+1}^{\phantom{\dagger}}
   \Bigg)
                  b_{2m,2m+1}^{\phantom{\dagger}}
                  S_{2m+1}^z S_1^z
   \prod_{k=1}^{n}b_{2k-1,2k}^{\dagger}
   \Bigg\vert 0 \right\rangle   \nonumber \\
&=&\left\langle 0  \Bigg\vert
   \Bigg(
   \prod_{k=1}^{m-1}b_{2k,2k+1}^{\phantom{\dagger}}
   \Bigg)
   \Bigg(
   \prod_{k=m+1}^{n}b_{2k,2k+1}^{\phantom{\dagger}}
   \Bigg)
                  \left(
                  \frac{1}{2}\widetilde{b}_{2m+1,2m}^{\phantom{\dagger}}
                  +\delta_{2m+1,2m}^{\phantom{\dagger}}
                  \right)
                  S_1^z
   \prod_{k=1}^{n}b_{2k-1,2k}^{\dagger}
   \Bigg\vert 0 \right\rangle \nonumber \\
&=&\left\langle 0  \Bigg\vert
   \Bigg(
   \prod_{k=1}^{m-1}b_{2k,2k+1}^{\phantom{\dagger}}
   \Bigg)
   \Bigg(
   \prod_{k=m+1}^{n}b_{2k,2k+1}^{\phantom{\dagger}}
   \Bigg)
                  \left(
                - \frac{1}{2}\widetilde{b}_{2m,2m+1}^{\phantom{\dagger}}
                  \right)
                  S_1^z
   \prod_{k=1}^{n}b_{2k-1,2k}^{\dagger}
   \Bigg\vert 0 \right\rangle \nonumber \\
&=&-\left\langle 0  \Bigg\vert
   \Bigg(
   \prod_{k=1}^{m-1}b_{2k,2k+1}^{\phantom{\dagger}}
   \Bigg)
   \Bigg(
   \prod_{k=m+1}^{n}b_{2k,2k+1}^{\phantom{\dagger}}
   \Bigg)
                  b_{2m-1,2m}^{\phantom{\dagger}}
                  S_{2m}^z
                  S_1^z
   \prod_{k=1}^{n}b_{2k-1,2k}^{\dagger}
   \Bigg\vert 0 \right\rangle   \nonumber \\
&=&-\left\langle 0  \Bigg\vert
   \Bigg(
   \prod_{k=1}^{n}b_{2k,2k+1}^{\phantom{\dagger}}
   \Bigg)
                  S_1^z S_{2m}^z
   \prod_{k=1}^{n}b_{2k-1,2k}^{\dagger}
   \Bigg\vert 0 \right\rangle   \nonumber \\
&=&-\left\langle 0  \Bigg\vert
   \Bigg(
   \prod_{k=1}^{n}b_{2k,2k+1}^{\phantom{\dagger}}
   \Bigg)
                  S_1^z S_{2m}^z
                  b_{2m-1,2m}^{\dagger}
   \Bigg(
   \prod_{k=1}^{m-1}b_{2k-1,2k}^{\dagger}
   \Bigg)
   \prod_{k=m+1}^{n}b_{2k-1,2k}^{\dagger}
   \Bigg\vert 0 \right\rangle   \nonumber \\
&=&-\left\langle 0  \Bigg\vert
   \Bigg(
   \prod_{k=1}^{n}b_{2k,2k+1}^{\phantom{\dagger}}
   \Bigg)
                  S_1^z
                  \left(
                  \frac{1}{2}\widetilde{b}_{2m,2m-1}^{\dagger}
                  +\delta_{2m,2m-1}^{\phantom{\dagger}}
                  \right)
   \Bigg(
   \prod_{k=1}^{m-1}b_{2k-1,2k}^{\dagger}
   \Bigg)
   \prod_{k=m+1}^{n}b_{2k-1,2k}^{\dagger}
   \Bigg\vert 0 \right\rangle \nonumber \\
&=&-\left\langle 0  \Bigg\vert
   \Bigg(
   \prod_{k=1}^{n}b_{2k,2k+1}^{\phantom{\dagger}}
   \Bigg)
                  S_1^z
                  \left(
                - \frac{1}{2}\widetilde{b}_{2m-1,2m}^{\dagger}
                  \right)
   \Bigg(
   \prod_{k=1}^{m-1}b_{2k-1,2k}^{\dagger}
   \Bigg)
   \prod_{k=m+1}^{n}b_{2k-1,2k}^{\dagger}
   \Bigg\vert 0 \right\rangle \nonumber \\
&=&(-1)^2\left\langle 0  \Bigg\vert
   \Bigg(
   \prod_{k=1}^{n}b_{2k,2k+1}^{\phantom{\dagger}}
   \Bigg)
                  S_1^z S_{2m-1}^z
                  b_{2m-1,2m}^{\dagger}
   \Bigg(
   \prod_{k=1}^{m-1}b_{2k-1,2k}^{\dagger}
   \Bigg)
   \prod_{k=m+1}^{n}b_{2k-1,2k}^{\dagger}
   \Bigg\vert 0 \right\rangle   \nonumber \\
&=&(-1)^2\left\langle 0  \Bigg\vert
   \Bigg(
   \prod_{k=1}^{n}b_{2k,2k+1}^{\phantom{\dagger}}
   \Bigg)
                  S_1^z S_{2m-1}^z
   \prod_{k=1}^{n}b_{2k-1,2k}^{\dagger}
   \Bigg\vert 0 \right\rangle   \nonumber \\
&=&(-)^2 S(2m-1)    \nonumber \\
&=& \cdots          \nonumber \\
&=&(-1)^{2m} S(1) \nonumber \\
&=&(-1)^{2m}\left\langle 0 \Bigg\vert
   \Bigg(
   \prod_{k=1}^{n}b_{2k,2k+1}^{\phantom{\dagger}}
   \Bigg)
                  S_1^z S_1^z
   \prod_{k=1}^{n}b_{2k-1,2k}^{\dagger}
   \Bigg\vert0 \right\rangle  \nonumber \\
&=&\frac{(-1)^{2m}}{4}\left\langle 0 \Bigg\vert
   \prod_{k=1}^{n}b_{2k,2k+1}^{\phantom{\dagger}}
   \prod_{k=1}^{n}b_{2k-1,2k}^{\dagger}
   \Bigg\vert0 \right\rangle.
\end{eqnarray}
Similarly, we have
\begin{eqnarray}
S(2m)
= (-1)^{2m-1}\frac{1}{4}\left\langle 0 \Bigg\vert
   \prod_{k=1}^{n}b_{2k,2k+1}^{\phantom{\dagger}}
   \prod_{k=1}^{n}b_{2k-1,2k}^{\dagger}
   \Bigg\vert0 \right\rangle,
\end{eqnarray}
Therefore, we obtain
\begin{eqnarray}
S(m)&=&\frac{1}{4}(-1)^{d(x,y)} w_j
\end{eqnarray}
where we used the relation $d(x,y)=m-1$
and the definition of $w_j$.

\section{}
\label{app:green}

We show (\ref{eq:proof2}).
Consider the line $W^{(x',y')} \cup {W'}^{(x',y')}$
with $2n$ $(n \geq 1)$ bonds.
We set
$W^{(x',y')}=\{\{2,3\},\{4,5\},\cdots,\{2n,2n+1(=y')\}\}$ and
${W'}^{(x',y')}=\{\{1(=x'),2\},\{3,4\},\cdots,\{2n-1,2n\}\}$. From
(\ref{eq:appB1}), the weight is
\begin{eqnarray}
\left\langle 0 \Bigg\vert
\Bigg(
\prod_{k=1}^{n}b_{2k,2k+1}^{\phantom{\dagger}}
\Bigg)
               c_{1,\sigma}^{\phantom{\dagger}}
               c_{2n+1,\rho}^{\dagger}
\prod_{k=1}^{n}b_{2k-1,2k}^{\dagger}
\Bigg\vert 0 \right\rangle
&=&L(n;\sigma,\sigma) \nonumber \\
&=&(-1)^n \nonumber \\
&=&(-1)^{l(x',y')/2}
\label{eq:appD1}
\end{eqnarray}
where we used the relation $l(x',y')=2n$.

\section{}
\label{app:number}

We show (\ref{eq:proof3}).
Consider the graph $W^{(x)} \cup {W'}^{(x)}$,
we set
$W^{(x)} =\{\{1(=x),2\},\{3,4\},\cdots,\{2n-1,2n\}\}$
and
${W'}^{(x)} =\{\{2,3\},\{4,5\},\cdots,\{2n,1\}\}$.
We have
\begin{eqnarray}
\lefteqn{
\left\langle 0 \left\vert
\Bigg(
   \prod_{k=1}^{n}b_{2k-1,2k}^{\phantom{\dagger}}
\Bigg)
                  n_{1,\sigma}^{\phantom{\dagger}}
   \prod_{k=1}^{n}b_{2k,2k+1}^{\dagger}
   \right\vert 0 \right\rangle
}
\nonumber \\
&=&
\left\langle 0\left\vert
\Bigg(
   \prod_{k=2}^{n}b_{2k-1,2k}^{\phantom{\dagger}}
\Bigg)
                    b_{1,2}^{\phantom{\dagger}}
                    \left(1-
                    c_{1,\sigma}^{\phantom{\dagger}}
                    c_{1,\sigma}^{\dagger}
                    \right)
                    b_{2n,1}^{\dagger}
   \prod_{k=1}^{n-1}b_{2k,2k+1}^{\dagger}
   \right\vert 0\right\rangle \nonumber \\
&=&
\left\langle 0\left\vert
   \prod_{k=1}^{n}b_{2k-1,2k}^{\phantom{\dagger}}
   \prod_{k=1}^{n}b_{2k,2k+1}^{\dagger}
   \right\vert 0 \right\rangle
-
   \left\langle 0 \left\vert
\Bigg(
   \prod_{k=1}^{n}b_{2k-1,2k}^{\phantom{\dagger}}
\Bigg)
                  c_{1,\sigma}^{\phantom{\dagger}}
                  c_{1,\sigma}^{\dagger}
   \prod_{k=1}^{n}b_{2k,2k+1}^{\dagger}
   \right\vert 0 \right\rangle.
\label{eq:appendixe1}
\end{eqnarray}From (\ref{eq:appB1}), the second term is
\begin{eqnarray}
\lefteqn{
   \left\langle 0 \left\vert
\Bigg(
   \prod_{k=1}^{n}b_{2k-1,2k}^{\phantom{\dagger}}
\Bigg)
                  c_{1,\sigma}^{\phantom{\dagger}}
                  c_{1,\sigma}^{\dagger}
   \prod_{k=1}^{n}b_{2k,2k+1}^{\dagger}
   \right\vert 0 \right\rangle
}
\nonumber \\
&=&\left\langle 0\left\vert
\Bigg(
   \prod_{k=2}^{n}b_{2k-1,2k}^{\phantom{\dagger}}
\Bigg)
                    b_{1,2}^{\phantom{\dagger}}
                    c_{1,\sigma}^{\phantom{\dagger}}
                    c_{1,\sigma}^{\dagger}
                    b_{2n,1}^{\dagger}
   \prod_{k=1}^{n-1}b_{2k,2k+1}^{\dagger}
   \right\vert 0\right\rangle \nonumber \\
&=&\left\langle 0\left\vert
\Bigg(
   \prod_{k=2}^{n}b_{2k-1,2k}^{\phantom{\dagger}}
\Bigg)
                   (-b_{1,1}^{\phantom{\dagger}}
                    c_{2,\sigma}^{\phantom{\dagger}})
                   (-c_{2n,\sigma}^{\dagger}
                    b_{1,1}^{\dagger})
   \prod_{k=1}^{n-1}b_{2k,2k+1}^{\dagger}
   \right\vert 0\right\rangle \nonumber \\
&=&\left\langle 0\left\vert
\Bigg(
   \prod_{k=2}^{n}b_{2k-1,2k}^{\phantom{\dagger}}
\Bigg)
                    c_{2,\sigma}^{\phantom{\dagger}}
                    c_{2n,\sigma}^{\dagger}
   \prod_{k=1}^{n-1}b_{2k,2k+1}^{\dagger}
   \right\vert 0\right\rangle \nonumber \\
&=&\left\langle 0\left\vert
\Bigg(
   \prod_{k=1}^{n-1}b_{2k,2k+1}^{\phantom{\dagger}}
\Bigg)
                    c_{1,\sigma}^{\phantom{\dagger}}
                    c_{2n-1,\sigma}^{\dagger}
   \prod_{k=1}^{n-1}b_{2k-1,2k}^{\dagger}
   \right\vert 0 \right\rangle \nonumber \\
&=&L(n-1;\sigma,\sigma) \nonumber \\
&=&(-1)^{n-1}  \nonumber \\
&=&(-1)^{l(x)/2-1}.
\label{eq:e2}
\end{eqnarray}
The fourth line is obtained by renaming the site $j$ to $j-1$.
We used the relation $l(x)=2(n-1)$. From (\ref{eq:a2})
and (\ref{eq:e2}), we have
\begin{eqnarray}
\left\langle 0 \left\vert
 \prod_{k=1}^{n}b_{2k-1,2k}^{\phantom{\dagger}}
                n_{1,\sigma}^{\phantom{\dagger}}
 \prod_{k=1}^{n}b_{2k,2k+1}^{\dagger}
 \right\vert 0 \right\rangle
&=&2(-1)^{l(x)/2-1}-(-1)^{l(x)/2-1}\nonumber \\
&=&(-1)^{l(x)/2-1}.
\end{eqnarray}

\section{}
\label{app:den}

We show (\ref{eq:proof4}).
For (i), the sets $W^{(x,y)}$ and ${W'}^{(x,y)}$ are empty.
The right hand side of (\ref{eq:dendenw}) is
\begin{eqnarray}
w_j(x,y;\sigma)
&=&
\frac{1}{2}
\left\{
\left\langle 0 \left\vert
       c_{x,\sigma}^{\phantom{\dagger}}
       c_{y,\sigma}^{\phantom{\dagger}}
       c_{y,\sigma}^{\dagger}
       c_{x,\sigma}^{\dagger}
\right\vert 0 \right\rangle
+
\left\langle 0 \left\vert
       c_{x,\sigma}^{\phantom{\dagger}}
       c_{y,-\sigma}^{\phantom{\dagger}}
       c_{y,-\sigma}^{\dagger}
       c_{x,\sigma}^{\dagger}
\right\vert 0 \right\rangle
\right\}
= 1,
\end{eqnarray}
where we used the equality
$\langle 0 \vert
       c_{x,\sigma}^{\phantom{\dagger}}
       c_{y,\sigma}^{\phantom{\dagger}}
       c_{y,\sigma}^{\dagger}
       c_{x,\sigma}^{\dagger}
\vert 0 \rangle$
$=$
$\langle 0 \vert
       c_{x,\sigma}^{\phantom{\dagger}}
       c_{y,-\sigma}^{\phantom{\dagger}}
       c_{y,-\sigma}^{\dagger}
       c_{x,\sigma}^{\dagger}
\vert 0 \rangle$
$=1$.

For (ii), we suppose that the site $x$ belongs to the graph. From
(\ref{eq:proof3}), the right hand side of (\ref{eq:dendenw}) is
\begin{eqnarray}
w_j(x,y;\sigma)
&=&
\frac{1}{2}
\left\{
\left\langle 0 \Bigg\vert
       c_{y,\sigma}^{\phantom{\dagger}}
       c_{y,\sigma}^{\dagger}
\Bigg\vert 0 \right\rangle
\left\langle 0 \Bigg\vert
\Bigg(
 \prod_{\{u',v'\}\in {W'}^{(x)}}
        b_{u',v'}^{\phantom{\dagger}}
\Bigg)
        c_{x,\sigma}^{\phantom{\dagger}}
        c_{x,\sigma}^{\dagger}
 \prod_{\{u,v\}\in W^{(x)}}
        b_{u,v}^{\dagger}
\Bigg\vert 0 \right\rangle \right.
\nonumber \\
& & \hspace{15mm}
\left. +
\left\langle 0 \Bigg\vert
       c_{y,-\sigma}^{\phantom{\dagger}}
       c_{y,-\sigma}^{\dagger}
\Bigg\vert 0 \right\rangle
\left\langle 0 \Bigg\vert
\Bigg(
 \prod_{\{u',v'\}\in {W'}^{(x)}}
\Bigg)
        b_{u',v'}^{\phantom{\dagger}}
        c_{x,\sigma}^{\phantom{\dagger}}
        c_{x,\sigma}^{\dagger}
 \prod_{\{u,v\}\in W^{(x)}}
        b_{u,v}^{\dagger}
\Bigg\vert 0 \right\rangle
\right\} \nonumber \\
&=& (-1)^{{l(x)}/2-1},
\end{eqnarray}
where we used the equality
$\langle 0 \vert
       c_{y,\sigma}^{\phantom{\dagger}}
       c_{y,\sigma}^{\dagger}
\vert 0 \rangle$
$=$
$\langle 0 \vert
       c_{y,-\sigma}^{\phantom{\dagger}}
       c_{y,-\sigma}^{\dagger}
\vert 0 \rangle$
$=1$.

For (iii), the right hand side of (\ref{eq:dendenw}) is
\begin{eqnarray}
w_j(x,y;\sigma)
&=&
\frac{1}{2}
\left\{
\left\langle 0 \Bigg\vert
\Bigg(
 \prod_{\{u',v'\}\in {W'}^{(x,y)}}b_{u',v'}^{\phantom{\dagger}}
\Bigg)
       c_{y,\sigma}^{\phantom{\dagger}}
       c_{y,\sigma}^{\dagger}
 \prod_{\{u,v\}\in W^{(x,y)}}b_{u,v}^{\dagger}
\Bigg\vert 0 \right\rangle  \right.
\nonumber \\
& & \hspace{50mm}
\left.
\left\langle 0 \Bigg\vert
\Bigg(
 \prod_{\{u',v'\}\in {W'}^{(x,y)}}b_{u',v'}^{\phantom{\dagger}}
\Bigg)
        c_{x,\sigma}^{\phantom{\dagger}}
        c_{x,\sigma}^{\dagger}
 \prod_{\{u,v\}\in W^{(x,y)}}b_{u,v}^{\dagger}
\Bigg\vert 0 \right\rangle \right.
\nonumber \\
& &
\left. \hspace{8mm} +
\left\langle 0 \Bigg\vert
\Bigg(
 \prod_{\{u',v'\}\in {W'}^{(x,y)}}b_{u',v'}^{\phantom{\dagger}}
\Bigg)
       c_{y,-\sigma}^{\phantom{\dagger}}
       c_{y,-\sigma}^{\dagger}
 \prod_{\{u,v\}\in W^{(x,y)}}b_{u,v}^{\dagger}
\Bigg\vert 0 \right\rangle  \right.
\nonumber \\
& & \hspace{50mm}
\left.
\left\langle 0 \Bigg\vert
\Bigg(
 \prod_{\{u',v'\}\in {W'}^{(x,y)}}b_{u',v'}^{\phantom{\dagger}}
\Bigg)
        c_{x,\sigma}^{\phantom{\dagger}}
        c_{x,\sigma}^{\dagger}
 \prod_{\{u,v\}\in W^{(x,y)}}b_{u,v}^{\dagger}
\Bigg\vert 0 \right\rangle
\right\} \nonumber \\
&=& (-1)^{{l(x)}/2-1}(-1)^{{l(y)}/2-1},
\end{eqnarray}
where we used (\ref{eq:proof3}).

For (iv),
consider the graph $W^{(x,y)} \cup {W'}^{(x,y)}$.
We set
$W^{(x,y)} =\{\{1(=x),2\},\{3,4\},\cdots,\{2n-1,2n\}\}$
and
${W'}^{(x,y)} =\{\{2,3\},\{4,5\},\cdots,\{2n,1\}\}$.
We denote one of the terms in the right hand side of (\ref{eq:dendenw})
by $G(n,j;\sigma,\tau)$ as
\begin{eqnarray}
G(n,j;\sigma,\tau)
&=& \left\langle 0 \Bigg\vert
\Bigg(
    \prod_{k=1}^{n} b_{2k-1,2k}^{\phantom{\dagger}}
\Bigg)
    c_{1,\sigma}^{\phantom{\dagger}}
    c_{j,\tau}^{\phantom{\dagger}}
    c_{j,\tau}^{\dagger}
    c_{1,\sigma}^{\dagger}
    \prod_{k=1}^{n} b_{2k,2k+1}^{\dagger}
    \Bigg\vert 0 \right\rangle.
\end{eqnarray}From (\ref{eq:identcb}), we have
\begin{eqnarray}
G(n,j;\sigma,\tau)
&=& \left\langle 0\Bigg\vert
\Bigg(
    \prod_{k=2}^{n} b_{2k-1,2k}^{\phantom{\dagger}}
\Bigg)
     b_{1,2}^{\phantom{\dagger}}
     c_{1,\sigma}^{\phantom{\dagger}}
     c_{j,\tau}^{\phantom{\dagger}}
     c_{j,\tau}^{\dagger}
     c_{1,\sigma}^{\dagger}
     b_{2n,1}^{\dagger}
    \prod_{k=1}^{n-1} b_{2k,2k+1}^{\dagger}
    \Bigg\vert 0\right\rangle \nonumber \\
&=& \left\langle 0\Bigg\vert
\Bigg(
    \prod_{k=2}^{n} b_{2k-1,2k}^{\phantom{\dagger}}
\Bigg)
    \Big(- b_{1,1}^{\phantom{\dagger}}
     c_{2,\sigma}^{\phantom{\dagger}}\Big)
     c_{j,\tau}^{\phantom{\dagger}}
     c_{j,\tau}^{\dagger}
    \Big(-c_{2n,\sigma}^{\dagger}
     b_{1,1}^{\dagger}\Big)
    \prod_{k=1}^{n-1} b_{2k,2k+1}^{\dagger}
    \Bigg\vert 0\right\rangle \nonumber \\
&=& \left\langle 0\Bigg\vert
\Bigg(
    \prod_{k=2}^{n} b_{2k-1,2k}^{\phantom{\dagger}}
\Bigg)
     c_{2,\sigma}^{\phantom{\dagger}}
     c_{j,\tau}^{\phantom{\dagger}}
     c_{j,\tau}^{\dagger}
     c_{2n,\sigma}^{\dagger}
    \prod_{k=1}^{n-1} b_{2k,2k+1}^{\dagger}
    \Bigg\vert 0\right\rangle \nonumber \\
&=& \left\langle 0\Bigg\vert
\Bigg(
    \prod_{k=1}^{n-1} b_{2k,2k+1}^{\phantom{\dagger}}
\Bigg)
     c_{1,\sigma}^{\phantom{\dagger}}
     c_{j-1,\tau}^{\phantom{\dagger}}
     c_{j-1,\tau}^{\dagger}
     c_{2n-1,\sigma}^{\dagger}
    \prod_{k=1}^{n-1} b_{2k-1,2k}^{\dagger}
    \Bigg\vert 0\right\rangle.
\label{eq:app:den1}
\end{eqnarray}
The last line is obtained by renaming of the lattice sites by the rule
$k \to k-1$.
For $j=2m$, we have
\begin{eqnarray}
G(n-1,2m;\sigma,\tau)
&=& \left\langle 0 \Bigg\vert
\Bigg(
    \prod_{k=1}^{n-1} b_{2k,2k+1}^{\phantom{\dagger}}
\Bigg)
     c_{1,\sigma}^{\phantom{\dagger}}
     c_{2m-1,\tau}^{\phantom{\dagger}}
     c_{2m-1,\tau}^{\dagger}
     c_{2(n-1)+1,\sigma}^{\dagger}
    \prod_{k=1}^{n-1} b_{2k-1,2k}^{\dagger}
    \Bigg\vert 0 \right\rangle \nonumber \\
&=& \left\langle 0 \Bigg\vert
\Bigg(
    \prod_{k=1}^{n-1} b_{2k,2k+1}^{\phantom{\dagger}}
\Bigg)
     c_{2m-1,\tau}^{\phantom{\dagger}}
     c_{2m-1,\tau}^{\dagger}
     c_{2(n-1)+1,\sigma}^{\dagger}
    \Big(- c_{1,\sigma}^{\phantom{\dagger}}
     b_{1,2}^{\dagger}\Big)
    \prod_{k=2}^{n-1} b_{2k-1,2k}^{\dagger}
    \Bigg\vert 0 \right\rangle \nonumber \\
&=& -{\rm sgn}(\sigma)
    \left\langle 0 \Bigg\vert
\Bigg(
    \prod_{k=1}^{n-1} b_{2k,2k+1}^{\phantom{\dagger}}
\Bigg)
     c_{2m-1,\tau}^{\phantom{\dagger}}
     c_{2m-1,\tau}^{\dagger}
     c_{2(n-1)+1,\sigma}^{\dagger}
     c_{2,-\sigma}^{\phantom{\dagger}}
    \prod_{k=2}^{n-1} b_{2k-1,2k}^{\dagger}
    \Bigg\vert 0 \right\rangle \nonumber \\
&=&-{\rm sgn}(\sigma)
    \left\langle 0 \Bigg\vert
\Bigg(
    \prod_{k=2}^{n-1} b_{2k,2k+1}^{\phantom{\dagger}}
\Bigg)
    \Big(b_{2,3}^{\phantom{\dagger}}
     c_{2,-\sigma}^{\phantom{\dagger}}\Big)
     c_{2m-1,\tau}^{\phantom{\dagger}}
     c_{2m-1,\tau}^{\dagger}
     c_{2(n-1)+1,\sigma}^{\dagger}
    \prod_{k=2}^{n-1} b_{2k-1,2k}^{\dagger}
    \Bigg\vert 0 \right\rangle \nonumber \\
&=&-\left\langle 0 \Bigg\vert
\Bigg(
    \prod_{k=2}^{n-1} b_{2k,2k+1}^{\phantom{\dagger}}
\Bigg)
     c_{3,\sigma}^{\phantom{\dagger}}
     c_{2m-1,\tau}^{\phantom{\dagger}}
     c_{2m-1,\tau}^{\dagger}
     c_{2(n-1)+1,\sigma}^{\dagger}
    \prod_{k=2}^{n-1} b_{2k-1,2k}^{\dagger}
    \Bigg\vert 0 \right\rangle.
\end{eqnarray}
We change the labeling of the lattice sites by the rule
$j \to j-2$ and obtain
\begin{eqnarray}
G(n-1,2m;\sigma,\tau)
&=&-\left\langle 0 \Bigg\vert
\Bigg(
    \prod_{k=1}^{n-2} b_{2k,2k+1}^{\phantom{\dagger}}
\Bigg)
     c_{1,\sigma}^{\phantom{\dagger}}
     c_{2(m-1)-1,\tau}^{\phantom{\dagger}}
     c_{2(m-1)-1,\tau}^{\dagger}
     c_{2(n-2)+1,\sigma}^{\dagger}
    \prod_{k=1}^{n-2} b_{2k-1,2k}^{\dagger}
    \Bigg\vert 0 \right\rangle \nonumber \\
&=&-G \left( n-1,2(m-1);\sigma,\tau \right)  \nonumber \\
&=& \cdots \nonumber \\
&=& (-1)^m G \left( n-m-1,2;\sigma,\tau \right)  \nonumber \\
&=& (-1)^m
    \left\langle 0 \Bigg\vert
\Bigg(
    \prod_{k=1}^{n-m-1} b_{2k,2k+1}^{\phantom{\dagger}}
\Bigg)
     c_{1,\sigma}^{\phantom{\dagger}}
     c_{1,\tau}^{\phantom{\dagger}}
     c_{1,\tau}^{\dagger}
     c_{2(n-m-1)+1,\sigma}^{\dagger}
    \prod_{k=1}^{n-m-1} b_{2k-1,2k}^{\dagger}
    \Bigg\vert 0 \right\rangle \nonumber \\
&=& (-1)^m
    \left\langle 0 \Bigg\vert
\Bigg(
    \prod_{k=1}^{n-m-1} b_{2k,2k+1}^{\phantom{\dagger}}
\Bigg)
     c_{1,\sigma}^{\phantom{\dagger}}
     c_{1,\tau}^{\phantom{\dagger}}
     c_{2(n-m-1)+1,\sigma}^{\dagger}
     \left(
    -c_{1,\tau}^{\dagger}
     b_{1,2}^{\dagger}
     \right)
    \prod_{k=2}^{n-m-1} b_{2k-1,2k}^{\dagger}
    \Bigg\vert 0 \right\rangle \nonumber \\
&=& (-1)^m
    \left\langle 0 \Bigg\vert
\Bigg(
    \prod_{k=1}^{n-m-1} b_{2k,2k+1}^{\phantom{\dagger}}
\Bigg)
     c_{1,\sigma}^{\phantom{\dagger}}
     c_{1,\tau}^{\phantom{\dagger}}
     b_{1,1}^{\dagger}
     c_{2(n-m-1)+1,\sigma}^{\dagger}
     c_{2,\tau}^{\dagger}
    \prod_{k=2}^{n-m-1} b_{2k-1,2k}^{\dagger}
    \Bigg\vert 0 \right\rangle \nonumber \\
&=& (-1)^m {\rm sgn}(\tau) \delta_{\sigma, -\tau}
    \left\langle 0 \Bigg\vert
\Bigg(
    \prod_{k=1}^{n-m-1} b_{2k,2k+1}^{\phantom{\dagger}}
\Bigg)
     c_{2(n-m-1)+1,\sigma}^{\dagger}
     c_{2,\tau}^{\dagger}
    \prod_{k=2}^{n-m-1} b_{2k-1,2k}^{\dagger}
    \Bigg\vert 0 \right\rangle \nonumber \\
&=& (-1)^{m} \delta_{\sigma, -\tau}
    \left\langle 0 \Bigg\vert
\Bigg(
    \prod_{k=2}^{n-m-1} b_{2k,2k+1}^{\phantom{\dagger}}
\Bigg)
     c_{3,-\tau}^{\dagger}
     c_{2(n-m-1)+1,\sigma}^{\dagger}
    \prod_{k=2}^{n-m-1} b_{2k-1,2k}^{\dagger}
    \Bigg\vert 0 \right\rangle.
\end{eqnarray}
We change the labeling of the lattice sites by the rule
$k \to k-2$ and obtain
\begin{eqnarray}
G(n-1,2m;\sigma,\tau)
&=& (-1)^{m} \delta_{\sigma, -\tau}
    \left\langle 0 \Bigg\vert
\Bigg(
    \prod_{k=1}^{n-m-2} b_{2k,2k+1}^{\phantom{\dagger}}
\Bigg)
     c_{1,-\tau}^{\dagger}
     c_{2(n-m-2)+1,\sigma}^{\dagger}
    \prod_{k=1}^{n-m-2} b_{2k-1,2k}^{\dagger}
    \Bigg\vert 0 \right\rangle \nonumber \\
&=& (-1)^{m} \delta_{\sigma, -\tau}
    L(n-m-2;\sigma, -\tau) \nonumber \\
&=& (-1)^{m} \delta_{\sigma, -\tau}
    \times (-1)^{n-m-2} \delta_{\sigma, -\tau} \nonumber \\
&=& (-1)^{n-1} \delta_{\sigma, -\tau},
\label{eq:app:den2}
\end{eqnarray}
where we used (\ref{eq:identL}).
For $j=2m-1$, the similar calculation leads
\begin{eqnarray}
G(2m-1;\sigma,\tau) &=& (-1)^{n-1} \delta_{\sigma, \tau}.
\label{eq:app:den3}
\end{eqnarray}
Combining (\ref{eq:app:den2}) and (\ref{eq:app:den3}),
(\ref{eq:app:den1}) is
\begin{eqnarray}
\left\langle 0 \Bigg\vert
\Bigg(
    \prod_{k=1}^{n} b_{2k,2k+1}^{\phantom{\dagger}}
\Bigg)
    c_{1,\sigma}^{\phantom{\dagger}}
    c_{y,\tau}^{\phantom{\dagger}}
    c_{y,\tau}^{\dagger}
    c_{1,\sigma}^{\dagger}
    \prod_{k=1}^{n} b_{2k-1,2k}^{\dagger}
    \Bigg\vert 0 \right\rangle
&=& (-1)^{n-1} \delta_{\sigma, (-1)^{d(x,y)} \tau}.
\label{eq:app:den4}
\end{eqnarray}
We obtain the weight
\begin{eqnarray}
w_j(x,y;\sigma)
&=&
\frac{1}{2}
\left[ G(n,y;\sigma,\sigma) +G(n,y;\sigma,-\sigma) \right]
\nonumber \\
&=&
(-1)^{l(x,y)/2-1}\times
 \frac{\delta_{\sigma, (-1)^{d(x,y)}\sigma}
+\delta_{\sigma, (-1)^{d(x,y)+1}\sigma}}{2} \nonumber \\
&=&
\frac{(-1)^{l(x,y)/2-1}}{2},
\end{eqnarray}
where we used the relation $n=l(x,y)/2$.

\newpage
\section{Derivation of transfer matrices for Model B}

In Model B, the sites which is identified in the cell construction
are $p$-sites
and there exists the sites with four bonds
in the geometric representation.
We need the procedures shown in Fig.\ \ref{fig:elimination2}
(b) and (c)
to calculate the contribution from the graph.
Since we do not need the procedure shown
in Fig.\ \ref{fig:elimination2} (a)
except for the correlation function
$\langle c_{i,\sigma}c_{j,\sigma}^{\dagger} \rangle$,
$m_j=0$ in (\ref{eq:norm3a}), (\ref{eq:spin9}), (\ref{eq:numa}),
(\ref{eq:dendenfirst}), and (\ref{eq:propz}).

\subsection{Norm of the ground state}
\label{app:transfb}

For the sake of convenience,
we draw the lattice shown in Fig.\ \ref{fig:modelBcell}(b)
as Fig.\ \ref{fig:modelBdemo}(a).
The ground state is written by (\ref{eq:gs2}) geometrically,
where an example of the valence-bond configuration $V$
is shown in Fig.\ \ref{fig:modelBdemo}(b).
The geometric representation of the norm is (\ref{eq:norm3a}),
where an example of the graph $V \cup V'$ is shown
in Fig.\ \ref{fig:modelBdemo}(c).
We first evaluate the contribution from a graph $V \cup V'$.
It can be decomposed into the subgraphs
$U_i \cup U'_i$ $(i=1$, $2$, $\cdots$, $n(V \cup V'))$.
The Loops which are extending over more than two cells are allowed.
(See Figs.\ \ref{fig:modelBdemo}(c) and (j)).
After the elimination of sites with four non-closed bonds
(by the procedure shown in Fig.\ \ref{fig:elimination2}(b) or (c))
we have one degenerate loop and self-closed bonds
(Fig.\ \ref{fig:modelBdemo}(d)).
Therefore, the loop (Fig.\ \ref{fig:modelBdemo} (j))
has weight $2$.

We derive the matrix ${\bf T}_n$ in (\ref{eq:mat:transf}).
A subgraph $U_i \cup U'_i$
is constructed by the five kinds of cells shown in
(Fig.\ \ref{fig:modelBnorm}).
The sum over the graph $V \cup V'$
is equivalent to that over all the combination of above five cells
under the restriction that a $p$-site has at most four
valence-bonds.
(The restriction means, for example, that the identification
of the right $p$-site in Fig.\ \ref{fig:modelBnorm}(b)
with the left $p$-site in Fig.\ \ref{fig:modelBnorm}(c)
is forbidden.)
We shall always take into account the restriction hereafter
and it should be understood implicitly.
We have to distinguish three cases due to the restriction.
Let $A_n$, $B_n$, and $C_n$ be the quantity defined
by the right-hand side of (\ref{eq:normg2})
on the lattice $\Lambda_n$.
For $A_n$, the sum is taken over all the combination of the cells
shown in Fig.\ \ref{fig:modelBnorm} with the restriction
that the $n$-th cell is represented
by Fig.\ \ref{fig:modelBnorm}(a) or (b).
For $B_n$ and $C_n$, the sum is taken as was done for $A_n$
with the restriction that the $n$-th cell is represented either
by Fig.\ \ref{fig:modelBnorm}(c) or (d) for $B_n$
and by (e) for $C_n$.
They are represented diagrammatically
\begin{eqnarray}
A_n &=&\\
B_n &=&\\
C_n &=&
\end{eqnarray}
The recursion relations are
\begin{eqnarray}
A_{n} &=& \nonumber\\
      &=& \lambda_2^4A_{n-1} +2\lambda_2^2A_{n-1}
          +2\lambda_2^2B_{n-1}\\
B_{n} &=& \nonumber \\
      &=& \lambda_1^2A_{n-1} +2\lambda_1^2B_{n-1} +\lambda_1^2C_{n-1}
     +\lambda_1^2\lambda_2^2A_{n-1} +\lambda_1^2\lambda_2^2B_{n-1}\\
C_{n} &=& \nonumber \\
      &=& \lambda_1^4A_{n-1} +2\lambda_1^4B_{n-1} +\lambda_1^4C_{n-1},
\label{eq:normrec}
\end{eqnarray}
where we adopt the following rule to assign weight 2
to the loop which is extending over more than two cells.
We assign the coefficient 1 for all the terms in $B_n$,
since the loop may continue to the $n+1$-th cell.
In the third term in $A_n$ and the second term in $B_n$ and $C_n$
we assign weight 2,
since we make sure that the loop finishes there.
The corresponding matrix is ${\bf T}_n$ in (\ref{eq:mat:transfb}).
Since any cell in Fig.\ \ref{fig:modelBnorm}
is allowed at the boundaries, the initial and the final vectors are
$\vec{I} \equiv (A_0, B_0, C_0)^T=(1, 0, 0)^T$ and
$\vec{F} \equiv (A_N, B_N, C_N)^T=(1, 2, 1)^T$, respectively.

\subsection{Expectation value of the number operator}
\label{app:matoccb}

The geometric representation of the expectation value
is (\ref{eq:numbergeometric}),
where an example of the graph $V \cup V'$ is shown
in Fig.\ \ref{fig:modelBdemo}(e).
We derive the matrix ${\bf N}_i^{(p)}$.
We have five kinds of graphs on the $i$-th cell
(Fig.\ \ref{fig:modelBocc}(a)-(e)).
Let $A^{(p)}_i$, $B^{(p)}_i$, and $C^{(p)}_i$ be the quantity
defined by the right-hand side of (\ref{eq:numbergeometric})
on the lattice $\Lambda_i$.
The restriction for the sum is that
the $i$-th cell is represented either
by Fig.\ \ref{fig:modelBocc}(a) or (b) for $A^{(p)}_i$,
by (c) or (d) for $B^{(p)}_i$, and by (e) for $C^{(p)}_i$.
The recursion relations are
\begin{eqnarray}
A^{(p)}_{i} &=&\nonumber \\
            &=& \lambda_2^2 A_{i-1} +2\lambda_2^2 B_{i-1}
               +\lambda_2^4 A_{i-1} \\
B^{(p)}_{i} &=&\nonumber \\
            &=& \frac{1}{2}\lambda_1^2\lambda_2^2 A_{i-1}
               +\lambda_1^2\lambda_2^2 B_{i-1}
               +\lambda_1^2 B_{i-1} +\lambda_1^2 C_{i-1} \\
C^{(p)}_{i} &=&\nonumber \\
            &=& \lambda_1^4B_{i-1},
\end{eqnarray}
where we used the same rule in subsection\ \ref{app:transfb}
for the subgraph without the number operator.
For the subgraph with the operator, we adopt the following rule
to assign the weight (\ref{eq:proof3}).
For the first and the third terms in $A^{(p)}_i$,
the third and fourth terms in $B^{(p)}_i$,
and $C^{(p)}_i$, we assigned coefficient 1.
For the second term in $A^{(p)}_i$,
we assigned weight 2,
since the loop is decomposed into self-closed bonds
and a degenerate loop which finishes there.
For the second term in $B^{(p)}_i$,
we assigned coefficient 1,
since the loop is decomposed into self-closed bonds
and a degenerate loop which may continue to the $i+1$-th cell.
For the first term in $B^{(p)}_i$,
we assinged coefficient $1/2$ to cancel weight $2$
which is assigned at another end of the loop
where we regard it as the end of a loop without the number operator.
The corresponding matrix is ${{\bf N}}_i^{(p)}$ in (\ref{eq:Bn}).
Using the cells shown in Fig.\ \ref{fig:modelBocc}(f) and (g),
the similar calculation leads
${\bf N}_i^{(d)}$ in (\ref{eq:Bn}).

\subsection{Spin correlation function}
\label{app:matspinb}

The geometric representation of the expectation value
is (\ref{eq:spingeometric}),
where an example of the graph $V \cup V'$ is shown
in Fig.\ \ref{fig:modelBdemo}(f). From the procedure
in Fig.\ \ref{fig:elimination2}(b) or (c),
the graph with the spin operators is decomposed into
a degenerate loop with the sites $i$ and $j$
and the self-closed bonds.
Therefore, $d(i,j)=1$ in (\ref{eq:spingeometric}).
The degenerate loop has weight $-1 \times \frac{1}{4} \times 2$.
We assign $\frac{1}{2}$ at the $i$-th cell
and $-\frac{1}{2} \times 2$ at the $j$-th cell
in the recursion relation.
We derive the matrix ${\bf S}_{n}$ in (\ref{eq:twopointgne}).
Let $S_n$ be the quantity defined by a sum.
The sum is taken over $V \cup V'$ on the lattice $\Lambda_n$
such that the graph consists of loops shown
in Fig.\ \ref{fig:modelBnorm} on the $k$-th cell ($1 \le k \le i-1$)
and that shown in Fig.\ \ref{fig:modelBnorm}(c) on the $l$-th cell
($i \le l \le n$).
The recursion relation is
\begin{eqnarray}
S_n &=&\nonumber \\
    &=&\nonumber \\
    &=& \lambda_1^2 \lambda_2^2 S_{n-1},
\label{eq:spintr}
\end{eqnarray}
and the corresponding matrix is ${\bf S}_n$ in (\ref{eq:spintrb1}).

We derive the matrix ${\bf S}_{i}^{R,(p)}$.
The graph on the $i$-th cell is shown in Fig.\ \ref{fig:modelBspin}(a).
Let $S_i^{(p)}$ be the quantity defined by a sum.
The sum is taken over $V \cup V'$ on the lattice $\Lambda_i$
such that
the graph consists of loops shown in Fig.\ \ref{fig:modelBnorm}
on the $k$-th cell
($1 \le k \le i-1$) and that shown in Fig.\ \ref{fig:modelBspin}(a)
on the $i$-th cell.
We have
\begin{eqnarray}
S_i^{(p)}&=&\nonumber \\
&=& \frac{1}{2} \times \lambda_1^2 \lambda_2^2 A_{i-1}.
\label{eq:aspin1}
\end{eqnarray}
The corresponding matrix is
${\bf S}_i^{R,(p)}$ in (\ref{eq:spintrb1}).

We derive the matrix ${\bf S}_{i}^{L,(p)}$.
The graph on the $j$-th cell is shown in Fig.\ \ref{fig:modelBspin}(b).
Let $B^{(p)}_j$ and $C^{(p)}_j$ be the quantity defined
by the right-hand side of (\ref{eq:spingeometric})
on the lattice $\Lambda_j$.
The restriction for the sum is that
the $j$-th cell is represented either by Fig.\ \ref{fig:modelBnorm}(d)
for $B_j$ and by (e) for $C_j$.
The recursion relations are
\begin{eqnarray}
A^{(p)}_{j} &=& 0 \\
B^{(p)}_{j} &=&\nonumber \\
       &=& -\frac{1}{2} \times 2 \times \lambda_1^2 S_{j-1} \\
C^{(p)}_{j} &=&\nonumber\\
       &=& -\frac{1}{2} \times 2 \times \lambda_1^4 S_{j-1}.
\label{eq:aspin2}
\end{eqnarray}
The corresponding matrix is
${\bf S}_j^{L,(p)}$ in (\ref{eq:spintrb1}). From the graph shown
in Fig.\ \ref{fig:modelBspin}(c) ((d)),
the similar calculation leads
the matrix ${\bf S}_{i}^{R,(d)}$ (${\bf S}_{j}^{L,(d)}$)
in (\ref{eq:spintrb1}).

\subsection{Singlet-pair correlation function}
\label{app:matsingleb}

The geometric representation of the expectation value
is (\ref{eq:norm3a}),
where an example of the graph $V \cup V'$ is shown
in Figs.\ \ref{fig:modelBdemo}(g) and (h)
for (\ref{eq:classifysinglet})-(i) and (iii), respectively.
In the Figures,
a double solid (broken) line represent the operator
$b_{i,j}^{\dagger}$ $(b_{k,l})$.
We derive the transfer matrix ${\bf H}_n$ in
(\ref{eq:twopointgne}).
Let $H_n$ be the quantity defined by a sum.
The sum is taken over $V \cup V'$ on the lattice $\Lambda_n$
such that the graph consists of loops shown
in Fig.\ \ref{fig:modelBnorm} on the $k$-th cell ($1 \le k \le i-1$)
and graph shown in Fig.\ \ref{fig:modelBsingle}(a)
on the $l$-th cell ($i < l \le n$).
The recursion relation is
\begin{eqnarray}
H_n &=&\nonumber \\
    &=&\nonumber \\
    &=& \lambda_1^2 \lambda_2^2 {\bf H}_{n-1},
\end{eqnarray}
and the corresponding matrix is ${\bf H}_n$
in (\ref{eq:singletrb}).

We derive the matrix ${\bf H}_i^{R,(pp)}$.
(We consider the case (\ref{eq:classifysinglet})-(i).)
The graph on the $i$-th cell is shown
in Fig.\ \ref{fig:modelBsingle}(b).
Let $H_i^{(pp)}$ be the quantity defined by a sum.
The sum is taken over $V \cup V'$ on the lattice $\Lambda_i$
such that the graph consists of loops shown
in Fig.\ \ref{fig:modelBnorm}
on the $n$-th cell ($1 \le n \le i-1$)
and that shown in Fig.\ \ref{fig:modelBsingle}(b)
on the $i$-th cell.
The recursion relation is
\begin{eqnarray}
H_i^{(pp)} &=&\nonumber \\
    &=& \lambda_1^2 \lambda_2^2 A_{i-1}.
\label{eq:bsingle1}
\end{eqnarray}
The corresponding matrix is ${\bf H}_i^{R,(pp)}$
in (\ref{eq:singletrb}).

We derive the matrix ${\bf H}_k^{L,(pp)}$.
(We consider the case (\ref{eq:classifysinglet})-(i).)
We have two kinds of graphs on the $k$-th cell
(Fig.\ \ref{fig:modelBsingle}(c) and (d)).
Let $B^{(pp)}_k$ and $C^{(pp)}_k$
be the quantity defined by the right-hand side of (\ref{eq:singletgeo})
on the lattice $\Lambda_k$.
The sum is taken over $V \cup V'$
with the restriction that
the $k$-th cell is represented by Fig.\ \ref{fig:modelBsingle}(c)
for $B^{(pp)}_k$ and (d) for $C^{(pp)}_k$.
The recursion relations are
\begin{eqnarray}
A^{(pp)}_{k} &=& 0 \\
B^{(pp)}_{k} &=&\nonumber \\
       &=& 2 \lambda_1^2 H_{k-1} \\
C^{(pp)}_{k} &=&\nonumber \\
       &=& 2 \lambda_1^4 H_{k-1},
\label{eq:bsingle2}
\end{eqnarray}
The corresponding matrix is ${\bf H}_k^{(pp)}$
in (\ref{eq:singletrb}).

We consider the case (\ref{eq:classifysinglet})-(iii).
The graphs on the $i$-th and the $k$-th cells are shown
in Fig.\ \ref{fig:modelBsingle}(e) and (f), respectively.
The similar calculation leads the matrices
${\bf H}_i^{R,(dp)}$ and ${\bf H}_k^{L,(dp)}$ in (\ref{eq:singletrb}).

\subsection{
Correlation function
$\langle c_{i,\sigma}^{\phantom{\dagger}}c_{j,\sigma}^{\dagger} \rangle$
}
\label{app:propb}

The geometric representation of the expectation value is
(\ref{eq:greengeometric}),
where an example of the graph $V \cup V'$ is shown
in Fig.\ \ref{fig:modelBdemo}(i).
We derive the matrix ${\bf G}_n$ in (\ref{eq:twopointgne}).
A line is constructed
by combining three kinds of cells shown
in Figs.\ \ref{fig:modelBprop}(a)-(c).
To calculate the weight of the graph,
it is necessary to use the procedure shown in
Fig.\ \ref{fig:elimination2}(a).
If we use the procedure $m$ times for the graph with $2n$ bonds,
we obtain a line with $2n-2m$ bonds and $m$ self-closed bonds.
The weight of the line is $(-1)^{(2n-2m)/2} \times (-1)^m$,
where the former power comes from (\ref{eq:proof2})
and the latter one comes from the procedure
shown in Fig.\ \ref{fig:elimination2}(a).
We use the same rule as that in (\ref{eq:gtpp})
and the minus sign in Fig.\ \ref{fig:elimination2}(a)
is automatically taken into account.
Let $D_n$ and $E_n$ be the quantity defined by a sum.
The sum is taken over $V \cup V'$ on the lattice $\Lambda_n$
such that the graph consists of loops shown
in Fig.\ \ref{fig:modelBnorm} on the $k$-th cell  ($1 \le k \le i-1$)
and the graphs shown in Fig.\ \ref{fig:modelBprop}(a)-(c)
on the $l$-th cell ($i \le l \le n$).
The restriction for the sum is that
the $n$-th cell is represented by (a) or (b) for $D_n$
and (c) for $E_n$.
The recursion relations are
\begin{eqnarray}
D_n &=&\nonumber \\
    &=&\nonumber \\
    &=& -\lambda_1 \lambda_2 D_{n-1}
        -\lambda_1 \lambda_2 E_{n-1}
        -\lambda_1 \lambda_2^3 D_{n-1}
\\
E_n &=&\nonumber \\
    &=&\nonumber \\
    &=& -\lambda_1^3 \lambda_2 D_{n-1}
        -\lambda_1^3 \lambda_2 E_{n-1}.
\end{eqnarray}
The corresponding matrix is ${\bf G}_n$ in (\ref{eq:bgreen1}).

We derive the matrix ${\bf G}_i^{R,(p)}$.
We have three kinds of graphs on the $i$-th cell
(Fig.\ \ref{fig:modelBprop}(d)-(f)).
Let $D_i^{(p)}$ and $E_i^{(p)}$
be the quantity defined by a sum.
The sum is taken over $V \cup V'$ on the lattice $\Lambda_i$
such that the graph consists of loops shown
in Fig.\ \ref{fig:modelBnorm} on the $k$-th cell ($1 \le k \le i-1$)
and those shown in Fig.\ \ref{fig:modelBprop}(d)-(f) on $i$-th cell.
The restriction for the sum is that
the $i$-th cell is represented either
by Fig.\ \ref{fig:modelBprop} (d) or (e) for $D_i^{(p)}$ and
by (f) for $E_i^{(p)}$.
The recursion relations are
\begin{eqnarray}
D_i^{(p)} &=&\nonumber\\
          &=&  - \lambda_1 \lambda_2 A_{i-1}
               - \lambda_1 \lambda_2 B_{i-1}
               - \lambda_1 \lambda_2^3 A_{i-1}
 \\
E_i^{(p)} &=&\nonumber\\
          &=& - \lambda_1^3 \lambda_2 A_{i-1}
              - \lambda_1^3 \lambda_2 B_{i-1}.
\end{eqnarray}
The corresponding matrix is ${\bf G}_i^{R,(p)}$ in (\ref{eq:bgreen1}).

We derive the matrix ${\bf G}_j^{L,(p)}$.
We have four kinds of graphs on the $j$-th cell
(Fig.\ \ref{fig:modelBprop}(g)-(j)).
Let $A^{(p)}_{j}, B^{(p)}_{j}$,
and $C^{(p)}_{j}$ be the quantity defined by the right-hand side
of (\ref{eq:greengeometric})
on the lattice $\Lambda_j$.
The sum is taken over $V \cup V'$
such that the graph consists of loops shown
in Fig.\ \ref{fig:modelBnorm}
on the $k$-th $(1 \le k \le i-1)$ cells
and a line shown in Fig.\ \ref{fig:modelBprop}(a)-(c)
on the $l$-th $(i \le l \le j-1)$ cells.
The restriction for the sum is that the $j$-th cell
is represented by either
by Fig.\ \ref{fig:modelBprop}(g) for $A^{(p)}_{j}$,
by (h) or (i) for $B^{(p)}_{j}$, and (j) for $C^{(p)}_{j}$.
The recursion relations are
\begin{eqnarray}
A_{j}^{(p)} &=&\nonumber\\
            &=& \lambda_2^2 D_{j-1}\\
B_{j}^{(p)} &=&\nonumber\\
            &=&  \frac{1}{2}\lambda_1^2\lambda_2^2D_{j-1}
                +\lambda_1^2D_{j-1}
                +\lambda_1^2E_{j-1}
 \\
C_{j}^{(p)} &=&\nonumber\\
            &=& \lambda_1^4D_{j-1} +\lambda_1^4E_{j-1}.
\end{eqnarray}
The coefficient $1/2$ in the first term in $B_i^{(p)}$
is to cancel weight $2$ which is assigned
at the left end of the line
where we regard it as the end of a loop.
The corresponding matrix is ${\bf G}_j^{L,(p)}$ in (\ref{eq:bgreen1}).

We derive the matrix ${\bf G}_i^{R,(d)}$.
We have two kinds of graphs on the $i$-th cell
(Fig.\ \ref{fig:modelBprop}(k) and (l)).
Let $D_i^{(d)}$ and $E_i^{(d)}$
be the quantity defined by a sum.
The sum is taken over $V \cup V'$ on the lattice $\Lambda_i$
such that the graph consists of loops shown
in Fig.\ \ref{fig:modelBnorm} on the $k$-th cell ($1 \le k \le i-1$)
and that shown in Fig.\ \ref{fig:modelBprop}(k) or (l) on $i$-th cell.
The restriction for the sum is that
the $y$-th cell is represented either
by Fig.\ \ref{fig:modelBprop}(k) for $D_y^{(d)}$
and by (l) for $E_y^{(d)}$.
The recursion relations are
\begin{eqnarray}
D_i^{(d)} &=&\nonumber\\
          &=& -\lambda_1\lambda_2^2A_{i-1}
              -\lambda_1\lambda_2^2B_{i-1} \\
E_i^{(d)} &=&\nonumber\\
          &=& -\lambda_1^3  A_{i-1}
              -2\lambda_1^3  B_{i-1}
              -\lambda_1^3 C_{i-1}.
\hspace{30mm}
\end{eqnarray}
The corresponding matrix is ${\bf G}_i^{R,(d)}$ in (\ref{eq:bgreen1}).

We derive the matrix ${\bf G}_{j}^{L,(d)}$.
We have two kinds of graphs on the $j$-th cell
(Fig.\ \ref{fig:modelBprop}(m) and (n)).
Let $A_j^{(d)}$ and $B_j^{(d)}$ be
the quantity defined by the right-hand side
of (\ref{eq:greengeometric})
on the lattice $\Lambda_j$.
The sum is taken over $V \cup V'$
such that the graph consists of loops shown
in Fig.\ \ref{fig:modelBnorm}
on the $k$-th $(1 \le k \le i-1)$ cells
and a line shown in Fig.\ \ref{fig:modelBprop}(a)-(c)
on the $l$-th $(i \le l \le j-1)$ cells.
The restriction for the sum is that the $j$-th cell is represented either
by Fig.\ \ref{fig:modelBprop}(m) for $A_j^{(d)}$ and by (n)
for $B_j^{(d)}$.
The recursion relations are
\begin{eqnarray}
A_{j}^{(d)} &=&\nonumber\\
            &=& -\lambda_2^3 D_{j-1} \\
B_{j}^{(d)} &=&\nonumber\\
            &=& -\frac{1}{2}\lambda_1^2 \lambda_2 D_{j-1}
                -\frac{1}{2}\lambda_1^2 \lambda_2 E_{j-1} \\
C_{j}^{(d)} &=& 0.
\end{eqnarray}
The coefficient $1/2$ in $B_j^{(d)}$ is
to cancel weight $2$ which is assigned
at the end of the line
where we regard it as the end of a loop.
The corresponding matrix is ${\bf G}_j^{L,(d)}$ in (\ref{eq:bgreen1}).

\newpage
\section{Derivation of transfer matrices for Model C}

In Model C, the sites which is identified in the cell construction
are $d$-sites
and there is no site with four bonds in the geometric representation.
Therefire, we do not need the procedures shown
in Figs.\ \ref{fig:elimination2},
$m_j=0$ in (\ref{eq:norm3a}), (\ref{eq:spin9}), (\ref{eq:propz}),
(\ref{eq:numa}), (\ref{eq:dendenfirst}), and (\ref{eq:singletgeo}).

\subsection{Norm of the ground state}
\label{app:transfc}

The ground state is written by (\ref{eq:gs2}) geometrically,
where an example of the valence-bond configuration $V$ is shown
in Fig.\ \ref{fig:modelCdemo}(a).
The geometric representation of the norm is (\ref{eq:norm3a}),
where an example of the graph $V \cup V'$ is shown
in Fig.\ \ref{fig:modelCdemo}(b).
We first evaluate the contribution from a graph $V \cup V'$.
It can be decomposed into the subgraphs
$U_i \cup U'_i$ $(i=1$, $2$, $\cdots$, $n(V \cup V'))$.

We derive the matrix ${\bf T}_n$ in (\ref{eq:mat:transf}).
A subgraph $U_i \cup U'_i$
is constructed by 27 kinds of cells shown in
Fig.\ \ref{fig:modelCnorm}.
The sum over the graph $V \cup V'$
is equivalent to that over all the combination of
above 27 kinds of cells under the restriction
that a $d$-site has at most two valence-bonds.
We shall always take into account the restriction hereafter
and it should be understood implicitly.
Due to the restriction, we have to distinguish four cases
with respect to two $d$-sites at left end of the lattice $\Lambda_n$:
(i) they do not belong to valence bond;
(ii) one of them belongs to two bonds;
(iii) they each belong to two distinct bonds;
(iv) they belong to a single degenerate loop.
Let $A_n$-$D_n$ be the quantity defined by the right-hand side
of (\ref{eq:normg2}) on the lattice $\Lambda_n$.
The sum is taken over $V \cup V'$
such that the graph consists of the subgraphs shown
in Fig.\ \ref{fig:modelCnorm}.
The restriction for the sum is that the $n$-th cell is classified
by the condition
(i) for $A_n$, (ii) for $B_n$, (iii) for $C_n$, and (iv) for $D_n$.
They are represented diagrammatically
\begin{eqnarray}
A_n &=&  \\
B_n &=&      =   \\
C_n &=&      =  \\
D_n &=&
\end{eqnarray}
For the weight of a non-degenerate loop with $2n$ bonds,
we assign the sign part $(-1)^{n-1}$ and the coefficient $2$
separately.
The loop belongs to $n$ cells, since a cell has $2$ bonds.
We assign $-1$ to each $n-1$ cells except for the cell
in the right end of the loop.
Accordingly, every recursion from $C_{n-1}$ gives a minus sign.
At left end of the loop, we assign the weight $2$.
Therefore, when we attach $C_{n-1}$ to the $n$-th cell,
we adopt the following rule.
If the $n$-th cell satisfy the condition (i) or (ii),
we make sure that the loop finishes on the $n$-th cell.
Therefore, we assign coefficient $-1 \times 2$.
In other cases, we assign coefficient $-1$.
The recursion relations are
\begin{eqnarray}
A_n &=&\nonumber \\
   &=& 2A_{n-1} +2\lambda^2A_{n-1} +2\lambda^2B_{n-1} +2\lambda^2A_{n-1}
        +2\lambda^2B_{n-1}
        -2\lambda^2C_{n-1} -2\lambda^2C_{n-1}  \nonumber \\
    & & +\lambda^4A_{n-1} +\lambda^4B_{n-1} +\lambda^4B_{n-1}
        +\lambda^4D_{n-1} \\
B_n &=&\nonumber \\
    &=& 2\lambda^2A_{n-1} +2\lambda^2B_{n-1} +2\lambda^2B_{n-1}
       +2\lambda^2D_{n-1}
       +2A_{n-1} +2B_{n-1} +2A_{n-1} +2B_{n-1} \nonumber \\
    & &-2C_{n-1} -2C_{n-1} \\
C_n &=&\nonumber \\
    &=& \lambda^2A_{n-1} +\lambda^2B_{n-1}
       +\lambda^2B_{n-1} +\lambda^2D_{n-1}
       +A_{n-1} +B_{n-1} +A_{n-1} +B_{n-1}
\nonumber \\
    & &       -C_{n-1} -C_{n-1} \\
D_n &=&\nonumber \\
    &=& 2\lambda^2A_{n-1} +2\lambda^2B_{n-1} +2\lambda^2B_{n-1}
       +2\lambda^2D_{n-1},
\label{eq:Cnpr}
\end{eqnarray}
where we assigned $\lambda^{N_b}$ to the valence-bonds
on the $n$-th cell as the contribution from $\lambda(U)\lambda(U')$.
Here $N_b$ is the number of the bonds which share a $p$-site.
The corresponding matrix is
\begin{eqnarray}
 \left( \begin{array} {cccc}
  2 +4\lambda^2 +\lambda^4 & 4\lambda^2 +2\lambda^4
& -4\lambda^2 & \lambda^4  \\
  4 +2\lambda^2            & 4 +4\lambda^2
& -4          & 2\lambda^2 \\
  2 +\lambda^2             & 2 +2\lambda^2
& -2          & \lambda^2  \\
  2                        & 4
&  0          & 2          \\
 \end{array}
\right).
\end{eqnarray}
The initial and final vectors are
$\vec{I}$ $\equiv (A_0, B_0, C_0, D_0)^T$ $=(1,0,0,0)^T$
and
$\vec{F}$ $\equiv (A_N, B_N, C_N, D_N)^T$ $=(1,2,0,1)^T$,
respectively.
We find $B_n$ $=2C_n$.
Therefore, the matrix is reduced to ${\bf T}_n$ in (\ref{eq:mat:transf})
with the base $(A_n, C_n, D_n)^T$.
The corresponding initial and final vectors are
$\vec{I}$ $\equiv (A_0, C_0, D_0)^T$ $=(1,0,0)^T$
and
$\vec{F}$ $\equiv (A_N, C_N, D_N)^T$ $=(1,4,1)^T$.

\subsection{Expectation value of the number operator}
\label{app:matocc}

The geometric representation of the expectation value
is (\ref{eq:numbergeometric}),
where an example of the graph $V \cup V'$ is shown
in Fig.\ \ref{fig:modelCdemo}(c).
We first derive the matrix ${\bf N}_i^{(p)}$.
We have 9 kinds of graphs on the $i$-th cell
(Fig.\ \ref{fig:modelCnorm}
(a1)-(a4), (a6), (b3), (b8), (c1), and (c2)).
Let $A_i^{(p)}$-$D_i^{(p)}$
be the quantities classified by the conditions (i)-(iv)
in Appendix\ \ref{app:transfc}, respectively.
When we attach $A_{i-1}^{(p)}$-$D_{i-1}^{(p)}$ to the $i$-th cell,
we adopt the following rule to assign weight
$(-1)^{l/2-1}$ to the valence bonds which share the $p$-site.
When we attach $C_{i-1}$ to the $i$-th cell,
we assign coefficient $-1$.
For the degenerate loop on $A_i$ and $B_i$,
we assign weight $1$.
For the valence bonds on $C_i$,
we assign coefficient $1/2$ to cancel weight $2$
which is assinged at another end of the loop
where we regard it as the end of a loop without the number operator.
The recursion relations are
\begin{eqnarray}
A_i^{(p)} &=& \nonumber \\
    &=& \lambda^2A_{i-1} +\lambda^2B_{i-1} +\lambda^2A_{i-1}
       +\lambda^2B_{i-1} -\lambda^2C_{i-1} -\lambda^2C_{i-1}
       +\lambda^4A_{i-1} +\lambda^4B_{i-1}
\nonumber \\
    & &+\lambda^4B_{i-1} +\lambda^4D_{i-1}
\label{eq:app:tr1} \\
B_i^{(p)} &=& \nonumber \\
    &=& \lambda^2A_{i-1} +\lambda^2B_{i-1}
+\lambda^2B_{i-1} +\lambda^2D_{i-1}
\label{eq:app:tr2} \\
C_i^{(p)} &=& \nonumber \\
    &=& \frac{1}{2}\lambda^2A_{i-1} +\frac{1}{2}\lambda^2B_{i-1}
       +\frac{1}{2}\lambda^2B_{i-1} +\frac{1}{2}\lambda^2D_{i-1}
\label{eq:app:tr3} \\
D_i^{(p)} &=& 0.
\end{eqnarray}
The corresponding matrix is
\begin{eqnarray}
 \left( \begin{array} {cccc}
  2\lambda^2 +\lambda^4 & 2\lambda^2 +2\lambda^4
& -2\lambda^2 & \lambda^4            \\
  \lambda^2             & 2\lambda^2
&  0          & \lambda^2            \\
  \frac{1}{2}\lambda^2  & \lambda^2
&  0          & \frac{1}{2}\lambda^2 \\
  0                     & 0
&  0          & 0                    \\
 \end{array}
\right).
\end{eqnarray}From the equality $B_n$ $=2C_n$,
it is reduced to ${\bf N}_i^{(p)}$ in (\ref{eq:Cn})
with the base $(A_n, C_n, D_n)^T$.

We derive the matrix ${\bf N}_i^{(d)}$.
We have 16 kinds of graphs on the $i$-th cell
(Fig.\ \ref{fig:modelCnorm}
(a1)-(a3), (a5), (b1), (b2), (b4), (b6), (b7), (b10), and (c3)-(c8)).
Let $A_i^{(d)}$-$D_i^{(d)}$
be the quantities classified by the conditions (i)-(iv)
in Appendix\ \ref{app:transfc}, respectively. From the same rule
in the derivation of ${\bf N}_i^{(p)}$,
the recursion relations are
\begin{eqnarray}
A_i^{(d)} &=& \nonumber \\
    &=& A_{i-1} +\lambda^2A_{i-1} +\lambda^2B_{i-1} +\lambda^2B_{i-1}
        -\lambda^2C_{i-1} -\lambda^2C_{i-1}
       +\frac{1}{2}\lambda^4B_{i-1} +\frac{1}{2}\lambda^4D_{i-1} \\
B_i^{(d)} &=& \nonumber \\
    &=& \lambda^2B_{i-1} +\lambda^2D_{i-1} +B_{i-1} +A_{i-1} +B_{i-1}
        -C_{i-1} -C_{i-1} \\
C_i^{(d)} &=& \nonumber \\
    &=& \frac{1}{2}\lambda^2B_{i-1} +\frac{1}{2}\lambda^2D_{i-1}
        +\frac{1}{2}A_{i-1} +\frac{1}{2}B_{i-1} +\frac{1}{2}B_{i-1}
        -\frac{1}{2}C_{i-1} -\frac{1}{2}C_{i-1} \\
D_i^{(d)} &=& \nonumber \\
    &=& \lambda^2B_{i-1} +\lambda^2D_{i-1}.
\end{eqnarray}
The corresponding matrix is
\begin{eqnarray}
 \left( \begin{array} {cccc}
  1 +\lambda^2  & 2\lambda^2 +\frac{1}{2}\lambda^4
& -2\lambda^2 & \frac{1}{2}\lambda^4  \\
  1             & 2 +\lambda^2
& -2          & \lambda^2             \\
  \frac{1}{2}   & 1 +\frac{1}{2}\lambda^2
& -1          & \frac{1}{2}\lambda^2  \\
  0             & 1
&  0          & 1                     \\
 \end{array}
\right).
\end{eqnarray}From the equality $B_n$ $=2C_n$,
it is reduced to ${\bf N}_i^{(d)}$ in (\ref{eq:Cn}).

\subsection{Spin correlation function}
\label{app:matspinc}

The geometric representation of the expectation value
is (\ref{eq:spingeometric}),
where an example of the graph $V \cup V'$ is shown
in Fig.\ \ref{fig:modelCdemo}(d).
We evaluate the weight of the graph with the spin operators.
Since there are two valence bonds in a cell,
$d(i,j)=\vert i-j+1 \vert + l_e$ in (\ref{eq:spingeometric})
where $l_e$ is the number of the bonds which are not
on the $n$-th cell ($i \le n \le j$).
(For example, $l_e=6$ in Fig.\ \ref{fig:modelCdemo}(d)).
We use the same rule in Appendix\ \ref{app:transfc} for $C_n$
and that in \ref{app:matspinc} for the assignment of the coefficient
$\frac{1}{4}$.
Accordingly, the coefficient
$(-1)^{d(i,j)}\frac{1}{4}$ in (\ref{eq:spingeometric})
is automatically taken into account.
We derive the matrix ${\bf S}_n$.
We have four kinds of graphs on the $n$-th cell
($i+1 \le n \le j-1$)
(Fig.\ \ref{fig:modelCspin}(a)-(d)).
Let $S_n$ be the quantity defined by a sum.
The sum is taken over $V \cup V'$ on the lattice $\Lambda_n$
such that the graph consists of the cells shown
in Fig.\ \ref{fig:modelCnorm} on the $k$-th cell ($1 \le k \le i-1$)
and those shown in Fig.\ \ref{fig:modelCspin}(a)-(d) on the $l$-th cell
($1 \le l \le n$).
The recursion relation is
\begin{eqnarray}
S_n &=& \nonumber \\
    &=& -S_{n-1} -S_{n-1},
\end{eqnarray}
and we have ${\bf S}_n =-2$.

We derive the matrix ${\bf S}_i^{R,(p)}$.
We have two kinds of graphs on the $i$-th cell
(Fig.\ \ref{fig:modelCspin}(e) and (f)).
The recursion relation is
\begin{eqnarray}
S_i^{(p)} &=&\nonumber \\
    &=& \frac{1}{2}\times \lambda^2 A_{i-1}
       +\frac{1}{2}\times \lambda^2 B_{i-1}
       +\frac{1}{2}\times \lambda^2 B_{i-1}
       +\frac{1}{2}\times \lambda^2 D_{i-1}.
\end{eqnarray}
The corresponding matrix is
$(\frac{1}{2}\lambda^2,\lambda^2,\frac{1}{2}\lambda^2,0)$
and is reduced to ${\bf S}_i^{R,(p)}$
in (\ref{eq:spintrc1}) with the base $(A_n, C_n, D_n)^T$.

We derive the matrix ${\bf S}_j^{L,(p)}$.
We have two kinds of graph on the $j$-th cell
(Fig.\ \ref{fig:modelCspin}(g) and (h)).
The recursion relation is
\begin{eqnarray}
{S'}_j^{(p)} &=&\nonumber \\
    &=& -\frac{1}{2}\times 2\lambda^2 S_{j-1}
        -\frac{1}{2}\times 2\lambda^2 S_{j-1}.
\end{eqnarray}
The recursion relations to the $j+1$-th cell are
\begin{eqnarray}
A_{j+1} &=&\nonumber \\
    &=& 2{S'}_j^{(p)} +2\lambda^2{S'}_j^{(p)} +2\lambda^2{S'}_j^{(p)}
        +\lambda^4{S'}_j^{(p)} \\
B_{j+1} &=& \nonumber \\
    &=& 2\lambda^2{S'}_j^{(p)} +2{S'}_j^{(p)} +2{S'}_j^{(p)}\\
C_{j+1} &=& \nonumber \\
    &=& \lambda^2{S'}_j^{(p)}  +{S'}_j^{(p)} +{S'}_j^{(p)} \\
D_{j+1} &=& \nonumber \\
    &=& 2\lambda^2{S'}_j^{(p)},
\end{eqnarray}
and the corresponding matrix is
\begin{eqnarray}
{\bf S'}_{j+1}^{(p)}&=&
\left(
 \begin{array} {c}
   2+ 4\lambda^2 +\lambda^4 \\
   4+ 2\lambda^2 \\
   2+ \lambda^2 \\
   2
 \end{array}
\right).
\end{eqnarray}
A simple calculation leads that the product
${\bf S'}_{j+1}^{(p)} {\bf S'}_j^{(p)}$
is reduced to
${\bf T}_{j+1}^{(p)} {\bf S}_j^{(p)}$
with the base $(A_n, C_n, D_n)^T$
and we obtain ${\bf S}_j^{L,(p)}$ in (\ref{eq:spintrc1}).

We derive the matrix ${\bf S}_i^{R,(d)}$.
We have six kinds of the graphs on the $i$-th cell
(Fig.\ \ref{fig:modelCspin}(i)-(n)).
The recursion relation is
\begin{eqnarray}
S_i^{(d)} &=&\nonumber \\
    &=& \frac{1}{2}\times A_{i-1}
       +\frac{1}{2}\times B_{i-1}
       -\frac{1}{2}\times C_{i-1}
       -\frac{1}{2}\times C_{i-1}.
\end{eqnarray}
The corresponding matrix is
$ (\frac{1}{2}\lambda^2,\frac{1}{2}\lambda^2,-\lambda^2,0)$
and is reduced to ${\bf S}_i^{R,(d)}$ in (\ref{eq:spintrc1}).

We derive the matrix ${\bf S}_i^{L,(d)}$.
We have six kinds of graphs on the $j-1$-th cell
(Fig.\ \ref{fig:modelCspin}(o)-(t)).
The recursion relation is
\begin{eqnarray}
A_{j-1}^{(d)} &=& 0 \\
B_{j-1}^{(d)} &=&\nonumber \\
    &=& -\frac{1}{2}\times 2 S_{j-2} -\frac{1}{2}\times 2 S_{j-2} \\
C_{j-1}^{(d)} &=&\nonumber \\
    &=& -\frac{1}{2}\times S_{j-2} -\frac{1}{2}\times S_{j-2} \\
C_{j-1}^{(d)} &=& 0.
\end{eqnarray}
The corresponding matrix is ${\bf S'}_{j-1}^{(d)}$ $= (0,-2,-1,0)$.
The recursion relations to the $j$-th cell are
\begin{eqnarray}
A_j^{(d)} &=& \nonumber \\
    &=& 2 \lambda^2B_{j-1} -2\lambda^2C_{j-1}
        -\lambda^2C_{j-1} \lambda^4B_{j-1}
\\
B_j^{(d)} &=& \nonumber \\
    &=& 2 \lambda^2B_{j-1} +2B_{j-1} -2 C_{j-1} -2 C_{j-1} \\
C_j^{(d)} &=& \nonumber \\
    &=& \lambda^2B_{j-1} +B_{j-1} -C_{j-1} -C_{j-1} \\
D_j^{(d)} &=& \nonumber \\
    &=& 2\lambda^2B_{j-1},
\end{eqnarray}
and the corresponding matrix is
\begin{eqnarray}
{\bf S'}_j^{(d)}&=&
 \left( \begin{array} {cccc}
         0 & 2\lambda^2 +\lambda^4 & -4 \lambda^2 & 0 \\
         0 & 2+ 2\lambda^2         & -4     & 0 \\
         0 & 1+ \lambda^2          & -2     & 0 \\
         0 & 2                     &  0     & 0
             \end{array} \right).
\end{eqnarray}
A simple calculation leads that the product
${\bf S'}_{j}^{(d)} {\bf S'}_{j-1}^{(d)}$
is reduced to
${\bf S}_{j}^{(d)} {\bf S}_{j-1}$
with the base $(A_n, C_n, D_n)^T$.
We obtain ${\bf S}_{j}^{L,(d)}$ in (\ref{eq:spintrc1}).

\subsection{Singlet-pair correlation function}
\label{app:matsinglec}

The geometric representation of the expectation value
is (\ref{eq:norm3a}),
where an example of the graph $V \cup V'$ is shown
in Figs.\ \ref{fig:modelCdemo}(e) and (f)
for (\ref{eq:classifysinglet})-(i) and (iii), respectively.
We derive the matrix ${\bf H}_n$.
On the $n$-th cell we have one kind of graph
(Fig.\ \ref{fig:modelCsingle}(a)).
Let $H_n$ be the quantity defined by a sum.
The sum is taken over $V \cup V'$ on the lattice $\Lambda_n$
such that the graph consists of the cells shown
in Fig.\ \ref{fig:modelCnorm} on the $k$-th cell ($1 \le k \le i-1$)
and that shown in Fig.\ \ref{fig:modelCsingle}(a) on the $l$-th cell
($i \le l \le n$).
It is represented diagrammatically
\begin{eqnarray}
H_n = ,
\end{eqnarray}
and the recursion relation is
\begin{eqnarray}
H_n &=&\nonumber \\
    &=& 2 H_{n-1}.
\end{eqnarray}
Therefore, we obtain ${\bf H}_n=2$.

We derive the matrix ${\bf H}_i^{R,(p)}$.
(We consider the case (\ref{eq:classifysinglet})-(i).)
We have one kind of graph on the $i$-th cell
(Fig.\ \ref{fig:modelCsingle}(b)).
The recursion relation is
\begin{eqnarray}
H_i^{(p)} &=&\nonumber \\
    &=& \lambda^2 A_{i-1} +\lambda^2 B_{i-1} +\lambda^2 B_{i-1}
        +\lambda^2 D_{i-1}.
\label{eq:Csp1}
\end{eqnarray}
The corresponding matrix is $\lambda^2 (1,2,1,0)$
and is reduced to ${\bf H}_i^{R,(p)}$ in (\ref{eq:spintrc1})
with the base $(A_n, C_n, D_n)^T$.

We derive the matrix ${\bf H}_{k+1}^{L,(p)}$.
(We consider the case (\ref{eq:classifysinglet})-(i).)
We have one kind of graph on the $k$-th cell
(\ref{fig:modelCsingle}(c)).
The recursion relation is
\begin{eqnarray}
{H'}_k^{(p)} &=&\nonumber \\
    &=& -2\lambda^2 H_{k-1}.
\end{eqnarray}
The recursion relations to the $k+1$-th cell are
\begin{eqnarray}
A_{k+1} &=& \nonumber \\
    &=& 2{H'}_k^{(p)} +2\lambda^2{H'}_k^{(p)}
       +2\lambda^2{H'}_k^{(p)} +\lambda^4{H'}_k^{(p)} \\
B_{k+1} &=& \nonumber \\
    &=& 2\lambda^2{H'}_k^{(p)} +2{H'}_k^{(p)} +2{H'}_k^{(p)}\\
C_{k+1} &=& \nonumber \\
    &=& \lambda^2{H'}_k^{(p)}  +{H'}_k^{(p)} +{H'}_k^{(p)} \\
D_{k+1} &=& \nonumber \\
    &=& 2\lambda^2{H'}_k^{(p)}.
\label{eq:Csp2}
\end{eqnarray}
Therefore, we have ${\bf H'}_{k+1}^{(p)}$ $={\bf S'}_{k+1}^{(p)}$.
A simple calculation leads that the product
${\bf H'}_{k+1}^{(p)} {\bf H'}_k^{(p)}$
is reduced to
${\bf H}_{k+1}^{L,(p)} {\bf H}_k$
with the base $(A_n, C_n, D_n)^T$.
We obtain ${\bf H}_{k+1}^{L,(p)}$ in (\ref{eq:singletrc}).

We derive the matrix ${\bf H}_i^{R,(dp)}$.
(We consider the case (\ref{eq:classifysinglet})-(iii).)
We have one kind of graph on the $i$-th cell
(Fig.\ \ref{fig:modelCsingle}(d)).
The transfer matrix is obtained by replacing the graph in (\ref{eq:Csp1})
by Fig.\ \ref{fig:modelCsingle}(d).
We find ${\bf H'}_{i}^{(dp)}={\bf H'}_{i}^{(p)}$
and obtain ${\bf H}_i^{R,(dp)}$ in (\ref{eq:singletrc}).
Similarly, we find
${\bf H'}_{k+1}^{(dp)}={\bf H'}_{k+1}^{(p)}$
using the graph Fig.\ \ref{fig:modelCsingle}(e).
We obtain ${\bf H}_{k+1}^{L,(d)}$ in (\ref{eq:singletrc}).

\subsection{
Correlation function
$\langle c_{i,\sigma}^{\phantom{\dagger}}c_{j,\sigma}^{\dagger} \rangle$
}
\label{app:prpc1}

The geometric representation of the expectation value is
(\ref{eq:greengeometric}),
where an example of the graph $V \cup V'$
for $\langle c_{i,\sigma}^{p}{c_{j,\sigma}^{p}}^{\dagger} \rangle$
is shown in Fig.\ \ref{fig:modelCdemo}(g).
For a line, we have 12 kinds of graphs
on the $n$-th cell ($i+1 \le n \le j-1$) (Fig.\ \ref{fig:modelCprop}).
We distinguish four cases
with respect to two $d$-sites at left end of the lattice $\Lambda_n$:
(i) the upper site belongs to a line and is an end of the line;
(ii) the lower site belongs to a line and is an end of the line;
(iii) both sites belong to a line and the lower site
is an end of the line;
(iv) both sites belong to a line and the upper site
is an end of the line.
Let $K_n$-$N_n$ be the quantity defined by a sum.
The sum is taken over $V \cup V'$ on the lattice $\Lambda_n$
such that the graph consists of loops shown
in Fig.\ \ref{fig:modelCnorm} on the $k$-th cell ($1 \le k \le i-1$)
and a line shown in Fig.\ \ref{fig:modelCprop}
on the $l$-th cell ($i \le k \le n$).
The restriction for the sum is that
the $n$-th cell is classified to (i) for $K_n$, (ii) for $L_n$,
(iii) for $M_n$, (iv) for $N_n$.
They are represented diagrammatically
\begin{eqnarray}
K_n &=& \\
L_n &=& \\
M_n &=& \\
N_n &=& .
\end{eqnarray}
We use the same rule as that in (\ref{eq:gtpp}),
when we attach $K_{n-1}$-$N_{n-1}$ to the $n$-th cell.
The recursion relations are
\begin{eqnarray}
K_n &=& \nonumber \\
    &=& -L_{n-1} -K_{n-1} -\lambda^2 L_{n-1} -\lambda^2 M_{n-1}
        -\lambda^2 K_{n-1} -\lambda^2 N_{n-1}\\
L_n &=& \nonumber \\
    &=& -K_{n-1} -L_{n-1} -\lambda^2 K_{n-1} -\lambda^2 N_{n-1}
        -\lambda^2 L_{n-1} -\lambda^2 M_{n-1}\\
M_n &=& \nonumber \\
    &=& -L_{n-1} -M_{n-1} -K_{n-1} -N_{n-1}\\
N_n &=& \nonumber \\
    &=& -K_{n-1} -N_{n-1} -L_{n-1} -M_{n-1}.
\end{eqnarray}
The corresponding matrix is
\begin{eqnarray}
-\left( \begin{array} {cccc}
  1 +\lambda^2  & 1 +\lambda^2  & \lambda^2  & \lambda^2 \\
  1 +\lambda^2  & 1 +\lambda^2  & \lambda^2  & \lambda^2 \\
  1             & 1             & 1          & 1         \\
  1             & 1             & 1          & 1         \\
 \end{array}
\right).
\end{eqnarray}From the equalities $K_n$ $=L_n$ and $M_n$ $=N_n$,
it is reduced to ${\bf G}_n$ in (\ref{eq:cgreen1}).

We derive ${\bf G}_i^{R,(p)}$.
We have eight kinds of graphs on the $i$-th cell
(Fig.\ \ref{fig:modelCprop}(pk1)-(pm6)).
Let $K_i^{(p)}$ and $M_i^{(p)}$ be the quantity defined by a sum.
The sum is taken over $V \cup V'$ on the lattice $\Lambda_i$
such that the graph consists of the cells shown
in Fig.\ \ref{fig:modelCnorm} on the $k$-th cell ($1 \le k \le i-1$).
The restriction for the sum is that
the $i$-th cell is represented by
Fig.\ \ref{fig:modelCprop}(pk1) or (pk2) for $K_n$
and by (pm1)-(pm6) for $M_n$.
They are represented diagrammatically
\begin{eqnarray}
K_i^{(p)} &=&  =  \\
M_i^{(p)} &=&  = .
\end{eqnarray}
We use the same rule described above (\ref{eq:gtpp})
except for that for $C_i$.
When we attach the graph $C_{i-1}$ to the $i$-th cell,
We assign one more coefficient $-1$ to cancel the coefficient $+1$
at the beginning of the line
where we regarded the right end of the line as that of a loop.
The recursion relations are
\begin{eqnarray}
K_i^{(p)} &=& \nonumber \\
          &=& -\lambda A_{i-1} -\lambda B_{i-1} -\lambda B_{i-1}
              -\lambda D_{i-1} \\
M_i^{(p)} &=& \nonumber \\
&=& -\lambda A_{i-1} -\lambda B_{i-1} -\lambda A_{i-1} -\lambda B_{i-1}
              +\lambda C_{i-1} +\lambda C_{i-1}.
\end{eqnarray}
The corresponding matrix is
\begin{eqnarray}
-\left( \begin{array} {cccc}
  \lambda   & 2\lambda  & 0         & \lambda  \\
  2\lambda  & 2\lambda  & -2\lambda & 0        \\
 \end{array}
\right),
\end{eqnarray}
and is reduced to ${\bf G}_i^{R,(p)}$ in (\ref{eq:cgreen1})
with the base $(A_n, C_n, D_n)^T$.

We derive ${\bf G}_j^{L,(p)}$.
We have ten kinds of graphs on the $j$-th cell
(Fig.\ \ref{fig:modelCprop}(a1)-(c4)).
Let $A^{(p)}_j$, $B^{(p)}_j$, and $C^{(p)}_j$ be the quantity defined
by the right-hand side of (\ref{eq:greengeometric})
on the lattice $\Lambda_j$.
The sum is taken over the graph $V \cup V'$
such that the graph consists of loops shown in Fig.\ \ref{fig:modelCnorm}
on the $k$-th cell $(1 \le k \le i-1)$
and a line shown in Fig.\ \ref{fig:modelCprop} (g1)-(g12)
on the $l$-th cell $(i \le l \le j)$.
The restriction for the sum is that the $j$-th cell
is represented by either
Fig.\ \ref{fig:modelCprop}(pa1) or (pa2) for $A^{(p)}_j$,
(pb1)-(pb4) for $B^{(p)}_j$,
and (pc1)-(pc4) for $C^{(p)}_j$.
They are represented diagrammatically
\begin{eqnarray}
A_j^{(p)} &=& \\
B_j^{(p)} &=&  =  \\
C_j^{(p)} &=&  =  .
\end{eqnarray}
When we attach the $j-1$-th graph to the $j$-th cell in $C^{(p)}$,
we assign coefficient $1/2$ to cancel weight $2$
which is assinged at end of the line
where we regard it as the left end of a loop.
The recursion relations are
\begin{eqnarray}
A_j^{(p)} &=& \nonumber \\
          &=& -\lambda K_{j-1} -\lambda K_{j-1}\\
B_j^{(p)} &=& \nonumber \\
&=& -\lambda K_{j-1} -\lambda M_{j-1} -\lambda K_{j-1} -\lambda M_{j-1}\\
C_j^{(p)} &=& \nonumber \\
          &=& -\frac{1}{2} \lambda K_{j-1} -\frac{1}{2}\lambda M_{j-1}
              -\frac{1}{2} \lambda K_{j-1} -\frac{1}{2}\lambda M_{j-1}\\
D_j^{(p)} &=& 0
\end{eqnarray}
The corresponding matrix is
\begin{eqnarray}
-\left( \begin{array} {cc}
  2\lambda  & 0        \\
  2\lambda  & 2\lambda \\
  \lambda   & \lambda  \\
  0         & 0        \\
 \end{array}
\right),
\end{eqnarray}
and is reduced to ${\bf G}_j^{L,(p)}$ in (\ref{eq:cgreen1}).

We derive ${\bf G}_i^{R,(d)}$.
Let $A_i^{(d)}$-$C_i^{(d)}$ be the quantity defined by a sum.
The sum is taken over $V \cup V'$ on the lattice $\Lambda_i$
such that the graph consists of the cells shown
in Fig.\ \ref{fig:modelCnorm} on the $k$-th cell ($1 \le k \le i-1$).
The restriction for the sum is that
the $i$-th cell is classified by the condition (i) for $A_i^{(p)}$
(ii) for $B_i^{(d)}$, and (iii) for $C_i^{(d)}$
in Appendix\ \ref{app:transfc}, respectively.
They are represented diagrammatically
\begin{eqnarray}
A_i^{(d)} &=& \\
B_i^{(d)} &=& \\
C_i^{(d)} &=& .
\end{eqnarray}
After a simple calculation we find $A_i^{(d)}$ $=A_i$,
$B_i^{(d)}$ $=B_i$, $C_i^{(d)}$ $=C_i$, and $D_i=0$.
We use the same rule for the derivation of ${\bf G}_i^{R,(p)}$.
Using the graph shown in
Fig.\ \ref{fig:modelCprop}(dk1)-(dn2)),
the recursion relations to the $i+1$-th cell are
\begin{eqnarray}
K_{i+1} &=&\nonumber \\
 &=& -\lambda^2 A_i^{(d)} -\lambda^2 B_i^{(d)} -C_i^{(d)} -A_i^{(d)}\\
L_{i+1} &=& \nonumber \\
 &=& -\lambda^2 A_i^{(d)} -\lambda^2 B_i^{(d)} -C_i^{(d)} -A_i^{(d)}\\
M_{i+1} &=& \nonumber \\
    &=& -A_i^{(d)} -B_i^{(d)} +C_i^{(d)}\\
N_{i+1} &=& \nonumber \\
    &=& -A_i^{(d)} -B_i^{(d)} +C_i^{(d)}.
\end{eqnarray}
The corresponding matrix is
\begin{eqnarray}
-\left( \begin{array} {cccc}
  1 +\lambda^2  & \lambda^2  & \lambda^2  & 0 \\
  1 +\lambda^2  & \lambda^2  & \lambda^2  & 0 \\
  1             & 1          & -1         & 0 \\
  1             & 1          & -1         & 0 \\
 \end{array}
\right).
\end{eqnarray}From the equalities $K_{i+1}=L_{i+1}$
and $M_{i+1}=N_{i+1}$,
we obtain ${\bf G}_i^{R,(d)}$ in (\ref{eq:cgreen1}).

We derive ${\bf G}_j^{L,(p)}$.
Let $K^{(d)}_j$, $L^{(d)}_j$, and $N^{(d)}_j$ be the quantity defined
by the right-hand side of (\ref{eq:propz}) on the lattice $\Lambda_j$.
The sum is taken over the graph $V \cup V'$
such that the graph consists of loops shown
in Fig.\ \ref{fig:modelCnorm}
on the $k$-th cell $(1 \le k \le i-1)$
and a line shown in Fig.\ \ref{fig:modelCprop} (g1)-(g12)
on the $l$-th cell $(i \le l \le j)$.
The restriction for the sum is that the $j$-th cell
is classified by the condition (i) for $K_i^{(p)}$
(ii) for $L_i^{(d)}$, and (iv) for $N_i^{(d)}$
in Appendix\ \ref{app:prpc1}, respectively.
They are represented diagrammatically
\begin{eqnarray}
K_j^{(d)} &=& \\
L_j^{(d)} &=& \\
N_j^{(d)} &=& .
\end{eqnarray}
We use the same rule in the derivation of ${\bf G}_j^{L,(p)}$.
Using the graph shown in
Fig.\ \ref{fig:modelCprop}(da1)-(dd1)),
the recursion relations are
\begin{eqnarray}
A_{j+1} &=& \nonumber \\
    &=& 2\lambda^2 K_j^{(d)} -\lambda^2 L_j^{(d)}
       +\lambda^4 K_j^{(d)}   +\lambda^4 N_j^{(d)}\\
B_{j+1} &=& \nonumber \\
    &=& 2\lambda^2 K_j^{(d)} +2\lambda^2 N_j^{(d)}
       +2 K_j^{(d)} - L_j^{(d)} \\
C_{j+1} &=& \nonumber \\
    &=& \lambda^2 K_j^{(d)} +\lambda^2 N_j^{(d)}
       +K_j^{(d)} -\frac{1}{2} L_j^{(d)} \\
D_{j+1} &=& \nonumber \\
    &=& 2 K_j^{(d)} +2 N_j^{(d)}.
\end{eqnarray}
The corresponding matrix is
\begin{eqnarray}
 \left( \begin{array} {cccc}
  2\lambda^2 +\lambda^4  & -\lambda^2    & 0  & \lambda^4   \\
  2 +2\lambda^2          & -1            & 0  & 2\lambda^2  \\
  1 +\lambda^2           & -\frac{1}{2}  & 0  & \lambda^2   \\
  2                      & 0             & 0  & 2           \\
 \end{array}
\right).
\end{eqnarray}From the equality $B_n$ $=2C_n$,
the matrix is reduced to ${\bf G}_j^{L,(p)}$ in (\ref{eq:cgreen1})
with the base $(A_n, C_n, D_n)^T$.

\begin{figure}
\caption{
The correspondence between the sites $x$ ($\in$ $\Lambda_N$)
and $r$ ($\in$ $C_n$).
\label{fig:sitemap}
}
\end{figure}

\begin{figure}
\caption{
Diagrammatic representation of the valence bonds.
The solid (broken) line represents a valence bond in the ket (bra).
We distinguish four kinds of bonds
and call them as non-closed ``ket-bond'' (a), non-closed ``bra-bond'' (b),
self-closed ``ket-bond'' (c), and self-closed ``bra-bond'' (d).
A $d$-site with $U=\infty$ is denoted by a solid circle,
and a $p$-site with $U=0$ is denoted by a circle.
\label{fig:valencebond}
}
\end{figure}

\begin{figure}
\caption{
Examples of valence-bond configurations (a) $V$, (b) $V'$,
and (c) their overlap $V \cup V'$.
The lattice is constructed from a cell with one $d$-site
and four $p$-sites.
$P$-sites in adjacent cells actually is a single $p$-site.
The thin broken lines represent hoppings of electrons.
The graph $V \cup V'$ is factorized
into four connected subgraphs (d)-(g).
\label{fig:bondconfig}
}
\end{figure}

\begin{figure}
\caption{
Diagrammatic representation of the identity
$b_{x,y}^{\dagger}b_{y,z}^{\dagger}=-b_{y,y}^{\dagger}b_{x,z}^{\dagger}$.
\label{fig:elimination1}
}
\end{figure}

\begin{figure}
\caption{
Examples of the elimination of a $p$-site with four bonds,
which exhaust all the cases.
For (a), we apply the identity
$b_{x,y}^{\dagger}b_{y,z}^{\dagger}=-b_{y,y}^{\dagger}b_{x,z}^{\dagger}$
once.
For (b) and (c), we apply it two times.
\label{fig:elimination2}
}
\end{figure}

\begin{figure}
\caption{
Example of the procedure of the elimination of the $p$-sites
with four bonds.
We have self-closed bonds (a), degenerate loop (b),
and non-degenerate loop (c) after the procedure.
\label{fig:elimination3}
}
\end{figure}

\begin{figure}
\caption{
Example of graph
where sites $x$ and $y$ belong to a single subgraph.
\label{fig:geospin}
}
\end{figure}

\begin{figure}
\caption{
(a), (b) Type (i) configuration.
(c), (d) Type (ii) configuration.
\label{fig:geoprop}
}
\end{figure}

\begin{figure}
\caption{
Diagrammatic  representation of the identity
$c_{x,\sigma}^{\dagger}b_{x,y}^{\dagger}
=-c_{y,\sigma}^{\dagger}b_{y,y}^{\dagger}$ (d).
A pentagon represents the operator $c_{x,\sigma}^{\dagger}$.
\label{fig:elimination4}
}
\end{figure}

\begin{figure}
\caption{
(a) and (b) Examples of the procedure
of the elimination of type (ii) configuration.
(c) Example of a line.
\label{fig:elimination5}
}
\end{figure}

\begin{figure}
\caption{
(a) Type (i) configuration.
(b) Type (ii) configuration.
\label{fig:geonumber}
}
\end{figure}

\begin{figure}
\caption{
(a) Type (i) configuration.
(b) Type (ii) configuration.
(c) Type (iii) configuration.
(d) Type (iv) configuration.
\label{fig:geoden}
}
\end{figure}

\begin{figure}
\caption{
Model A.
(a) A cell compose of two $d$-sites and one $p$-site.
(b) The lattice constructed from the cell with cell labelings.
(c) The same lattice as (b)
drawn differently with unit cell labelings.
\label{fig:modelAcell}
}
\end{figure}

\begin{figure}
\caption{
The dispersion relations $E_+$ and $E_-$.
(a) The parameters are $\lambda_1=\lambda_2=1$.
(b) The parameters are $\lambda_1=\lambda_2=\frac{1}{\sqrt{2}}$.
\label{fig:modelaband}
}
\end{figure}

\begin{figure}
\caption{
Examples of configuration of the valence-bonds
in the geometric representation
for (a) the ground state,
(b) the norm of the ground state,
(c) the expectation value of the number operator $n_{i,\sigma}^p$,
(d) the density correlation function for $p$-sites,
(e) the correlation functions
$\langle c_{i,\sigma}^p c_{j\sigma}^{p \dagger}\rangle$
and (f) the correlation function
$\langle c_{i,\sigma}^d c_{j\sigma}^{d \dagger}\rangle$.
\label{fig:modelAgeorep}
}
\end{figure}

\begin{figure}
\caption{
Configurations of the valence-bonds on a cell.
The weights including the
contributions from $\lambda(U)\lambda(U')$ are (a) $2\lambda_2^2$,
(b) $1$, (c) $2\lambda_1^2 \lambda_2^2$, and (d) $2\lambda_1^2$.
\label{fig:modelAnorm}
}
\end{figure}

\begin{figure}
\caption{
Configurations of the valence-bonds on the $i$-th cell
in the geometric representation
for $\langle n_{i,\sigma}^p \rangle$ (a)-(c)
and $\langle n_{i,\sigma}^d \rangle$ (d) and (e).
The weights including the contributions from $\lambda(U)\lambda(U')$
are (a) $\lambda_2^2$, (b) 1, (c) $\lambda_1^2$,
(d) $\lambda_2^2$, (e) $\lambda_1^2 \lambda_2^2$,
and (f) $\lambda_1^2$.
\label{fig:modelAoccp}
}
\end{figure}

\begin{figure}
\caption{
Occupation on the $p$-site (a) and the $d$-site (b)
for $\lambda_2=$ $0.1$, $0.5$, $1$, $2$, and $4$ (solid line)
and $\lambda_2=\lambda_1$ (broken line).
\label{fig:modelA:occp:res}
}
\end{figure}

\begin{figure}
\caption{
Density correlation function for the nearest-neighbor $p$-sites
for $\lambda_2=$ $0.5$, $1$, $2$, $4$, and $8$ (solid line)
and $\lambda_2=\lambda_1$ (broken line).
\label{fig:modelA:denp:res}
}
\end{figure}

\begin{figure}
\caption{
Density correlation function for the nearest-neighbor $d$-sites
for $\lambda_2=$ $0.5$, $1$, $2$, and $4$ (solid line)
and $\lambda_2=\lambda_1$ (broken line).
\label{fig:modelA:dend:res}
}
\end{figure}

\begin{figure}
\caption{
Configurations of the valence-bonds  on the $n$-th cell (a) ($i+1 \le n \le
j-1$)
in the geometric representation for
$\langle c_{i,\sigma}^{\alpha}c_{j,\sigma}^{\alpha \dagger}\rangle$.
Those
on the $i$-th cell (b) and (c), and the $j$-th cell (d) and (e)
for $\alpha = p$,
and on the $i$-th cell (f) and the $j-1$-th cell (g)
for $\alpha = d$.
The contributions from $\lambda(U)\lambda(U')$
are (a) $\lambda_1 \lambda_2$, (b) $\lambda_1$, (c) $\lambda_1 \lambda_2^2$,
(d) $\lambda_2$, (e) $\lambda_1^2 \lambda_2$, (f) $\lambda_1 \lambda_2$,
and (g) $\lambda_1 \lambda_2$.
\label{fig:modelAprop}
}
\end{figure}

\begin{figure}
\caption{
Momentum distribution functions for the $p$-site (a)
and the $d$-site for (b)
for $\lambda_1=\lambda_2=$ $0.1$, $0.2$, $0.5$, $1$, $2$, and $100$.
\label{fig:modelA:res:mom1}
}
\end{figure}

\begin{figure}
\caption{
Momentum distribution functions for the $p$-site (a)
and the $d$-site (b)
for $\lambda_2=$ $0.01$, $0.1$, $0.5$, $1$, $2$, and $100$
with $\lambda_1=1$.
\label{fig:modelA:res:mom2}
}
\end{figure}

\begin{figure}
\caption{
Correlation lengths of the correlation function
$\langle c_{i,\sigma}^{\phantom{\dagger}}c_{j,\sigma}^{\dagger} \rangle$ (a)
and that of the density correlation function (b)
for $\lambda_1=\lambda_2$.
\label{fig:modelA:corrleng}
}
\end{figure}

\begin{figure}
\caption{
Model B.
(a) A cell compose of two $p$-sites and one $d$-site.
(b) The lattice constructed from the cell with cell labelings.
\label{fig:modelBcell}
}
\end{figure}

\begin{figure}
\caption{
The dispersion relations $E_+$ and $E_-$.
The parameters are $\lambda_1=\lambda_2=1$.
\label{fig:modelbband}
}
\end{figure}

\begin{figure}
\caption{
Occupation on the $p$-site (a)
and the $d$-site (b)
for $\lambda_2=$ $1$, $2$, $4$, and $8$ (solid line)
and $\lambda_2=\lambda_1$ (broken line).
\label{fig:modelB:occ:res}
}
\end{figure}

\begin{figure}
\caption{
Density correlation function for the nearest-neighbor $p$-sites
for $\lambda_2=$ $0.5$, $1$, $2$, $4$, and $8$ (solid line)
and $\lambda_2=\lambda_1$ (broken line).
\label{fig:modelB:den:res1}
}
\end{figure}

\begin{figure}
\caption{\addtolength{\baselineskip}{-1mm}
Density correlation function for the nearest-neighbor $d$-sites
for $\lambda_2=$ $0.5$, $1$, $2$, $4$, and $8$ (solid line)
and $\lambda_2=\lambda_1$ (broken line).
\label{fig:modelB:den:res2}
}
\end{figure}

\begin{figure}
\caption{\addtolength{\baselineskip}{-1mm}
Spin correlation functions for the nearest-neighbor $p$-sites
(a) for $\lambda_2=$ $2$, $4$, and $8$ (solid line),
$d$-sites (b) for $\lambda_2=$ $1$, $2$, and $4$ (solid line),
and $p$- and $d$-sites (c)
for $\lambda_2=$ $1$, $2$, $4$, and $8$ (solid line).
The broken lines are for $\lambda_2=\lambda_1$.
\label{fig:modelB:spin:res}
}
\end{figure}

\begin{figure}
\caption{
Singlet-pair correlation function for the nearest-neighbor $p$-sites
for $\lambda_2=$ $1$, $2$, $4$, and $8$ (solid line)
and $\lambda_2=\lambda_1$ (broken line).
\label{fig:modelB:single:res}
}
\end{figure}

\begin{figure}
\caption{
Momentum distribution functions for the $p$-site (a) and the $d$-site (b)
for $\lambda_1=\lambda_2=$ $0.01$, $1$, $10$, and $100$.
\label{fig:modelB:mom:res1}
}
\end{figure}

\begin{figure}
\caption{
Momentum distribution functions for the $p$-site (a) and the $d$-site (b)
for $\lambda_2=$ $0.01$, $0.1$, $1$, $5$, and $10$
with $\lambda_1=1$.
\label{fig:modelB:mom:res2}
}
\end{figure}

\begin{figure}
\caption{
Correlation lengths of the correlation function
$\langle c_{i,\sigma}^{\phantom{\dagger}}c_{j,\sigma}^{\dagger} \rangle$ (a),
that of the density correlation function (b),
and that of the spin and the singlet-pair correlation functions (c)
for $\lambda_1=\lambda_2$.
\label{fig:modelB:corrleng}
}
\end{figure}

\begin{figure}
\caption{
Model C.
(a) A cell compose of one $p$-sites and four $d$-site.
(b) The lattice constructed from the cell with cell labelings.
\label{fig:modelCcell}
}
\end{figure}

\begin{figure}
\caption{
The dispersion relations $E_1$, $E_2$, and $E_3$.
The parameters are (a) $\lambda=1$,
(b) $\lambda=\sqrt{2}$,
and (c) $\lambda=2$.
\label{fig:modelcband}
}
\end{figure}

\begin{figure}
\caption{
Occupation on the $p$-site (a)
and the $d$-site (b).
\label{fig:modelC:occ:res}
}
\end{figure}

\begin{figure}
\caption{
Density correlation functions for the nearest-neighbor
$p$-sites (a) and $d$-sites (b).
\label{fig:modelC:den:res}
}
\end{figure}

\begin{figure}
\caption{
Spin correlation functions for the nearest-neighbor
$p$-sites (a), $p$- and $d$-sites (b), and $d$-sites (c).
\label{fig:modelC:spin:res}
}
\end{figure}

\begin{figure}
\caption{
Singlet-pair correlation function for the nearest-neighbor $p$-sites.
\label{fig:modelC:single:res}
}
\end{figure}

\begin{figure}
\caption{
Momentum distribution functions for the $p$-site (a) and the $d$-site (b)
for $\lambda=$ $0.1$, $0.5$, $1$, $2$, and $10$.
\label{fig:modelC:mom:res1}
}
\end{figure}

\begin{figure}
\caption{
Correlation lengths of the correlation function
$\langle c_{i,\sigma}^{\phantom{\dagger}}c_{j,\sigma}^{\dagger} \rangle$ (a),
that of the density correlation function (b),
and that of the spin
and the singlet-pair correlation functions (c).
\label{fig:modelC:corrleng}
}
\end{figure}

\begin{figure}
\caption{
(a) The same lattice as Fig.\ \ref{fig:modelBcell}(b)
drown differently with unit cell labelings.
Examples of the configurations of the valence-bonds
in the geometric representation for
(b) the ground state,
(c) the norm of the ground state,
(d) the norm of the ground state after the elimination
    of $p$-sites with four bonds,
(e) the expectation value of the number operator $n_{i,\sigma}^p$,
(f) the spin correlation function
    $\langle S_i^{z,p} S_j^{z,p} \rangle$,
(g) the singlet-pair correlation functions
    $\langle b_{i,i}^{\dagger} b_{k,k}^{\phantom{\dagger}} \rangle$
and (h) $\langle b_{i,j}^{\dagger} b_{k,l}^{\phantom{\dagger}} \rangle$,
and (i)the correlation function
    $\langle c_{i,\sigma}^p c_{j,\sigma}^{p \dagger}\rangle$.
(j) Examples of the loops extending over more than two cells.
\label{fig:modelBdemo}
}
\end{figure}

\begin{figure}
\caption{
Configurations of the valence-bond on a cell.
The contributions from $\lambda(U)\lambda(U')$ are
(a) $\lambda_2^2$, (b) $\lambda_2^4$, (c) $\lambda_1^2 \lambda_2^2$,
(d) $\lambda_1^2$, and (e) $\lambda_1^4$.
\label{fig:modelBnorm}
}
\end{figure}

\begin{figure}
\caption{
Configurations of the valence-bonds on the $i$-th cell
in the geometric representation for $\langle n_{i,\sigma}^p \rangle$
(a)-(d) and of $\langle n_{i,\sigma}^d \rangle$ (e)-(g).
The contributions from $\lambda(U)\lambda(U')$ are
(a) $\lambda_2^2$, (b) $\lambda_2^4$, (c) $\lambda_1^2 \lambda_2^2$,
(d) $\lambda_1^4$,
(e) $\lambda_2^2$, (f) $\lambda_1^2 \lambda_2^2$,
and (g) $\lambda_1^2$.
\label{fig:modelBocc}
}
\end{figure}

\begin{figure}
\caption{
Configurations of the valence-bonds on the $i$-th cell (a) and the $j-1$-th
cell (b)
in the geometric representation for
$\langle S_i^{z, p} S_j^{z, p} \rangle$
and on the $i$-th cell (c) and the $j$-th cell (d)
for $\langle S_i^{z, d} S_j^{z, d} \rangle$.
The contributions from $\lambda(U)\lambda(U')$ are
(a) $\lambda_1^2 \lambda_2^2$, (b) $\lambda_1^2 \lambda_2^2$,
(c) $\lambda_1^2$, and (d) $\lambda_2^2$.
\label{fig:modelBspin}
}
\end{figure}

\begin{figure}
\caption{
Configurations of the valence-bonds on the $n$-th cell (a) ($i+1 \le n \le
j-1$)
in the geometric representation for the singlet-pair correlation functions.
Those on the $i$-th cell (b) and the $k$-th cell (c) and (d) for
$\langle b_{i,i}^{\dagger} b_{k,k}\rangle$,
and on the $i$-th cell (e) and the $k$-th cell (f)
in $\langle b_{i,j}^{\dagger} b_{k,l}\rangle$.
The contributions from $\lambda(U)\lambda(U')$ are
(a) $\lambda_1^2 \lambda_2^2$,
(b) $\lambda_1^2 \lambda_2^2$,
(c) $\lambda_1^2$,
(d) $\lambda_1^4$,
(e) $\lambda_1^2 \lambda_2$, and
(f) $\lambda_1 \lambda_2^2$.
\label{fig:modelBsingle}
}
\end{figure}

\begin{figure}
\caption{
Configurations of the valence-bonds on the $n$-th cell (a)-(c)
($i+1 \le n \le j-1$) in the geometric representation for
$\langle c_{i,\sigma}^{\alpha}c_{j,\sigma}^{\alpha \dagger}\rangle$.
Those on the $i$-th cell (d)-(f)
and the $j$-th cell (g)-(j)
for $\alpha = p$,
and on the $i$-th cell (k) and (l)
and on the $j$-th cell (m) and (n)
for $\alpha = d$.
The contributions from $\lambda(U)\lambda(U')$ are
(a) $\lambda_1 \lambda_2$, (b) $\lambda_1^3 \lambda_2$,
(c) $\lambda_1 \lambda_2^3$, (d) $\lambda_1 \lambda_2$,
(e) $\lambda_1 \lambda_2^3$, (f) $\lambda_1^3 \lambda_2$,
(g) $\lambda_2^2$, (h) $\lambda_1^2 \lambda_2^2$,
(i) $\lambda_1^2$, (j) $\lambda_1^4$,
(k) $\lambda_1 \lambda_2^2$, (l) $\lambda_1^3$,
(m) $\lambda_2^3$, and (n) $\lambda_1^2 \lambda_2$.
\label{fig:modelBprop}
}
\end{figure}

\begin{figure}
\caption{
Examples of the configurations of the valence-bonds
in the geometric representation for
(a) the ground state,
(b) the norm of the ground state,
(c) the expectation value of the number operator $n_{i,\sigma}^p$,
(d) the spin correlation function
    $\langle S_i^{z, p} S_j^{z, d} \rangle$,
the singlet-pair correlation functions
    (e) $\langle b_{i,i}^{\dagger} b_{k,k}^{\phantom{\dagger}} \rangle$ and
    (f) $\langle b_{i,j}^{\dagger} b_{k,l}^{\phantom{\dagger}} \rangle$,
and
(g) the correlation function
    $\langle c_{i,\sigma}^p c_{j,\sigma}^{p \dagger}\rangle$.
\label{fig:modelCdemo}
}
\end{figure}

\begin{figure}
\caption{
Configurations of the valence-bonds on a cell.
The contributions from $\lambda(U)\lambda(U')$ are
$1$ for the valence bonds which do not belong to a $p$-site,
$\lambda^2$ for the valence bonds which belong to a $p$-site
(except for (a6)), and
$\lambda^4$ for (a6).
\label{fig:modelCnorm}
}
\end{figure}

\begin{figure}
\caption{
Configurations of the valence-bonds on the $n$-th cell
(a)-(d) ($i+1 \le n \le j-1$)
in the geometric representation for the spin correlation functions.
Those on the $i$-th cell (e) and (f) and on the $j$-th cell (g) and (h)
for $\langle S_i^{z, p} S_j^{z, p} \rangle$,
and on the $i$-th cell (i)-(n)
and on the $j$-th cell (o)-(t) for $\langle S_i^{z, d} S_j^{z, d} \rangle$.
\label{fig:modelCspin}
}
\end{figure}

\begin{figure}
\caption{
Configurations of the valence-bonds on the $n$-th cell
(a) ($i+1 \le n \le j-1$)
in the geometric representation for the singlet-pair correlation functions.
Those on the $i$-th cell (b) and the $k$-th cell (c)
for $\langle b_{i,i}^{\dagger} b_{k,k}^{\phantom{\dagger}} \rangle$,
and
on the $i$-th cell (d) and the $u$-th cell (e)
for $\langle b_{i,j}^{\dagger} b_{k,l}^{\phantom{\dagger}} \rangle$.
\label{fig:modelCsingle}
}
\end{figure}

\begin{figure}
\caption{
Configurations of the valence-bonds on the $n$-th cell ($i+1 \le n \le j-1$)
in the geometric representation for the correlation function
$\langle c_{i,\sigma}^{\phantom{\dagger}}c_{j,\sigma}^{\dagger} \rangle$.
Those on the $i$-th cell (pk1)-(pm6) and on the $j$-th cell (pa1)-(pc4)
for $\langle c_{i,\sigma}^p c_{j,\sigma}^{p \dagger} \rangle$,
and
on the $i$-th cell (dk1)-(dn2) and on the $j$-th cell (da1)-(dd2)
for $\langle c_{i,\sigma}^d c_{j,\sigma}^{d \dagger} \rangle$,
\label{fig:modelCprop}
}
\end{figure}

\begin{table}
\caption{
The cell construction of the models
in Refs.\ \cite{BG92,Mi92,St93,Ta93a,Ta93b}.  Parameters (a) are those
of Refs.\ \cite{BG92,Mi92,St93}.  When we choose $\lambda_j$'s, which
are defined in the cell, those in (b), the parameters (c) represents
those in column (a).  The model with boundary condition in the last
column satisfy the uniqueness condition in Ref.\ \cite{Ta93b}.  Here
$O$ and $P$ denote open and periodic boundary conditions,
respectively.  The symbol $D$ denotes that the boundary condition
depends on the model.
}
\label{table:models}
\begin{minipage}{165mm}
\begin{tabular}{lccclcc}
\scriptsize
author             &
cell               &
dimension          &
                   &
$~~~~~$parameters  &
                   &
uniqueness         \\
&
&
&
(a)&
(b)&
(c)&  \\
%
\hline
U. Brandt                          &
  &
$d \ge 2$                                         &
$(t)$                                             &
$\lambda_j=\lambda$ $(j=1,2,3,4)$                 &
$(-\lambda^2)$                                    &
$O$ $(d\ge3)$ \\
and A. Giesekus                                   &
 &
$d \ge 2$                                         &
$\frac{V}{t}$                                     &
$\lambda_j=\frac{1}{\sqrt{2}}$ $(j=1,2,3,4)$, $\lambda_5=\sqrt{2}\lambda$  &
$-2\lambda$                                       &
$O$, $P$ \\
\cline{2-7}
A. Mielke                                         &
see text                                          &
$d \ge 1$                                         &
$(t)$                                             &
$\lambda_j=\lambda$                               &
$(-\lambda^2)$                                    &
$D$                         \\
\cline{2-7}
R. Strack                                         &
  &
$d=1$                                             &
$\frac{V}{t}$                                     &
$\lambda_1=\lambda_2=1$, $\lambda_3=\lambda i$, $\lambda_4=-\lambda i$  &
$\lambda$                                         &
$O$                       \\
                                                  &
 &
$d=1$                                             &
$\frac{V}{t}$                                     &
$\lambda_1=\lambda$, $\lambda_2=-\lambda$, $\lambda_3=1$ &
$\frac{1}{\lambda}$                               &
$O$                        \\
                                                  &
  &
$d=2$                                             &
$\frac{V}{t}$                                     &
$\lambda_1=\lambda$, $\lambda_2=-\lambda$, $\lambda_3=1$ &
$\frac{1}{\lambda}$                               &
$O$, $P$                          \\
\cline{2-7}
H. Tasaki                                         &
$d$-sites                                         &
$d \geq 1$                                        &
$\lambda_j$                                       &
$\lambda_j$                                       &
$\lambda_j$                                       &
$D$                           \\
\cline{2-7}H. Tasaki                              &
$d$- and $p$-sites                                &
$d \geq 1$                                        &
$\lambda_j$                                       &
$\lambda_j$                                       &
$\lambda_j$                                       &
$D$                           \\
%
\hline
Model A                                           &
  &
$d = 1$                                           &
                                                  &
$\lambda_1$, $\lambda_2$, $\lambda_3=1$           &
                                                  &
$O$, $P$                           \\
\cline{2-7}
Model B                                           &
  &
$d = 1$                                           &
                                                  &
$\lambda_1$, $\lambda_2$, $\lambda_3=1$           &
                                                  &
$O$                           \\
\cline{2-7}
Model C                                           &
  &
$d = 1$                                           &
                                                  &
$\lambda_j=1$ $(j=1,2,3,4)$, $\lambda_5=\lambda$  &
                                                  &
$O$, $P$                           \\
%
%
\end{tabular}
\end{minipage}
\vspace{12cm}
\end{table}

\end{document}